\documentclass[11pt,a4paper]{article}
\pdfoutput=1

\usepackage{amsmath,amssymb}
\usepackage{epsfig,graphicx}
\usepackage{subfigure}
\usepackage{graphicx}
\usepackage{rotating}
\usepackage{cancel}
\usepackage{bm}
\usepackage{color}
\usepackage{comment}
\usepackage{cite}
\usepackage{psfrag}

\usepackage{amsmath,amsfonts,braket,slashed,amssymb,tikz,bm,psfrag,graphicx,color,dsfont,tikz}
\usepackage{multicol, multirow}
\usepackage{ctable}
\usepackage{setspace,longtable,tikz,times,booktabs}

\usepackage{cite}

\renewcommand\({\left(}

\renewcommand\[{\left[}

\usepackage{pifont}

\usepackage[
	colorlinks=true,
	urlcolor=blue,
	linkcolor=red,
	citecolor=blue
]{hyperref}

\usepackage{tikz-cd}

\usepackage{xcolor,colortbl}
\definecolor{Gray}{gray}{0.9}

\begin{document}

\vspace{1.0cm}

\begin{center}
\Large\bf\boldmath
Hunting All the Hidden Photons
\end{center}

\vspace{0.2cm}
\begin{center}
Martin Bauer, Patrick Foldenauer and Joerg Jaeckel, \\
\vspace{0.7cm} 
{\sl Institut f\"ur Theoretische Physik, Universit\"at Heidelberg,
Philosophenweg 16, 69120 Heidelberg, Germany}\\[.4cm]
\textbf{Abstract}\\[3mm] 
\parbox{0.9\textwidth}{
We explore constraints on gauge bosons of a weakly coupled $U(1)_{B-L}$, $U(1)_{L_\mu-L_e}$, $U(1)_{L_e-L_\tau}$ and $U(1)_{L_\mu-L_\tau}$. To do so we apply the full constraining power of experimental bounds derived for a hidden photon of a secluded $U(1)_{X}$ and translate them to the considered gauge groups. In contrast to the secluded hidden photon that acquires universal couplings to charged Standard Model particles through kinetic mixing with the photon, for these gauge groups the couplings to the different Standard Model particles can vary widely. We take finite, computable loop-induced kinetic mixing effects into account, which provide additional sensitivity in a range of experiments. 
In addition, we collect and extend limits from neutrino experiments as well as astrophysical and cosmological observations and include new constraints from white dwarf cooling. We discuss the reach of future experiments in searching for these gauge bosons.
}
\end{center}

\vspace{0.cm}

\section{Introduction}\label{sec:intro}

The Standard Model (SM) provides a complete and very successful description of particle physics at the electroweak scale. Lorentz invariance and gauge symmetry strongly constrain the number of renormalizable interactions between SM fields and a possible New Physics sector. One of these renormalizable interactions is induced by kinetic mixing between a new $U(1)_X$ gauge boson $X_\mu$ and the hypercharge $U(1)_Y$ gauge boson $B_\mu$ through the operator \cite{Holdom:1985ag}
\begin{align}
\mathcal{L}=-\frac{\epsilon'}{2}F_{\mu\nu}X^{\mu\nu}\,,
\end{align}
connecting the corresponding field strength tensors, $X^{\mu\nu}$ and $F^{\mu\nu}$, respectively. Upon redefining the fields and rotating to the mass eigenstates, the new gauge boson acquires a coupling to the hypercharge current proportional to $\epsilon'$. For small $\epsilon'$ and light $U(1)_X$ gauge bosons, its couplings mostly align with the couplings of the SM photon and are suppressed by a factor $\epsilon'$. This motivates the name "hidden" or "dark photon" for $X_\mu$. 

Yet, there is also the possibility that one of the remaining global symmetries of the SM is gauged. Since only anomaly-free symmetries can be gauged, the number of possible additional gauge groups in the SM without the introduction of additional fermions charged under $SU(3)_C\times SU(2)_L\times U(1)_Y$ is limited. Out of the four independent global symmetries of the SM Lagrangian, $U(1)_B$, $U(1)_{L_e}$, $U(1)_{L_\mu}$, $U(1)_{L_\tau}$ three combinations are anomaly-free without any additional particles,  $U(1)_{L_\mu-L_e}$, $U(1)_{L_e-L_\tau}$ and $U(1)_{L_\mu-L_\tau}$ \cite{Foot:1990mn, He:1990pn, He:1991qd}. 
The difference between baryon and lepton number $U(1)_{B-L}$ is also anomaly-free if right-handed neutrinos are introduced. Differences between baryon family numbers, e.g. $U(1)_{B_1-B_3}$ or combinations, e.g. $U(1)_{B_3-L_\tau}$, are also anomaly-free, but result in an unviable CKM matrix. The addition of right-handed neutrinos allows to reproduce a phenomenologically viable lepton mixing matrix without charged lepton flavour changing couplings for the $U(1)_{L_\mu-L_e}$, $U(1)_{L_e-L_\tau}$, and $U(1)_{L_\mu-L_\tau}$ gauge groups \cite{Heeck:2011wj}.

In this paper we focus on the four anomaly-free groups, $U(1)_{L_\mu-L_e}$, $U(1)_{L_e-L_\tau}$, $U(1)_{L_\mu-L_\tau}$  and $U(1)_{B-L}$.
The phenomenology of a possible gauge boson of these gauge groups can be very different from that of a secluded hidden photon. For example, at tree level the considered gauge bosons would not couple to the $W^\pm$ gauge bosons. Moreover, the gauge bosons of charged lepton family number differences would purely couple to the respective charged leptons and neutrinos and not to baryons.  
As a consequence, constraints on hidden photons for which universal couplings are assumed do not directly translate into constraints on such additional gauge bosons and new constraints from neutrino experiments arise.

In the absence of kinetic mixing, constraints on light $U(1)_{B-L}$ gauge bosons have been discussed in 
\cite{
Kopp:2012dz,
Heeck:2014zfa,
Bilmis:2015lja,
Jeong:2015bbi,
Ilten:2018crw
}. 
The groups $U(1)_{L_\mu-L_e}$, $U(1)_{L_e-L_\tau}$  are considered in~\cite{Wise:2018rnb} and limits on $U(1)_{L_\mu-L_\tau}$ gauge bosons have been derived in 
\cite{
Altmannshofer:2014cfa,
Altmannshofer:2014pba,
Kamada:2015era,
Araki:2015mya,
Araki:2017wyg,
Kaneta:2016uyt,
Gninenko:2018tlp
} 
(the last two papers take into account kinetic mixing).
However, since in all four cases SM fermions are charged under both the new U(1) as well as under hypercharge, a kinetic mixing term between the new gauge boson and the hypercharge boson is automatically induced at one-loop, even if $\epsilon'=0$ at tree-level. In the case of $U(1)_{L_\mu-L_e}$, $U(1)_{L_e-L_\tau}$, and $U(1)_{L_\mu-L_\tau}$  this mixing term is finite and has significant impact on the experimental sensitivities. We will discuss this mixing in more detail in Section~\ref{sec:models}.

The central aim of this paper is to use experiments and observations searching for hidden photons to derive limits on $U(1)_{L_\mu-L_e}$, $U(1)_{L_e-L_\tau}$, $U(1)_{L_\mu-L_\tau}$   and $U(1)_{B-L}$ gauge bosons. 
We use a large set of experiments ranging from beam-dump and fixed target experiments~
\cite{
Bergsma:1985is,
Riordan:1987aw, 
Bjorken:1988as, 
Bross:1989mp,
Davier:1989wz,
MeijerDrees:1992kd,
Athanassopoulos:1997er,
Aguilar:2001ty,
Bjorken:2009mm,
Essig:2010gu,
Essig:2010xa,
Merkel:2011ze,
Blumlein:2011mv,
Abrahamyan:2011gv,
Andreas:2012mt,
Gninenko:2012eq,
Wojtsekhowski:2012zq,
Kahn:2012br,
Blumlein:2013cua,
Merkel:2014avp,
Battaglieri:2014hga,
Balewski:2014pxa,
Batley:2015lha,
Anelli:2015pba, 
Alekhin:2015byh,
Gardner:2015wea,
Banerjee:2016tad
},
$e^+e^-$ colliders~
\cite{
Aubert:2009cp,
Abe:2010gxa,
Archilli:2011zc,
Babusci:2012cr,
Babusci:2014sta,
Curtin:2014cca,
Gninenko:2014pea,
Lees:2014xha,
Ilten:2015hya,
Ilten:2016tkc,
Inguglia:2016acz,
Anastasi:2016ktq,
Aaij:2017rft,
Lees:2017lec,
Gninenko:2018tlp
},  
to lepton precision experiments~
\cite{
Bertl:1985mw,
Alam:1995mt,
Bouchendira:2010es,
Davoudiasl:2012ig,
Endo:2012hp,
Blondel:2013ia,
Echenard:2014lma
}.
In addition we consider the experimental measurements of solar neutrinos with Borexino~\cite{Bellini:2011rx,Harnik:2012ni,Kaneta:2016uyt}, laboratory neutrino experiments such as, e.g. CHARM-II~\cite{Vilain:1994qy,Vilain:1993kd}, COHERENT~\cite{Akimov:2015nza, Akimov:2017ade} and TEXONO~\cite{Deniz:2009mu} as well as tests of neutrino trident production~\cite{Altmannshofer:2014cfa,Geiregat:1990gz,Mishra:1991bv,Adams:1998yf}. Furthermore, we include new astrophysical limits from the energy loss of white dwarfs~\cite{Dreiner:2013tja}.
We also discuss the parameter space where the measured deviation in the anomalous magnetic moment of the muon can be explained~\cite{Pospelov:2008zw,Altmannshofer:2014pba}. 
Looking into the future we consider projections for planned and proposed experiments for the case of the universal hidden photon, the $U(1)_{L_\mu-L_e}$, $U(1)_{L_e-L_\tau}$, and $U(1)_{L_\mu-L_\tau}$  gauge bosons as well as the $U(1)_{B-L}$ gauge boson, respectively. \\
In spirit our paper is similar to the recent recasting framework provided in~\cite{Ilten:2018crw}, where in particular also $U(1)_{B-L}$ is considered. However, we consider in addition the theoretically as well as phenomenologically particularly interesting case of the three lepton family groups. In particular, additional  $U(1)_{L_\mu-L_\tau}$ gauge bosons have recently received increasing attention. A light $U(1)_{L_\mu-L_\tau}$ gauge boson provides one of the few not yet excluded light new physics explanations for the discrepancy between the SM prediction and the experimental determination of the anomalous magnetic moment of the muon \cite{Ma:2001md, Harigaya:2013twa}, can explain the spectrum of the Icecube high-energy neutrino events \cite{Araki:2014ona, DiFranzo:2015qea}, and has the right quantum numbers to explain the hints of lepton flavour non-universality reported by LHCb \cite{Altmannshofer:2014cfa, Crivellin:2015lwa, Altmannshofer:2016jzy }.

For  $U(1)_{L_\mu-L_e}$, $U(1)_{L_e-L_\tau}$, and $U(1)_{L_\mu-L_\tau}$,  we explicitly take into account the unavoidable kinetic mixing generated by the Standard Model particles, that has significant effects on the sensitivities. Technically, where possible and necessary, we recreated the analysis of the experiments thereby making use of more detailed information such as the energy spectrum of the particles in the experiments. 

The remainder of this paper is organized as follows: In Section~\ref{sec:models} we introduce  the considered hidden gauge bosons and discuss their phenomenological features. In Section~\ref{sec:experiments} we discuss the strategies employed to recast or rederive existing limits and projections for future experiments. We present the results of our analysis in Section~\ref{sec:results} and conclude in Section~\ref{sec:conclusions}. In the Appendices we provide additional details. In Appendix~\ref{rotation} we discuss in detail the rotation to the mass eigenstates as well as Higgs interactions. A detailed discussion of different implementations of beam dump limits as well as a comparison between rescaling limits and recreating the analysis is given in Appendix~\ref{app:comparison}. Finally, Appendix~\ref{app:longtable} provides information on the different experiments, the relevant processes and couplings.

\section{Hidden Gauge Bosons}
\label{sec:models}

We consider four different extensions of the SM by an additional gauge boson $X_\mu$ given by the Lagrangian
\begin{align}\label{eq:lag}
\mathcal{L}=-\frac{1}{4}\hat F_{\mu\nu} \hat F^{\mu\nu}-\frac{\epsilon'}{2}\hat F_{\mu\nu}\hat X^{\mu\nu}-\frac{1}{4}\hat X_{\mu\nu} \hat X^{\mu\nu}-g'\,j_\mu^Y \hat B^\mu -g_x\,j_\mu^x \hat X^\mu+\frac{1}{2}\hat M_X^2 \hat X_\mu \hat X^\mu\,, 
\end{align}
where $\hat B_\mu$ denotes the hypercharge gauge boson with the field strength tensor $\hat F_{\mu\nu}$, $g'$ the hypercharge gauge coupling and $j_\mu^Y$ the hypercharge current. Here, $\hat X_{\mu\nu}$, $g_x$ and $\hat M_{X}$ are the field strength tensor, coupling and mass term of the new $U(1)$ gauge boson and $j_\mu^x$ the corresponding current. A potential mass mixing term in \eqref{eq:lag} has been omitted.
The hatted fields indicate that the kinetic terms in \eqref{eq:lag} are not canonically normalized and the
corresponding gauge fields need to be redefined. 

The current depends on the considered gauge group,
\begin{align}
j^X_\mu&= 0\,, \qquad && U(1)_X \notag \,, \\
j^{\, i-j}_\mu&= \bar L_i \gamma_\mu L_i 
          + \bar \ell_i\gamma_\mu \ell_i 
          - \bar L_j \gamma_\mu L_j -\bar\ell_j\gamma_\mu \ell_j\,,
            \qquad && U(1)_{L_i-L_j} \,,\notag \\ 
j^{B-L}_\mu&=  \frac{1}{3}\bar Q \gamma_\mu Q 
          + \frac{1}{3}\bar u_R\gamma_\mu u_R 
          + \frac{1}{3}\bar d_R\gamma_\mu d_R
          - \bar L \gamma_\mu L 
          - \bar \ell\gamma_\mu \ell  
          - \bar \nu_R\gamma_\mu \nu_R \,,
            \qquad && U(1)_{B-L} \; ,
\label{eq:fermion_currents}
\end{align}
with $i\neq j=e, \mu, \tau$. 

\bigskip
In the following we will focus on masses in the MeV to multi GeV region. This is mostly for phenomenological reasons. For the case of the hidden photon this region was suggested by a (now essentially ruled out) explanation of the $(g-2)_{\mu}$ anomaly~\cite{Pospelov:2008zw} as well as dark matter applications (cf., e.g.,~\cite{ArkaniHamed:2008qn}). Consequently, many of the recent experimental activities focussed on this region. That said, as we will see, the explanation of $(g-2)_{\mu}$ with a weakly coupled $U(1)_{L_\mu-L_\tau}$~\cite{Altmannshofer:2014pba} is still viable (cf. Fig.~\ref{fig:LmLt}).
Such masses can arise from a Higgs or a Stueckelberg mechanism~(See, e.g.~\cite{Cicoli:2011yh} for string models realizing Stueckelberg masses in the considered range). However, in the former case extra effects due to the additional Higgs boson are likely.

Let us also briefly comment on the smallness of the gauge couplings. The smallness of the kinetic mixing parameter of the hidden photon is naturally suggested if it is loop-induced~\cite{Holdom:1985ag}. However, for the gauge groups we want to consider here we are dealing with the gauge couplings themselves. Such small gauge couplings arise for example in the context of LARGE volume compactifications of string theory~\cite{Burgess:2008ri}\footnote{Using hidden gauge groups that have such small couplings, one can also naturally obtain kinetic mixings significantly below the loop suggested value of $\epsilon\sim 10^{-3}$~\cite{Cicoli:2011yh}.}.
These models naturally suggest gauge couplings in the region
\begin{equation}
\alpha_{X}=\frac{g^{2}_{x}}{4\pi}\sim 10^{-9},
\end{equation}
but somewhat smaller values are also possible.
Nevertheless, this provides a theoretically interesting target area.

\subsection{Interactions of the canonically normalized fields}
After rotation and proper normalization (cf. Appendix~\ref{rotation}) we obtain the interactions of the now unhatted gauge fields and currents,
\begin{align}\label{eq:currentcouplings}
{\mathcal{L}}_\mathrm{int}=&\left(ej_\text{EM} , \frac{e}{\sin \theta_w \cos \theta_w} j_Z, g_{x}j_{x}\right) \,K\,\begin{pmatrix}A\\ Z\\  A'\end{pmatrix} \,, 
\end{align}
with
\begin{align}
K= \begin{pmatrix}
1 & 0 & -\epsilon \phantom{e}\\
0 & 1& 0 \\
0 & \epsilon \tan\theta_w&  1
\end{pmatrix} +\mathcal{O}(\epsilon \delta, \epsilon^{ 2})\,,
\label{eq:KK}
\end{align}
and
\begin{equation}
\epsilon=\epsilon^{\prime}\cos(\theta_w),\quad\delta=\frac{\hat M^{2}_{A'}}{\hat M^{2}_{Z}}.
\end{equation}
To leading order in $\epsilon$ the masses before and after the basis change are equal,
\begin{equation}
M_{A'}^2=\hat{M}_{X}^2(1 +\mathcal{O}(\epsilon^2)).
\end{equation}
The same procedure also gives the interactions of the new gauge fields with the Higgs boson. This is also detailed in Appendix~\ref{rotation}.

\subsection{ Kinetic mixing}
\label{sec:kinmixing}
Gauge groups such as $U(1)_{{L_\mu}-{L_\tau}}$ do not feature direct interactions with
the first generation of Standard Model particles that make up most of the ordinary matter
and therefore most of the experimental apparatuses that we consider. They are therefore automatically much harder to probe. Nevertheless, since $\mu$ and $\tau$ are also charged under the electromagnetic $U(1)$, there exists an unavoidable kinetic mixing at the loop level.
This allows us to probe these gauge groups also in experiments with first generation particles. Similarly, this also allows to probe the purely leptonic gauge groups in experiments with baryonic particles.
Let us now consider this loop-induced kinetic mixing in more detail. 

\begin{figure}
\begin{center}
\includegraphics[width=.7\textwidth]{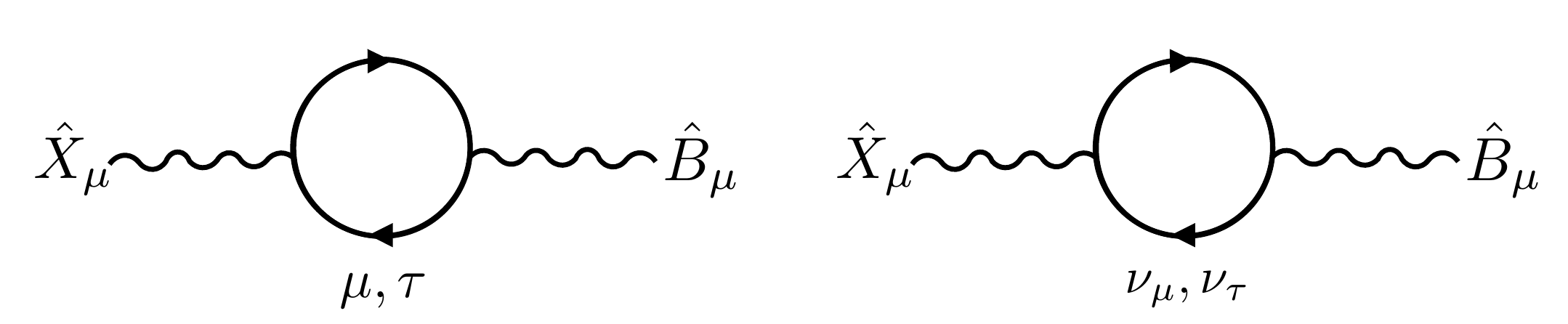}
\end{center}
\caption{\label{fig:kinmix} Diagrams contributing to the kinetic mixing between the hypercharge gauge boson $\hat B_\mu$ and the $U(1)_{L_\mu-L_\tau}$ gauge boson $\hat X_\mu$. }
\end{figure}

If the abelian extension of the SM gauge group in \eqref{eq:fermion_currents} is not embedded in a non-abelian gauge group, kinetic mixing can be induced by a new fundamental parameter $\epsilon$. Kinetic mixing between non-abelian and abelian gauge groups is not possible at the renormalizable level.\footnote{Beyond the renormalizable level, kinetic mixing can arise from higher-dimensional operators involving the symmetry breaking Higgs fields, see, e.g.~\cite{Bruemmer:2009ky}. The loop effects discussed in the following can be viewed as generating such operators when integrating out fields.} However, if a non-abelian gauge group is broken $SU(N)\to U(1)$ at some high scale, loop effects from fields charged under both this new $U(1)$ and $U(1)_Y$ induce a kinetic mixing parameter in the broken phase. For the example of $U(1)_{{L_\mu}-{L_\tau}}$, the diagrams shown in Fig.~\ref{fig:kinmix} give
\begin{align}\label{eq:lmutaumix}
\epsilon_{\mu\tau}(q^2)=-\frac{e\, g_{\mu\tau}}{4\pi^2}\int_0^1 dx\,x(x-1)\bigg[3\log\bigg(\frac{m_\mu^2+q^2x(x-1)}{m_\tau^2+q^2x(x-1)}\bigg)+ \log\bigg(\frac{m_{\nu_2}^2+q^2x(x-1)}{m_{\nu_3}^2+q^2x(x-1)}\bigg) \bigg]\,,
\end{align}
where $q^2$ is the transferred momentum and $g_{\mu\tau}$ the gauge coupling of $U(1)_{{L_\mu}-{L_\tau}}$. The same result holds for $U(1)_{{L_\mu}-{L_e}}$ and $U(1)_{{L_e}-{L_\tau}}$ with the obvious replacements. For large momentum transfer $q^2\gg m_\tau^2$, this mixing parameter is power suppressed $\epsilon_{\mu\tau}\propto m_\mu^2/q^2-m_\tau^2/q^2$, whereas for low momentum transfer  $q^2\ll m_\mu^2$, the mixing can become relevant  $\epsilon_{\mu\tau}\propto \log (m_\mu^2/m_\tau^2)$. Since this loop-induced kinetic mixing for the lepton family number gauge groups is finite, we take it into account when we present the constraints on the corresponding gauge bosons in Section~\ref{sec:results}. 

As an interesting theoretical feature we note that the finiteness of \eqref{eq:lmutaumix} is not guaranteed by the fact that the symmetry is anomaly-free alone. In addition, it implies that 
the gauged lepton-family number difference $U(1)_{L_\mu-L_\tau}$ can be embedded in a $G_{L_\mu-L_\tau}$ which breaks to $U(1)_{L_\mu-L_\tau}$ without mixing between the corresponding neutral gauge boson and any component of the hypercharge gauge boson (and analogous for $U(1)_{L_e-L_\tau}$ and $U(1)_{L_\mu-L_e}$). One can construct a UV completion with a gauge group $G_\text{SM}$ in which the SM is embedded and the gauge group $G_{L_\mu-L_\tau} \supset U(1)_{L_\mu-L_\tau}$, such that
%
%
%
%
\begin{center}
\begin{tikzcd}
& G_\text{SM} \arrow[rightarrow]{d}{}  &\times  & G_{L_\mu-L_\tau} \arrow[rightarrow]{d}{}\\
& SU(3)_C\times SU(2)_L \times U_Y(1) & \times & U(1)_{L_\mu-L_\tau}
\end{tikzcd}\,.
\end{center}
This is for example not possible in the case of an embedding of the $U(1)_{B-L}$ gauge group which we discuss below. As a consequence, neither the scalar that breaks $G_{L_\mu-L_\tau}\to U(1)_{L_\mu-L_\tau}$ nor the scalar responsible for giving the $U(1)_{L_\mu-L_\tau}$ gauge boson a mass necessarily contributes to the loop-induced mixing $\epsilon_{\mu\tau}$. A straightforward way to embed $U(1)_{L_\mu-L_\tau}$ is to choose $G_{L_\mu-L_\tau} = SU(2)_{L_\mu-L_\tau}$, and break it  to the gauge boson corresponding to the third generator, which determines the couplings to the doublets $(L_\mu, L_\tau)$ through $\sigma_3=\text{diag} (1,-1)$ \cite{He:1991qd, Heeck:2011wj}.

For $U(1)_{B-L}$, the result of the one-loop calculation analogous to \eqref{eq:lmutaumix} is not finite and its magnitude depends on the choice for the renormalization scale. This implies that the gauge couplings of $U(1)_Y$ and those of a possible non-abelian embedding of  $SU(N)_{B-L}\supset U(1)_{B-L}$ cannot be independent. Similar to the situation of the loop-induced kinetic mixing between the photon and the $Z$ boson in the SM \cite{Fleischer:1986au}, the renormalization scale dependence of three parameters, the wavefunctions for the $U(1)_{B-L}$ boson, the hypercharge boson as well as $\epsilon(\mu)$, need to be absorbed by the field renormalizations of the two original fields in the unbroken phase. We can therefore not determine the kinetic mixing parameter 
unambiguously and neglect it when we present the constraints on $g_{B-L}$ and $M_{A'}$ in Section~\ref{sec:results}. That said, since all SM particles relevant to the experiments and observations we consider carry charges, the effect of the kinetic mixing can be considered small.

For completeness let us note that in the case of a completely secluded $U(1)_{X}$, where all the SM particles carry no $X$ charge, kinetic mixing is not generated within the SM, instead additional beyond the Standard Model particles are necessary to generate contributions at one-loop.

\subsection{Flavour structure}
Both a secluded $U(1)_X$ and the $U(1)_{B-L}$ gauge boson couple universally to all SM quark flavours and lepton flavours, and hence lead to flavour-conserving vertices. In the case of a gauged lepton number difference, this is less obvious. Since the couplings to leptons are non-universal, flavour changing vertices can in principle arise upon rotating the leptons from the interaction to the mass eigenbasis. However, for gauged lepton family number differences, the lepton Yukawa couplings which respect this symmetry are diagonal already in the interaction eigenbasis
\begin{align}
\mathcal{L}_Y=&-\begin{pmatrix}\bar e_L& \bar \mu_L& \bar \tau_L\end{pmatrix}\,\begin{pmatrix}
y_{e}&0&0\\
0&y_{\mu}&0\\
0&0&y_{\tau} 
\end{pmatrix}\, \begin{pmatrix}e_R\\\mu_R\\\tau_R
\end{pmatrix}
\phi  \notag \\[3pt]
&-\begin{pmatrix}\bar \nu_e& \bar \nu_\mu& \bar \nu_\tau\end{pmatrix}\,\begin{pmatrix}
y_{\nu_e}&0&0\\
0&y_{\nu_\mu}&0\\
0&0&y_{\nu_\tau} 
\end{pmatrix}\, \begin{pmatrix}N_1\\N_2\\N_3
\end{pmatrix} \tilde \phi + \text{h.c.} \,,
\end{align}
where we have included three right-handed neutrinos $N_1, N_2, N_3$ that are singlets apart from lepton family number charges, and $\phi$ denotes the SM Higgs boson.  As a result, the couplings of the $A'$ gauge boson to leptons are diagonal. This Lagrangian also produces a diagonal lepton mixing matrix. However, Majorana masses that respect the lepton family symmetry as well as mass terms induced by the scalar $S$ that gives a mass to the  $A'$ gauge boson  contribute to neutrino masses
\begin{align}
\mathcal{L_M}=-\frac{1}{2}N_i^T\mathcal{C}^{-1}\big( M_R\big)_{ij} N_j\,.
\end{align}
The texture of the matrix $\mathcal{M}_R$ depends on the gauge group and the charge $Q_S$ of the scalar $S$ under this group,
\begin{center}
\begin{tabular}{cccc}
&$U(1)_{L_\mu-L_e}$&$U(1)_{L_e-L_\tau}$&$U(1)_{L_\mu-L_\tau}$\\[.3cm]
$\mathcal{M}_R^{|Q_S|=1}$&$\begin{pmatrix} 
0&m&M\\
m&0& M\\
M&M&m
\end{pmatrix}$&$\begin{pmatrix} 
0&M&m\\
M&m& M\\
m&M&0
\end{pmatrix}$&$\begin{pmatrix} 
m&M&M\\
M&0& m\\
M&m&0
\end{pmatrix}$\\[1cm]
$\mathcal{M}_R^{|Q_S|=2}$&$\begin{pmatrix} 
M&m&0\\
m&M& 0\\
0&0&m
\end{pmatrix}$&$\begin{pmatrix} 
M&0&m\\
0&m& 0\\
m&0&M
\end{pmatrix}$& $\begin{pmatrix} 
m&0&0\\
0&M& m\\
0&m&M
\end{pmatrix}$ 
\end{tabular}\,,
\end{center}
where we have fixed the magnitude of the charge of the leptons to $|Q_\ell|=1$ and for charges of the scalar $S$ other than $|Q_S|=1,2$ one obtains the above textures with $M \to 0$. Here, $m$ is the Majorana scale that can be fully independent of the mass scale induced by the vacuum expectation value of $M\propto \langle S\rangle= M_{A'}^2/(2 g_x^2 Q_S^2) $. For the hierarchy $m\gg M, v$, the texture $\mathcal{M}_R^{|Q_S|=1}$ for $U(1)_{L_\mu-L_\tau}$ has been discussed in detail in \cite{Heeck:2011wj} and we refrain from a discussion of the phenomenology in the neutrino sector here. The texture $\mathcal{M}_R^{|Q_S|=1}$ for $U(1)_{L_{\mu}-L_{\tau}}$ is strongly preferred by the structure of the neutrino mixing matrix and compatible with the global fit to the leptonic CP phase \cite{Araki:2012ip, Crivellin:2015lwa, Ezhela:2003pp}.\footnote{We thank Julian Heeck for pointing this out to us.}

 In addition, the  $A'$ boson acquires lepton-flavour violating couplings to neutrinos, but not to charged leptons. Notice that the introduction of right-handed neutrinos does not introduce additional contributions to the kinetic mixing in \eqref{eq:lmutaumix}. Flavour changing couplings arise only at the one-loop level for all three gauge structures, $U(1)_X$, $U(1)_{B-L}$, and $U(1)_{L_\mu-L_e}, U(1)_{L_e-L_\tau}, U(1)_{L_\mu-L_\tau}$ considered here.

\section{Searching for Hidden Photons}
\label{sec:experiments}
In Section~\ref{sec:results} we present and discuss the results of recomputing the limits from searches for secluded, hidden photons for  $U(1)_{B-L}$ and 
$U(1)_{L_\mu-L_e}, U(1)_{L_e-L_\tau}, U(1)_{L_\mu-L_\tau}$ gauge bosons. Before we do so, let us describe our strategy for recasting electron and proton beam dumps and fixed target experiments as well as collider searches. In addition, we consider bounds from white dwarfs and neutrino experiments.

%
\begin{figure}
\begin{center}
\includegraphics[width=.56\textwidth]{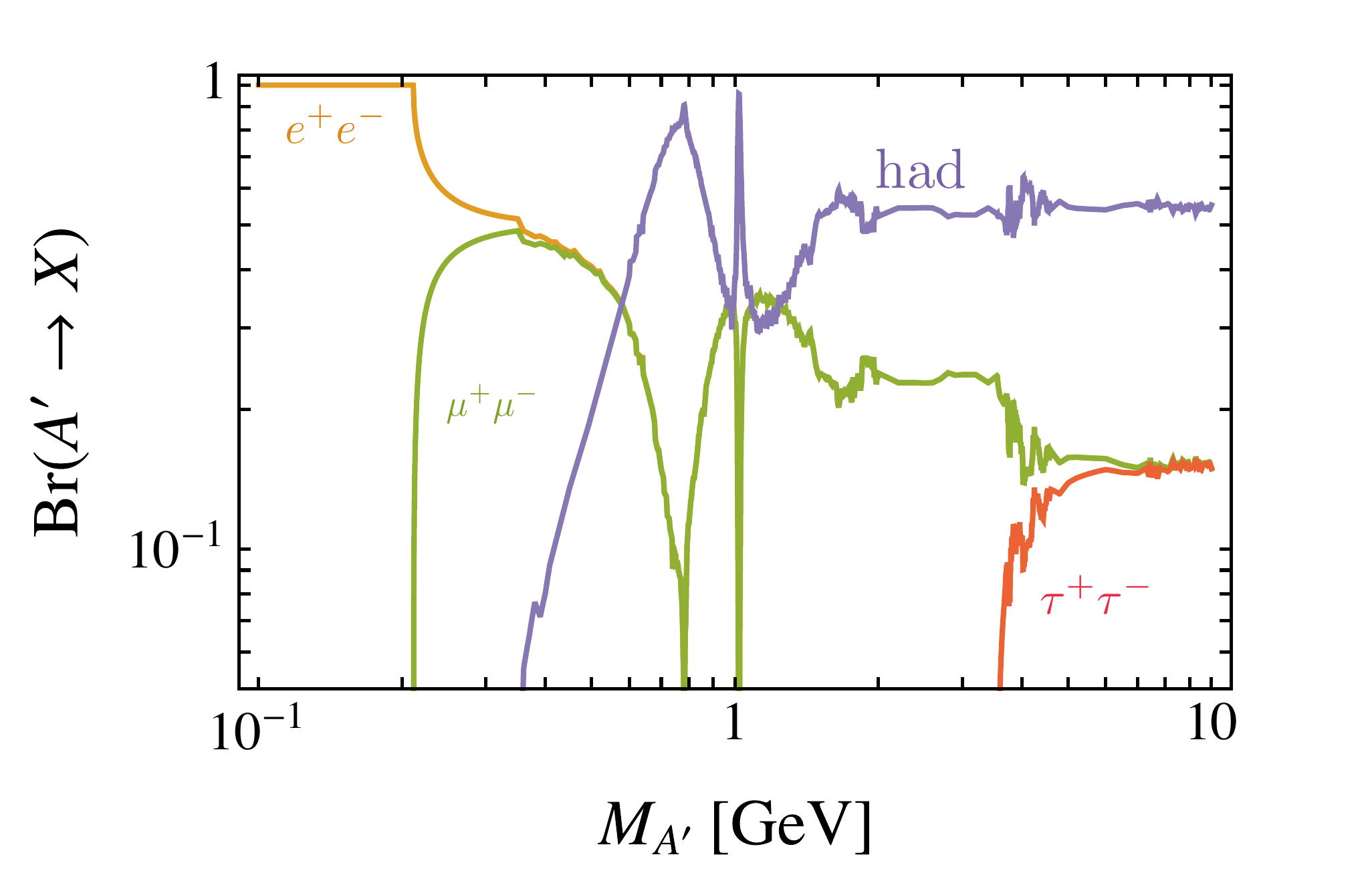}
\end{center}
\caption{\label{fig:BRuni} Branching ratios for the gauge bosons of a secluded $U(1)_X$ gauge group mixing with the SM hypercharge gauge boson. See text for details.  }
\end{figure}
%

%
\begin{figure}
\begin{center}
\includegraphics[width=1.02\textwidth]{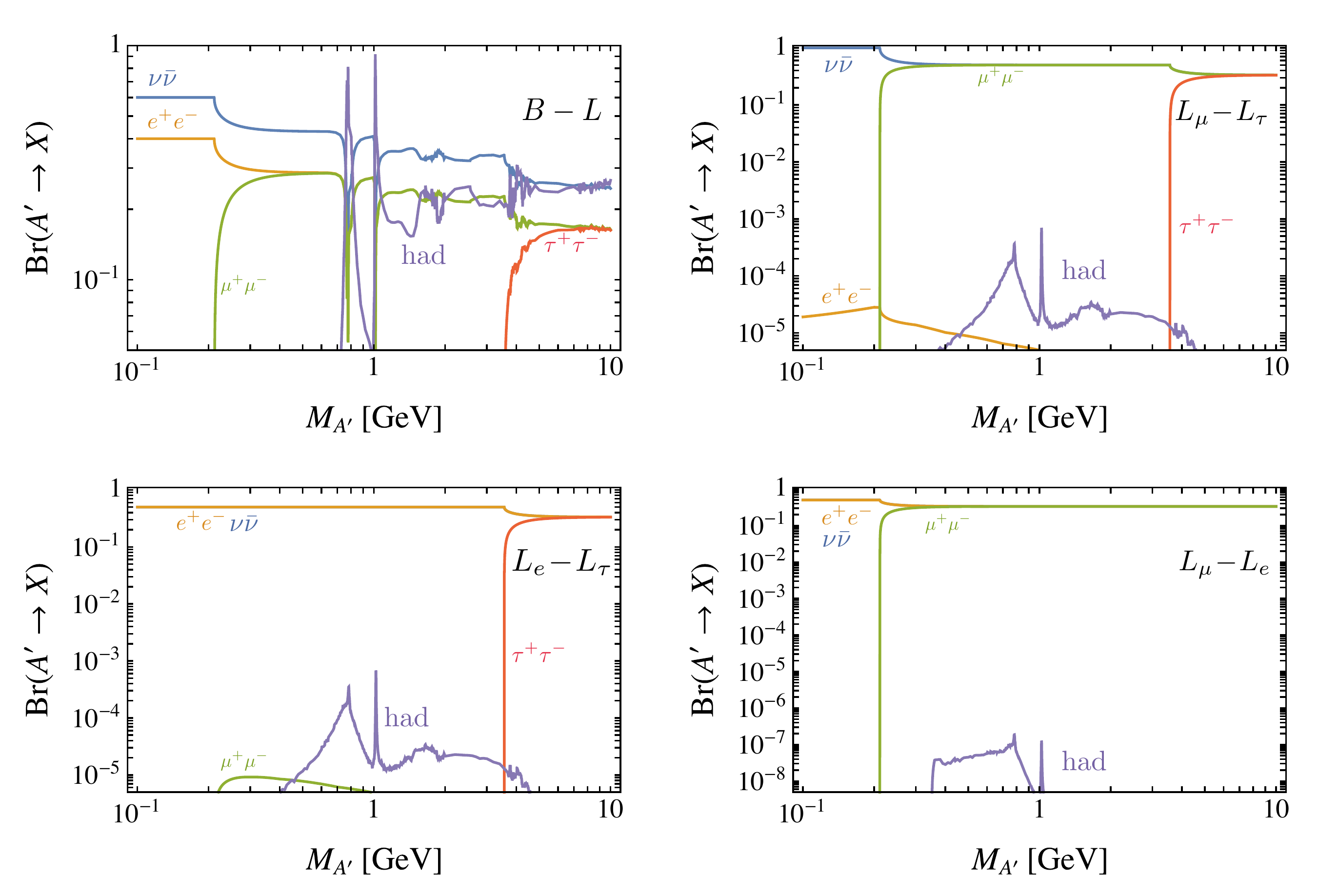}
\end{center}
\vspace{-.5cm}
\caption{\label{fig:BRsother} Branching ratios for the gauge bosons of a $U(1)_{B-L}$ (upper left), $ U(1)_{L_\mu-L_\tau}$ (upper right), $U(1)_{L_e-L_\tau}$ (lower left) and $ U(1)_{L_\mu-L_e}$ (lower right) gauge group. In the lower two panels, the branching ratio into neutrinos is indistinguishable from $\text{Br}(A'\to e^+e^-)$. See text for details.  }
\end{figure}
%

\subsection{Decay widths and branching ratios}\label{sec:BRs}
A crucial ingredient in all laboratory searches are the decay widths and branching ratios.
The decay widths for the gauge boson of a secluded $U(1)_X$ are purely determined by mixing with the hypercharge gauge boson. For charged SM leptons, the decay widths are straightforwardly computed  by replacing the coupling of a (massive) photon by $\alpha \to \alpha \epsilon^2$. Decays into hadrons can be determined with a data-driven approach by taking advantage of measurements of the ratio between the production cross section of hadronic final states and muon pairs in $e^+e^-$ colissions, $R(s)=\sigma(e^+e^-\to \text{hadrons})/\sigma(e^+e^-\to \mu^+\mu^-)$ \cite{Ezhela:2003pp, Patrignani:2016xqp}. The hadronic decay width is then given by
\begin{align}
\Gamma (A'\to \text{hadrons})=\epsilon^2\,\Gamma(\gamma^*\to \mu^+\mu^-) \,R(M_{A'}^2)\,\qquad \text{for} \quad U(1)_X\,,
\end{align}
where $\Gamma(\gamma^\ast\to \mu^+\mu^-)$ is the partial decay width for a virtual SM photon of mass $M_{A'}$.
 We show the results in Fig.~\ref{fig:BRuni}. 
 
 For gauge bosons of charged lepton family number differences, decays into hadronic final states are also only possible through kinetic mixing, and can be determined analogous to the universal gauge boson,
  \begin{align}
\Gamma (A'\to \text{hadrons})=\epsilon_{\mu\tau}(M_{A'}^2)^2\Gamma(\gamma^\ast\to \mu^+\mu^-) R(M_{A'}^2)\,\qquad \text{for} \quad U(1)_{L_\mu-L_\tau}\,,
\end{align}
where the kinetic mixing parameter is given by \eqref{eq:lmutaumix} and the obvious replacements hold for $U(1)_{L_\mu-L_e}$ and $U(1)_{L_e-L_\tau}$. The partial decay width into the leptons charged under the corresponding gauge group can be directly deduced from \eqref{eq:fermion_currents} and \eqref{eq:Zpwidth} from Appendix~\ref{rotation}. The respective uncharged lepton family can only couple through kinetic mixing with the photon. The branching ratios for the gauge boson of a gauged lepton family number difference are shown in Fig.~\ref{fig:BRsother}. Hadronic decays are suppressed in all cases and the different shape of $\Gamma(A'\to \text{hadronic})$ in the case of $U(1)_{L_\mu-L_e}$ can be explained by the approximate cancellation in $\epsilon_{\mu e}(M_{A'}^2)\approx m_\mu^2/M_{A'}^2-m_e^2/M_{A'}^2\approx 0$ for $M_{A'} > m_\mu$. 

For a $U(1)_{B-L}$ gauge boson, we take advantage of the analysis in~\cite{Ilten:2018crw}, where the couplings are computed using a data-driven method based on vector-meson dominance (VMD) for masses $M_{A'}$ below the QCD scale. The flavour-universal charges lead to an absence of $A'-\rho$ mixing. Therefore, hadronic decays only open up once the much narrower $\omega$-resonance turns on at $m_\omega = 782$ MeV, and below that scale, the leptonic decay rates dominate as is evident from the upper left panel of Fig.~\ref{fig:BRsother}. For masses of the $A'$ gauge boson above the QCD scale, the vector dominance model breaks down at around $M_A'\gtrsim 1.65$ GeV \cite{Ilten:2018crw}. We rescale the $R$-ratio with the $B-L$ charges above this value,
\begin{align}
\Gamma (A'\to \text{hadrons})=\frac{\sum_\text{q}\Gamma(A'\to q\bar q)}{\sum_\text{q}\Gamma(\gamma^*\to q\bar q) }\,\Gamma(\gamma^*\to \mu^+\mu^-) \,R(M_{A'}^2)\,\qquad \text{for} \quad U(1)_{B-L}\,,
\end{align}
in which the sum extends over all quarks with masses $m_q< M_{A'}/2$. This matching is good, given the expected precision of the VMD method of about $10\%-20\%$ \cite{Ilten:2018crw}.

%
\begin{figure}
\begin{center}
\includegraphics[width=.7\textwidth]{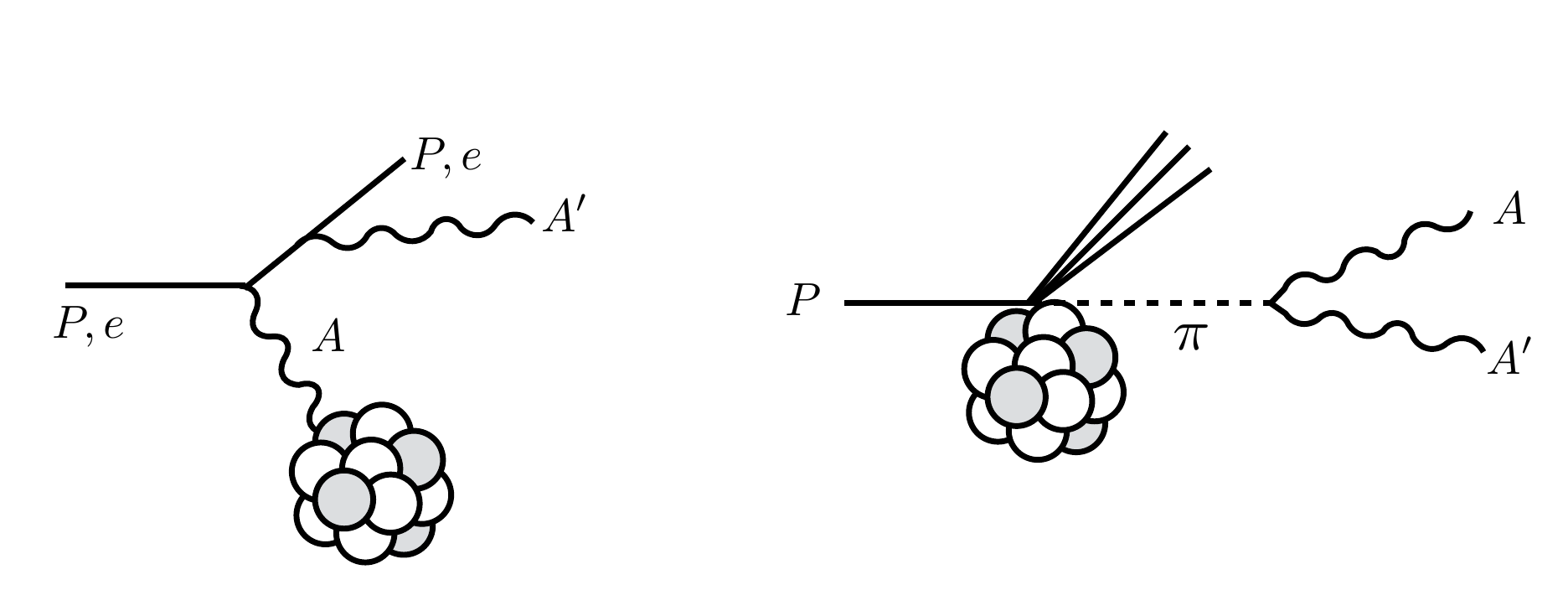}
\end{center}
\caption{\label{fig:hp_prod}  Hidden photon production through Bremsstrahlung (left) and meson production (right) in beam dump experiments. }
\end{figure}
%

\begin{table}
\begin{center}
 \begin{tabular}{c | c c c c c c c} 
 Experiment& Target $\left[ ^\mathrm{A} _\mathrm{Z} \mathrm{X} \right]$& $E_0$ [GeV]  & $L_\text{sh}$ [m]& $L_\text{dec}$ [m]& $N_\mathrm{in}$&$N_\text{obs}$&$N_{95\%}$\\[.3cm]
\hline
&&&&&\\[-.3cm]
CHARM       & $^{63.55}_{\phantom{00.}29}{\mathrm{Cu}}$ &400 &480 &35 & $2.4\times10^{18}$ & 0&3  \\[.3cm]
E137            & $^{26.98}_{\phantom{00.}13}{\mathrm{Al}}$ & 20&179 &204 &$1.87\times10^{20}$ &0 &3   \\[.3cm]
E141            & $^{183.84}_{\phantom{000.}74}{\mathrm{W}}$ & 9&0.12 &35 &$2\times10^{15}$ &$1126^{+1312}_{-1126}$ & 3419  \\[.3cm]
E774           & $^{183.84}_{\phantom{000.}74}{\mathrm{W}}$ & 275& 0.3& 2&$5.2\times10^{9}$ &$0^{+9}_{-0}$ &18  \\[.3cm]
LSND          & H$_2$O & 0.8 & 25.85 & 8.3 &$9.2\times10^{22}$ &25 &50  \\[.3cm]
Orsay          & $^{183.84}_{\phantom{000.}74}{\mathrm{W}}$ & 1.6& 1&2 &$2\times10^{16}$ &0&3  \\[.3cm]
U70/NuCal  &$ ^{55.85}_{\phantom{00.}26}{\mathrm{Fe}}$& 68.6&64 &23 & $1.71 \times10^{18}$& 5 &7.1/4.5  \\[.2cm]
\rowcolor{Gray}  \phantom{\huge(}SHiP            &$^{183.84}_{\phantom{000.}74}{\mathrm{W}}$ &400 &60 &50 & $2 \times10^{20}$&0 &10  \\[.3cm]
\rowcolor{Gray}  SeaQuest   &$ ^{55.85}_{\phantom{00.}26}{\mathrm{Fe}}$ &120 & 5& 0.95& $1.44\times10^{18}$& 0&3  \\[.3cm]
\rowcolor{Gray}  $ \begin{matrix} \text{FASER}\\ (\text{far})\end{matrix}$   &$ \begin{matrix}^{1}_{1}{\mathrm{H}}\\ (pp -\text{coll})\end{matrix}$ & $\begin{matrix}9\times10^7\\(\sqrt{s}=13 \text{ TeV}) \end{matrix}$ & 390& 10& $\begin{matrix} 2.3\times 10^{16}\\(300 \text{ fb}^{-1})\end{matrix}$& 0&3  \\[.3cm]
\end{tabular}
\end{center}
\caption{\label{tab:material} Material constants and specifications for the different beam dump experiments looking for very displaced vertices. Future experiments are indicated by grey shading.}
\end{table}

\subsection{Beam dump and fixed target experiments}

Beam dump and fixed target experiments provide the best sensitivity to hidden photons of a secluded $U(1)_X$ with masses $M_{A'} \lesssim 1$ GeV for almost the complete range of kinetic mixing parameters $\epsilon$. In the following, we discuss how the existing limits and projections for future experiments change for  
$U(1)_{L_\mu-L_e}, U(1)_{L_e-L_\tau}, U(1)_{L_\mu-L_\tau}$, and $U(1)_{B-L}$ gauge bosons. The material constants and specifications of the beam dump and fixed target experiments we discuss are collected in Table~\ref{tab:material}.  

\subsubsection{Electron beam dump experiments}

In electron beam dump experiments such as SLAC E137, SLAC E141~\cite{Bjorken:2009mm,Bjorken:1988as, Riordan:1987aw, Andreas:2012mt}, Fermilab E774 \cite{Bross:1989mp} and Orsay \cite{Davier:1989wz}, hidden photons can be produced in the Bremsstrahlung process shown on the left hand side of Fig.~\ref{fig:hp_prod}. They subsequently travel through a shielding material before decaying. In the case of a dielectron final state, the number of electrons produced from hidden photon decays in an electron beam dump experiment with an incident electron beam of $N_e$ electrons with energy $E_0$ is given by \cite{Bjorken:2009mm,Bjorken:1988as}
\begin{align}\label{eq:eevents}
N=N_e\frac{N_0X_0}{A}\int_{M_{A'}}^{E_0-m_e}dE_{A'}\,\int_{E_{A'}+m_e}^{E_0}dE_e\,&\int_0^{\rho \,L_\text{sh}/X_0}dt\,\bigg[ I_e(E_0,E_e,t)\frac{1}{E_e}\frac{d\sigma}{dx_e}\Big|_{x_e=\frac{E_{A'}}{E_e}}\,\notag\\
&\times e^{-\frac{L_\text{sh}}{\ell_{A'}}}\Big(1-e^{-\frac{L_\text{dec}}{\ell_{A'}}}\Big)\bigg]\,\text{Br}(A' \to e^+e^-)\,,
\end{align}
where $N_0\approx 10^{23}$ mole$^{-1}$ is Avogadro's number, $X_0$ and $A$ are the radiation length and mass number of the target material, and $x_e=E_{A'}/E_e$ is the fraction of the energy of the incoming electrons carried by the hidden photon. 
The function $I_e(E_0, E_e,t)$ describes the energy distribution of the incoming photons after passing through a medium of $t$ radiation lengths \cite{Mo:1968cg}.
The length of the decay volume $L_\text{dec}$ and of the target and shielding $L_\text{sh}=L_\text{target}+L_\text{shield}$ depend on the experiment, whereas the hidden photon branching ratio $\text{Br}(A' \to e^+e^-)$, the average hidden photon decay length $\ell_{A'}$, and the differential hidden photon production cross section $d\sigma/dx_e$ are model-dependent quantities. The general expression for the average hidden photon decay length reads
\begin{align}
\ell_{A'}=\frac{\beta\gamma}{\Gamma_\text{tot}}=\beta\frac{E_{A'}}{M_{A'}}\frac{1}{\Gamma_\text{tot}}\,,
\end{align}
with the relativistic boost factor $\gamma$, the velocity $\beta$ and the total hidden photon decay width $\Gamma_\text{tot}$.
The differential hidden photon production cross section is given by
\begin{align}\label{eq:diffcross}
\frac{d\sigma}{dx_e}=4\alpha^3\epsilon^2 \xi\,\sqrt{1-\frac{M_{A'}^2}{E_0^2}}\,\frac{1-x_e+\frac{x_e^2}{3}}{M_{A'}^2\frac{1-x_e}{x_e}+m_e^2 x_e}\,,
\end{align}
in which $\xi$ denotes the effective photon flux, which is a function of the beam energy, the properties of the material and the hidden photon mass $M_{A'}$ \cite{Andreas:2012mt}. 
Note that when the radiation length and the decay length are comparable, also the decay probability $\propto \exp(-L_\text{sh}/\ell_{A'})$ depends on the position where the particle is produced, i.e. $t$. We keep the full dependence in computing the limits in these cases. There is also a geometric factor depending on the detector shape and size as well as the angular distributions of the gauge bosons considered here. We neglect this factor for electron beam dumps, which in the case of the secluded hidden photon amounts to a correction of at most $10\%$~\cite{Andreas:2013xxa}. Efficiency factors for the experimental reconstruction and experimental cuts are not included in \eqref{eq:eevents} but applied when we compute the constraints. 
\bigskip

In adapting the bounds from searches for universal hidden photons for $U(1)_{B-L}$ and the gauge bosons of gauged lepton family number differences, we replace the branching ratios by the values computed in Section \ref{sec:BRs} and compute the corresponding average decay length with the respective total width $\Gamma_\text{tot}$. In the differential cross section \eqref{eq:diffcross}, the couplings are replaced by 
\begin{align}\label{eq:Ebeamdumps1}
\alpha^3\epsilon^2\quad  \to\quad \begin{cases} \alpha^2\alpha_{\mu e}\,, &\quad \text{for} \quad U(1)_{L_\mu-L_e}\,,\\
\alpha^2\alpha_{e \tau}\,, &\quad \text{for} \quad U(1)_{L_e-L_\tau}\,,\\
\alpha^2\alpha_{\mu \tau} \,\epsilon_{\mu\tau}(M_A)^2\,, &\quad \text{for} \quad U(1)_{L_\mu-L_\tau}\,,\\
\alpha^2\alpha_{B-L}\,, &\quad \text{for} \quad U(1)_{B-L}\,,
\end{cases}
\end{align}
where $\alpha_{ij}=g_{ij}^2/(4\pi)$ and $g_{ij}$ denotes the gauge coupling of $U(1)_{L_i-L_j}$, and $\alpha_{B-L}=g_{B-L}^2/(4\pi)$. In writing \eqref{eq:Ebeamdumps1} the fact that the $B-L$ charge of electrons is $-1$ has already been accounted for.  

Looking at \eqref{eq:Ebeamdumps1} it is clear that the electron beam dump constraints are expected to be important for $U(1)_{L_{\mu}-L_e}, U(1)_{L_e-L_\tau}$ and $U(1)_{B-L}$ gauge bosons, whereas for the $U(1)_{L_\mu-L_\tau}$ gauge boson the constraints are expected to be strongly attenuated. In contrast, experiments with muon beams would be particularly relevant for searches for $U(1)_{L_\mu-L_\tau}$ gauge boson. We comment on such a possibility below in the context of fixed target experiments.

\subsubsection{Electron (and future muon) fixed target experiments}
In addition to the beam dumps discussed above we also consider fixed target experiments~\cite{Essig:2010xa} such as APEX~\cite{Abrahamyan:2011gv}, A1/MAMI~\cite{Merkel:2014avp,Merkel:2011ze}, HPS~\cite{Battaglieri:2014hga}, NA64~\cite{Banerjee:2016tad},  VEPP-3~\cite{Wojtsekhowski:2012zq} and DarkLight~\cite{Kahn:2012br,Balewski:2014pxa}. Here, production is typically via Bremsstrahlung, too.
However, the signal is not a very displaced vertex, but kinematic features such as, e.g. a resonance bump in the invariant mass spectrum. This arises for example when the produced on-shell hidden photon decays into a pair of electrons. 

Most of the above experiments search for visible decay products ($e$, $\mu$). They target the region where the hidden photon decays promptly. This eliminates all the non-trivial geometric dependencies discussed above for the beam dumps.
To recast the limits we need to keep the product of production cross section times branching ratio constant. We do this by using the branching ratios computed in Section~\ref{sec:BRs} and for the production the replacements given in \eqref{eq:Ebeamdumps1}.

NA64, VEPP-3 and the invisible mode of DarkLight measure missing energy. This provides two search regions. The first is for prompt decays into invisible particles and the second is an essentially stable hidden photon. In the parameter region where these experiments provide sensitivity to our considered gauge groups we can assume prompt decays into neutrinos that are effectively invisible. We can then rescale as above but using the neutrino branching ratios from Section~\ref{sec:BRs}.\footnote{Any hidden photon that does not decay would also be considered invisible and therefore only increase the signal. Hence, our estimate is conservative.} 
As it turns out for DarkLight the visible mode~\cite{Balewski:2014pxa} is more sensitive for the gauge bosons considered here.

\bigskip
A future modified version of NA64 utilizing the upgraded muon beam at the CERN SPS delivering up to $10^{12}$ muons on target has been proposed \cite{Gninenko:2014pea,Gninenko:2018tlp}. Employing the same search strategy for missing energy this NA64$\mu$ setup will be able to set much more severe bounds, especially in the case of $U(1)_{L_\mu-L_\tau}$.

\subsubsection{Proton beam dump experiments}\label{sec:protBD}

In proton beam dump experiments, such as CHARM~\cite{Bergsma:1985is}, LSND~\cite{Athanassopoulos:1997er} and U70/Nu-Cal~\cite{Blumlein:2011mv,Blumlein:2013cua}, as well as fixed-target experiments, such as SINDRUM I~\cite{MeijerDrees:1992kd}, NA48/2~\cite{Batley:2015lha}, and the future SHiP facility~\cite{Anelli:2015pba, Alekhin:2015byh}, hidden photons are produced in Bremsstrahlung as well as in meson decays produced in proton collisions with the target material. Similarly, the recently proposed experiment FASER~\cite{Feng:2017uoz} \footnote{ We use this opportunity to thank Felix Kling for kindly providing the meson distributions that facilitate an easy calculation of the FASER limits.} searching for very displaced hidden photon decays at the LHC is making use of these production mechanisms. 
Other proposed experiments searching for long-lived particles (LLPs) at LHC are MATHUSLA~\cite{Chou:2016lxi,Evans:2017lvd} and CodexB~\cite{Gligorov:2017nwh}. We expect them to have sensitivity in a similar region. While clearly important, calculating their precise sensitivities requires a detailed study beyond the scope of this work. We therefore will showcase the projected FASER limits representative for this class of newly proposed experiments searching for LLPs at the LHC.

The Bremsstrahlung process in proton beam dump experiments is similar to electron Bremsstrahlung with the difference that the cross section for proton collisions with the target material is usually determined experimentally. The number of expected events for $N_P$ incoming protons with initial energy $E_P$ can then be written as
\begin{multline}\label{eq:PbeamCS}
N=N_P\int_{M_{A'}}^{E_P-M_P} dE_{A'} \frac{1}{E_{P}}\frac{\sigma_{PA}(2M_P(E_P-E_{A'}))}{\sigma_{PA}(2M_PE_P)} \int_0^{p_{\perp,\text{max}}^2}\omega_{A'P}(p_\perp^2) dp_\perp^2 \\
\times e^{-\frac{L_\text{sh}}{\ell_{A'}}}\Big(1-e^{-\frac{L_\text{dec}}{\ell_{A'}}}\Big)\,\text{Br}(A' \to e^+e^-)\,,
\end{multline}
where  $\omega_{A'P}(p_\perp^2)$ is a weighting function relating the cross section of the $2\to3$ process $\sigma(P+A\to P+A+A')$ to the hadronic cross section $\sigma_{PA}$ of the process $P+A\to P+A$, where $A$ denotes the mass number of the target nucleus. This takes into account the splitting $P\to P + A'$ in the initial state. The ratio $\sigma_{ P A}(2M_P(E_P-E_{A'}))/\sigma_{P A}(2M_PE_P)$ relates the hadronic scattering cross section  $\sigma(P\, A \to P\, A )$ at the reduced center-of-mass energy after radiation of the $A'$-boson to the one evaluated at the initial center-of-mass energy. Explicit expressions for these functions can be found in \cite{Blumlein:2013cua}.  The hadronic cross section $\sigma_{P A}$ is linked to the inelastic proton-proton cross section by a function $f(A)$ via $\sigma_{PA}=f(A)\, \sigma_{PP}$, which however drops out in the ratio in~\eqref{eq:PbeamCS}. The proton-proton cross section $\sigma_{PP}$ can then be extracted from experimental data \cite{Cudell:2001pn, Nakamura:2010zzi}\footnote{There is also a constraint for $M_{A'}<2m_e$ from hidden photon conversion in the detector material, which is however not relevant for the masses we consider \cite{Blumlein:2013cua}.}.

The upper limit of the integral over the momentum component perpendicular to the beam axis in \eqref{eq:PbeamCS} is in principle given by $p_{\perp, \text{max}}^2=\max(E_P^2,E_{A'}^2)$ in the approximation of an elastic emission $P \to P+A'$. In practice it is, however, constrained by the largest value for which the fit to the form factors going into the cross section $\sigma_{PP}$ is reliable. 

In full analogy
to the case of Bremsstrahlung in electron beam dump and fixed-target experiments it follows that the gauge coupling in $\omega_{AP}(p_\perp^2)\propto \alpha\,\epsilon^2$ is replaced by
\begin{align}\label{eq:Ebeamdumps}
\alpha\epsilon^2\quad  \to\quad \begin{cases} \alpha_{\mu e}\,\epsilon_{\mu e}(M_{A'})^2\,\,, &\quad \text{for} \quad U(1)_{L_\mu-L_e}\,,\\
\alpha_{e \tau}\,\epsilon_{e\tau}(M_{A'})^2\,\,, &\quad \text{for} \quad U(1)_{L_e-L_\tau}\,,\\
\alpha_{\mu \tau} \,\epsilon_{\mu\tau}(M_{A'})^2\,, &\quad \text{for} \quad U(1)_{L_\mu-L_\tau}\,,\\
\alpha_{B-L}\,, &\quad \text{for} \quad U(1)_{B-L}\,,
\end{cases}
\end{align}
which is suppressed for all flavours of gauged lepton family numbers, and the $B-L$ charge of the proton has been implicitly accounted for. In addition, the branching ratio as well as the lifetime are computed separately for the different models we consider. 
\bigskip

An additional source of hidden photons from beam dump experiments are decays of mesons produced in collisions of protons with the target material, as depicted on the right-hand side of Fig.~\ref{fig:hp_prod}. For a secluded hidden photon, the number of expected events is given by \cite{Blumlein:2011mv}
\begin{align}
N=\frac{N_P}{\sigma(PP\to X)}\, \int_{-1}^1dx_F \int_{0}^{p_{\perp, \text{max}}^2}\,dp_\perp^2\,A^{\alpha(x_F)}&\,\frac{d\sigma(P P\to M X) }{dx_F dp_\perp^2} \text{Br}(M \to A' \gamma)\,\notag\\
&\times e^{-\frac{L_\text{sh}}{\ell_{A'}}}
\Big(1-e^{-\frac{L_\text{dec}}{\ell_{A'}}}\Big)\,\text{Br}(A'\to e^+e^-)\,,
\end{align}
{where $x_F$ denotes the Feynman-x variable and $A^{\alpha(x_F)}=\sigma(PN\to M X)/\sigma(PP\to M X)$, with the mass  number A of the nuclei $N$ of the target material. Typically, the mesons contributing dominantly to the signal are the pseudo-scalars $M=\pi^0,\eta,\eta'$, which have a large cross section into photons. For example,  the differential meson production cross section in proton-nucleon collisions have been measured in \cite{Ammosov:1976zk,AguilarBenitez:1991yy} at 70 and 400 GeV, respectively.  

The cross section is normalized to the total inclusive cross section $\sigma(PP\to X)$, which itself depends on the center-of-mass energy.  As before, the decay length and branching ratios of $A'$ depend on the underlying gauge group, and 
\begin{align}\label{eq:BRpi}
\text{Br}(M\to A'\gamma)=2 \epsilon^2  \bigg(1-\frac{M_{A'}^2}{M_{M}^2}\bigg)^3 \text{Br}(M\to \gamma\gamma)\,,
\end{align}
for a secluded hidden photon. For the $U(1)_{L_\mu-L_e}$, $U(1)_{L_e-L_\tau}$and $U(1)_{L_\mu-L_\tau}$ gauge groups,  the corresponding expressions follow with the replacements
\begin{align}
\alpha\epsilon^2\quad  \to\quad \begin{cases} \alpha_{\mu e}\,\epsilon_{\mu e}(M_{A'})^2\,\,, &\quad \text{for} \quad U(1)_{L_\mu-L_e}\,,\\
\alpha_{e \tau}\,\epsilon_{e\tau}(M_{A'})^2\,\,, &\quad \text{for} \quad U(1)_{L_e-L_\tau}\,,\\
\alpha_{\mu \tau} \,\epsilon_{\mu\tau}(M_{A'})^2\,, &\quad \text{for} \quad U(1)_{L_\mu-L_\tau}\,.\\
\end{cases}
\end{align}

\begin{figure}
\begin{center}
\includegraphics[width=.7\textwidth]{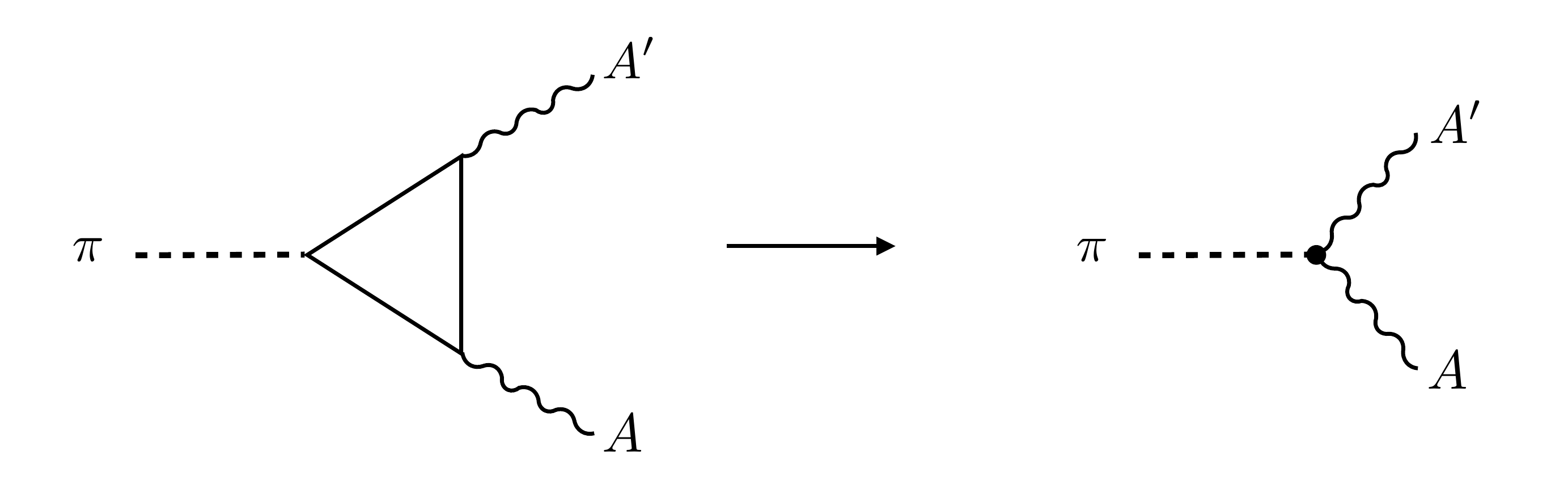}
\end{center}
\caption{\label{fig:pion_anomaly} Anomalous coupling of a photon and a hidden photon to a pion. }
\end{figure}

For $U(1)_{B-L}$, there is a contribution to $\text{Br}(\pi^0\to A'\gamma)$ from kinetic mixing (which we neglect), as well as a contribution from the mixed electromagnetic-$B-L$ anomaly (cf. Fig.~\ref{fig:pion_anomaly})
\begin{align}
\mathcal{L}=\frac{e\, g_{B-L}}{16\pi^2\, f_{\pi}}\text{Tr}\left[\sigma^3Q Q_{B-L}\right]\,F_{\mu\nu} \tilde F^{\prime \mu\nu} \pi^0
\end{align}
where  $\sigma^3=\text{diag}(1,-1)$, $Q=\text{diag}(2/3,-1/3)$, $Q_{B-L}=\text{diag}(1/3,1/3)$, $\tilde F^{\prime\,\mu\nu}=1/2\,\epsilon_{\mu\nu\alpha\beta}F^{\prime\,\alpha\beta}$ and $f_\pi$ is the pion decay constant. Since $\text{Tr}[\sigma^3Q Q_{B-L}]=\text{Tr}[\sigma^3Q^2]=1$ (including a color factor of 3), the branching ratio of pseudo-scalar mesons decaying into a photon and a $U(1)_{B-L}$ gauge boson follows from \eqref{eq:BRpi} by the replacement 
\begin{align}
\alpha\epsilon^2\quad  \to\quad  \alpha_{B-L}\,, &\quad \text{for} \quad U(1)_{B-L}\,.
\end{align}
 This result can be recovered in the VMD approach neglecting mass differences between the vector gauge bosons \cite{Ilten:2018crw}.\footnote{The branching ratio $\text{Br}(M\to A' A')=\mathcal{O}(\epsilon(M_{A'})^4)$ in the case of $U(1)_{L_\mu-L_\tau}$, $U(1)_{L_\mu-L_e}$ and $U(1)_{L_e-L_\tau}$, and vanishes identically for $U(1)_{B-L}$ in the absence of kinetic mixing, because $\text{Tr}[\sigma^3]=0$.}

\subsection{Collider experiments}\label{sec:colliders}

Hidden photons can also be produced in collider experiments, by s-channel production and meson decays. Searches have been performed by ATLAS and CMS at the LHC \cite{Curtin:2014cca}, by LHCb \cite{Aaij:2017rft, Ilten:2015hya, Ilten:2016tkc}, at the $e^+e^-$ colliders BaBar \cite{Aubert:2009cp, Lees:2014xha}, Belle \cite{Abe:2010gxa,Inguglia:2016acz} and KLOE \cite{Archilli:2011zc, Babusci:2012cr, Anastasi:2015qla, Anastasi:2016ktq}.

\subsubsection{Hadron colliders}
At the LHC, hidden photons can be produced directly through Drell-Yan production or through the decay of heavy resonances, e.g. $H \to Z A'$ \cite{Curtin:2014cca}. Feynman diagrams for these processes are shown on the left hand side of Fig.~\ref{fig:collider1}. 

\begin{figure}
\begin{center}
\includegraphics[width=1\textwidth]{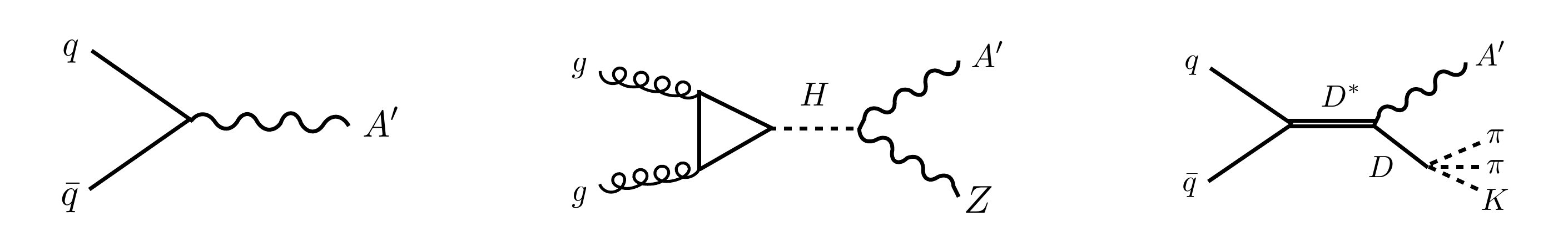}
\end{center}
\caption{\label{fig:collider1} Diagrams for the production of hidden photons at hadron colliders. Drell-Yan production (left), Higgs decays (center) and excited meson decays (right). }
\end{figure}

Limits from Drell-Yan production of hidden photons can be obtained for masses $M_A^2 > (12\ \text{GeV})^2$ because of cuts imposed to suppress backgrounds. The production cross section for an on-shell $A'$ for a given quark initial state can be brought into the usual form
\begin{align}
\sigma(q\bar q\to A' )=\frac{12\pi}{M_{A'}^2}\,\text{Br}(A'\to q\bar q)\,.
\end{align}

Production of hidden photons in Higgs decays are further constrained by the mass of the hidden gauge boson and the Higgs decay width is given in \eqref{eq:Hwidths} (see Appendix~\ref{rotation}). For the decays $H\to \gamma A'$, there is no suppression $\propto M_{A'}^2/M_Z^2$, but the partial decay width is loop-suppressed. Note that Higgs decays into hidden gauge bosons at tree-level are only possible for a non-zero kinetic mixing parameter, which we assume to vanish in the case of $U(1)_{B-L}$ and to be strongly suppressed for $U(1)_{L_\mu-L_e}$, $U(1)_{L_e-L_\tau}$ and $U(1)_{L_\mu-L_\tau}$, because of the power-suppression of $\epsilon(M_{A'}^2)\approx m_\ell^2/M_{A'}^2$. 

Another way to produce a hidden photon is by taking advantage of the large production cross section of heavy, excited mesons at LHCb, which decay into the ground state by radiating a photon, as illustrated on the right of Fig.~\ref{fig:collider1}. The authors of \cite{Ilten:2015hya} proposed a search through the neutral rare charm meson decay $D^*\to DA'$, for which the $\text{Br}(D^*\to D\gamma)\approx 35\%$ is particularly large, because the mass difference $\Delta M_D=M_{D^*}-M_D = 142.12\pm 0.07$ MeV leads to a phase space suppression of $\text{Br} (D^*\to D\pi^0)\approx 65\%$. The latter also contributes to the signal through subsequent pion decays $\pi^0\to \gamma A'$. For secluded $U(1)_X$ gauge bosons and a given luminosity $\mathfrak{L}_\mathrm{LHCb}$ the number of hidden photons produced from $D^*$ decays is therefore given by
\begin{align}\label{eq:exitedD}
N_{A'} = \mathfrak{L}_\mathrm{LHCb} \,  \sigma^\mathrm{prod}_{D*}
 \Big[  \text{Br}(D^*\to D\gamma)  \epsilon^2 \Big(1 &- \frac{M_{A'}^2}{\Delta M_D^2}\Big)^\frac{3}{2} \!\!\!+ \text{Br}(D^*\to D\pi^0) \text{Br}(\pi^0\to \gamma\gamma)2\epsilon^2 \Big(1-\frac{M_{A'}^2}{M_{\pi^0}^2}\Big)^2\Big] \,,
\end{align}
where $\sigma_{D^*}^\text{prod}=\sigma(pp \to D^* +X)$ denotes the $D^*$ production cross section. 
For the $U(1)_{L_\mu-L_e}$, $U(1)_{L_e-L_\tau}$, $U(1)_{L_\mu-L_\tau}$ and $U(1)_{B-L}$ gauge groups, the kinetic mixing factors $\epsilon^2$ in \eqref{eq:exitedD} are replaced by 
\begin{align}\label{eq:hadRescale}
\alpha\epsilon^2\quad  \to\quad \begin{cases} \alpha_{\mu e}\,\epsilon_{\mu e}(M_{A'})^2\,\,, &\quad \text{for} \quad U(1)_{L_\mu-L_e}\,,\\
 \alpha_{e \tau}\,\epsilon_{e\tau}(M_{A'})^2\,\,, &\quad \text{for} \quad U(1)_{L_e-L_\tau}\,,\\
\alpha_{\mu \tau} \,\epsilon_{\mu \tau}(M_{A'})^2\,, &\quad \text{for} \quad U(1)_{L_\mu-L_\tau}\,,\\ \alpha_{B-L}/4 \,,&\quad \text{for} \quad U(1)_{B-L} \,\,\text{and }\,\, \text{Br}(D^*\to D\gamma)\,,\\
\alpha_{B-L} \,,&\quad \text{for} \quad U(1)_{B-L} \,\,\text{and }\,\, \text{Br}(D^*\to D\pi^0)\,.
\end{cases}
\end{align}
\subsubsection*{A note on LHCb displaced searches}

The simple coupling rescaling procedure of \eqref{eq:hadRescale} is only applicable to limits obtained from searches of prompt decays. However, the authors of \cite{Ilten:2015hya} have also proposed a search for displaced $A'$ decays in $D^*\to D \gamma $ transitions. The derivation of these limits proceeds in anlaogy to the calculation of meson decay induced beam dump limits described in  Section~\ref{sec:protBD}. This, however, demands knowledge of the $D^*$  production spectra. To our knowledge these spectra have not yet been measured for LHC energies of $\sqrt{s}=13$ TeV and the Monte Carlo generated spectra used in the calculations in~\cite{Ilten:2015hya} have not been published. To obtain rigorous constraints from $D^*\to D \gamma $ transitions to the gauge groups discussed in this work, such Monte Carlo simulations of the $D^*$ production spectra need to be done. This is, however, beyond the scope of this letter and we will leave it to future work.

\subsubsection{$e^+e^-$ colliders}
At $e^+e^-$ colliders, hidden gauge bosons are produced through radiative return or through heavy meson decay. Feynman diagrams for the corresponding processes are shown in Fig.~\ref{fig:collider2}. In the latter case, the decay widths $\Gamma(\Phi\to \eta A')$ can be obtained in full analogy to
the $D^*\to D A'$ case, because the initial state plays no role. In the case of radiative return, the differential production cross section for the $U(1)_X$ gauge boson is given by \cite{Essig:2009nc}
 \begin{align}\label{eq:sigeegA}
\frac{\sigma{(e^+e^-\to \gamma A')}}{d\cos\theta} = \frac{2\pi \alpha^2 \epsilon^2}{s}\left(1-\frac{M_{A'}^2}{s}\right)\frac{1+\cos^2\theta+\frac{4M_{A'}^2\,s}{(s-M_{A'}^2)^2}}{1-\cos^2\theta}\,,
\end{align}
where $\theta$ is the angle between the beam line and the photon momentum.
The production cross section for gauged lepton flavour number and $U(1)_{B-L}$ follows from \eqref{eq:sigeegA} with the replacements
\begin{align}\label{eq:Ebeamdumps}
\alpha^2\epsilon^2\quad  \to\quad \begin{cases} \alpha\alpha_{\mu e}\,\,, &\quad \text{for} \quad U(1)_{L_\mu-L_e}\,,\\
\alpha\alpha_{e \tau}\,\,, &\quad \text{for} \quad U(1)_{L_e-L_\tau}\,,\\
\alpha\alpha_{\mu \tau} \,\epsilon_{\mu \tau}(M_{A'})^2\,, &\quad \text{for} \quad U(1)_{L_\mu-L_\tau}\,,\\
\alpha \alpha_{B-L}\,, &\quad \text{for} \quad U(1)_{B-L}\,,
\end{cases}
\end{align}
making this channel particularly relevant for all gauge groups apart from $U(1)_{L_\mu-L_\tau}$.

\begin{figure}
\begin{center}
\includegraphics[width=.7\textwidth]{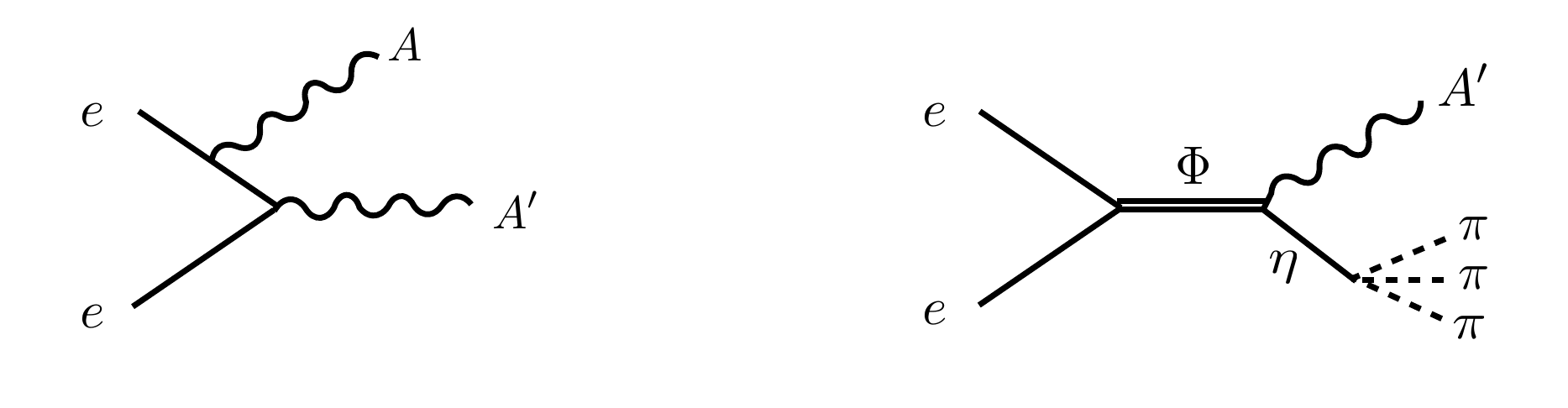}
\end{center}
\caption{\label{fig:collider2} Diagrams for the different production mechanisms of hidden photons at $e^+e^-$ colliders: Radiative return (left) and meson decays (right). }
\end{figure}

For the experiments BaBar \cite{Aubert:2009cp, Lees:2014xha}, Belle \cite{Abe:2010gxa,Inguglia:2016acz} and KLOE \cite{Archilli:2011zc, Babusci:2012cr, Anastasi:2016ktq, Anastasi:2015qla}, the decays are prompt for all relevant regions. Again we use the relevant branching fractions from Section~\ref{sec:BRs}.
As our hidden photons also feature invisible decays into neutrinos they can also be searched for in a mono photon (or mono-$\Phi$) search~\cite{Lees:2017lec}. Due to the lower SM background, the mono-photon searches can allow for more stringent limits than searches for $e^+e^-\to \gamma A' \to \gamma \ell^+\ell^-$  \cite{Ferber:2018}.

\subsection{Rare $\mu$ and $\tau$ decays and Mu3e}\label{sec:Mu3e}
Experiments designed to search for lepton flavour violation are particularly well suited to constrain the gauge groups we consider here, because all non-trivially anomaly-free gauge groups have gauge couplings to leptons. However, since none of the gauge bosons we consider have flavour-changing neutral couplings, decays of the type $\mu^+\to e^+e^-e^+$ or $\tau^+\to \mu^+e^-e^+$, etc., are not mediated at tree-level. 

More promising are the rare muon and $\tau$ decays into charged and neutral leptons, 
\linebreak $\mu^+\to e^+\nu_e\bar \nu_\mu A'(\to e^+e^-)$ or $\tau^+\to e^+\nu_e\bar \nu_\tau A'(\to e^+e^-)$. In presence of new gauge bosons this process can be mediated
by the diagrams shown in Fig.~\ref{fig:mu3e}.
However, these processes are also present in the SM via the process, $\mu^+\to e^+\nu_e\bar \nu_\mu \gamma^*(\to e^+e^-)$ or $\tau^+\to e^+\nu_e\bar \nu_\tau \gamma^*(\to e^+e^-)$, etc.
The best measurements of these processes have been performed by the SINDRUM \cite{Bertl:1985mw} and CLEO collaborations \cite{Alam:1995mt}, respectively,
\vspace{-.3cm}
\begin{align}
\text{Br}(\mu^- \to e^-e^+e^-\nu_e\nu_\mu) &=(3.4\pm0.4)\times 10^{-5}\,,\\
\text{Br}(\tau \to e\ e^+e^-  \nu_e \nu_\tau) &=(2.8\pm 1.5)\times 10^{-5}\,,\\
\text{Br}(\tau \to \mu\ e^+e^- \nu_\tau\nu_\mu) &\leq 3.2 \times 10^{-5}\,.
\end{align}
Below the muon threshold the SM background can be reduced by requiring the hidden photon to be on-shell.
The future {\it Mu3e} experiment will probe $10^{15}-10^{16}$ muon decays, providing three to four orders of magnitude more muons than SINDRUM \cite{Blondel:2013ia}. Projections for the expected limits from a search by the {\it Mu3e} experiment for secluded hidden photons contributing to the process $\mu^- \to e^-e^+e^-\nu_e\nu_\mu$ have been computed \cite{Echenard:2014lma}. These limits take advantage of a resonance in the invariant $e^+e^- -$ mass spectrum and are relevant for all gauge groups we consider, and particularly interesting for the  $U(1)_{L_\mu-L_e}$, $U(1)_{L_e-L_\tau}$ and $U(1)_{L_\mu-L_\tau}$. 

%
\begin{figure}
\begin{center}
\includegraphics[width=.6\textwidth]{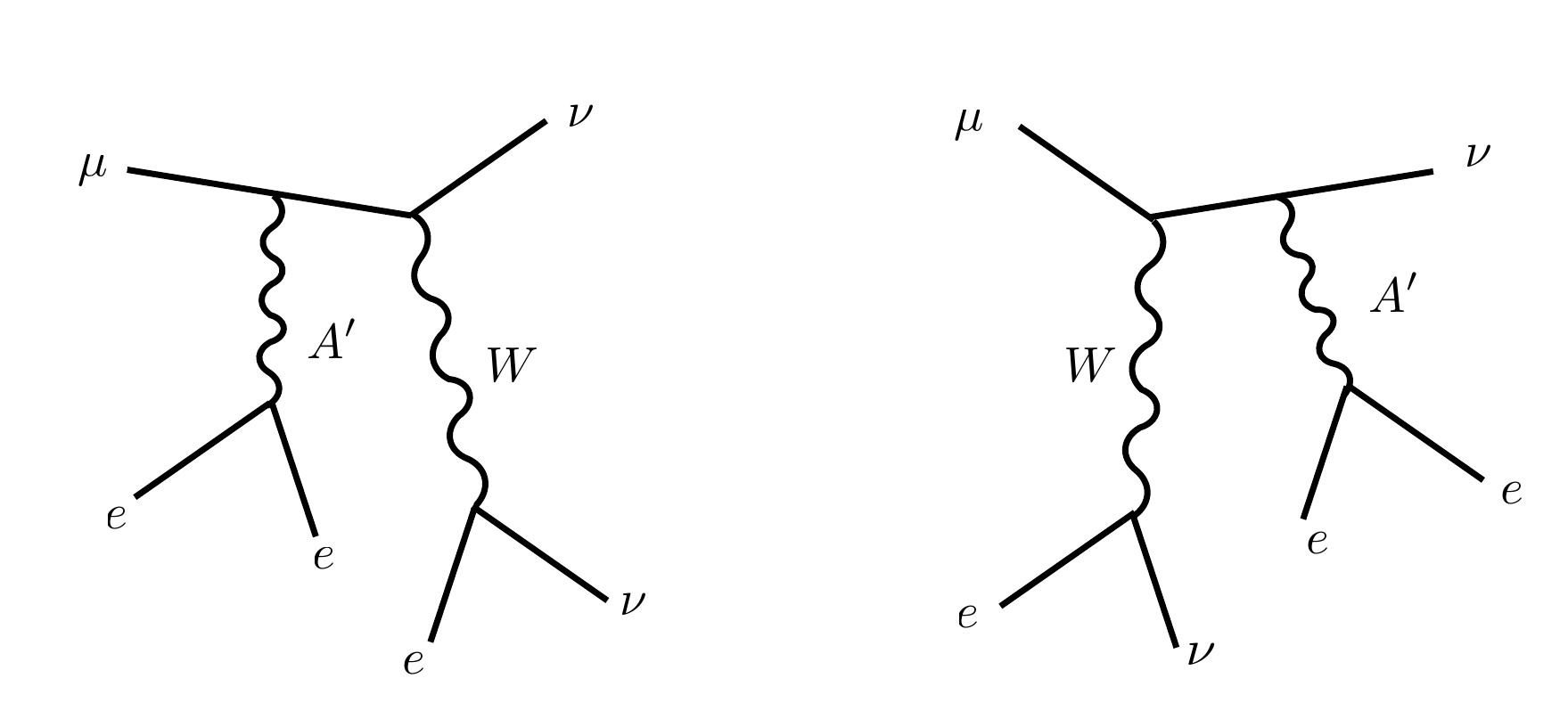}
\end{center}
\caption{\label{fig:mu3e} Diagram contributing to the $\mu^+\to e^-e^+e^-\nu_\mu\bar \nu_e$ signal mediated by the exchange of $A'$. }
\end{figure}
%
For the example of $U(1)_{L_\mu-L_\tau}$, the relevant diagrams are shown in Fig.~\ref{fig:mu3e}. Depending on the gauge group, there are also diagrams in which the hidden photon is radiated from the electron-leg, the electron-neutrino leg or from the $W^\pm$-propagator. The latter is suppressed with respect to inital- and final-state radiation by $m_\mu^2/M_W^2\approx 10^{-6}$ and can be neglected. Initial- and final-state radiation scale differently for the different gauge groups, due to the lepton-family specific couplings. We implement the model in \texttt{MadGraph5} \cite{Alwall:2014hca} to compute the branching ratio $\text{Br}(\mu^+\to e^+ \nu_e \bar \nu_\mu A')$ for the different gauge groups we consider, taking the appropriate scaling of initial and final-state radiation processes into account. We scan the coupling-mass parameter space and rescale the limits \cite{Echenard:2014lma} to derive projected limits for a future {\it Mu3e} search. 

\bigskip
Tau decays can similarly provide limits on gauge bosons with couplings to taus and electrons. To the best of our knowledge, no search for a resonance in rare leptonic tau decays $\tau\to e \nu_e\nu_\tau A'$ has been performed. We produced the corresponding branching ratio with \texttt{MadGraph} and estimated the current and future reach, assuming a sensitivity comparable to the error on the $\text{Br}(\tau \to e e^+e^-  \nu_e \nu_\tau)$ measurement by CLEO and the projected improvement by Belle II \cite{Arroyo-Urena:2017ihp}. The current and projected limits are rather weak. For a $U(1)_X$ or $U(1)_{B-L}$ gauge boson, current (future) searches for tau decays probe values of $\epsilon^2 \lesssim 5\times 10^{-4} (5\times 10^{-5})$, $\alpha_{B-L}/\alpha \lesssim 5\times 10^{-4} (5\times 10^{-5})$, respectively. For a $U(1)_{L_e-L_\tau}$ gauge boson, the current (projected) limits are slightly better,  $\alpha_{e-\tau}/\alpha \lesssim 3\times 10^{-4} (3\times 10^{-5})$. These constraints are not competitive with the other constraints discussed in this section.
The loop-induced contribution of an $U(1)_{L_\mu-L_\tau}$ gauge boson to the decay $\tau \to \mu \bar \nu_\mu \nu_\tau$ has been considered in \cite{Altmannshofer:2014cfa} in order to address the deviation in the measured $\text{Br}(\tau \to \mu \nu_\tau \bar\nu_\mu)$ compared to the SM prediction. For masses of $M_{A'}<10$ GeV an explanation of this deviation requires a coupling of $\alpha_{\mu-\tau}/\alpha \approx 10^{-2}$, which is safely excluded. 

\subsection{Limits from neutrino experiments}
\subsubsection{Neutrino trident production}
The neutrino trident process $\nu+Z\to \nu \mu^+\mu^-$ has been identified as an important probe for the gauge couplings of the $U(1)_{L_\mu-L_\tau}$ gauge boson \cite{Altmannshofer:2014pba}. The current limit has been obtained by combining the measurements of the CHARM-II collaboration \cite{Geiregat:1990gz}, the CCFR collaboration \cite{Mishra:1991bv}, and the NuTeV collaboration \cite{Adams:1998yf}. The weighted average normalized to the SM value is given by
\begin{align}
\frac{\sigma (\nu_{\mu}+Z\to \nu_{\mu} \mu^+\mu^-)}{\sigma_\text{SM} (\nu_{\mu}+Z\to \nu_{\mu} \mu^+\mu^-)}=0.83\pm 0.18\,.
\end{align}

In Fig.~\ref{fig:trident}, we show the $A'$-exchange diagram contributing to this process. For the $U(1)_{L_\mu-L_\tau}$, $U(1)_{L_\mu-L_e}$ and $U(1)_{B-L}$ gauge bosons, the $A'$ contribution to this process is not suppressed. The potential contribution from a $U(1)_X$ or a $U(1)_{L_e-L_\tau}$ gauge boson is completely negligible, because it only arises through mixing with the $Z$. 

In principle, there is an additional diagram for $U(1)_{L_\mu-L_e}$ and $U(1)_{B-L}$ gauge bosons from incoming electron neutrinos, but both the wide-band neutrino beam at CERN (CHARM-II) and Fermilab (CCFR/NuTeV) produce 2-3 orders of magnitude more muon than electron-neutrinos and we can safely neglect this contribution \cite{VandeVyver:1996qc, Romosan:1996nh}. 

We can therefore directly adopt the limits from \cite{Altmannshofer:2014pba}. In the future, these limits can be improved by measurements of the neutrino trident production cross section at LBNE \cite{Altmannshofer:2014pba}, with the INGRID detector at T2K \cite{Kaneta:2016uyt}, and by measurements of atmospheric neutrino trident production with Cherenkov telescopes \cite{Ge:2017poy}.

%
\begin{figure}
\begin{center}
\includegraphics[width=.3\textwidth]{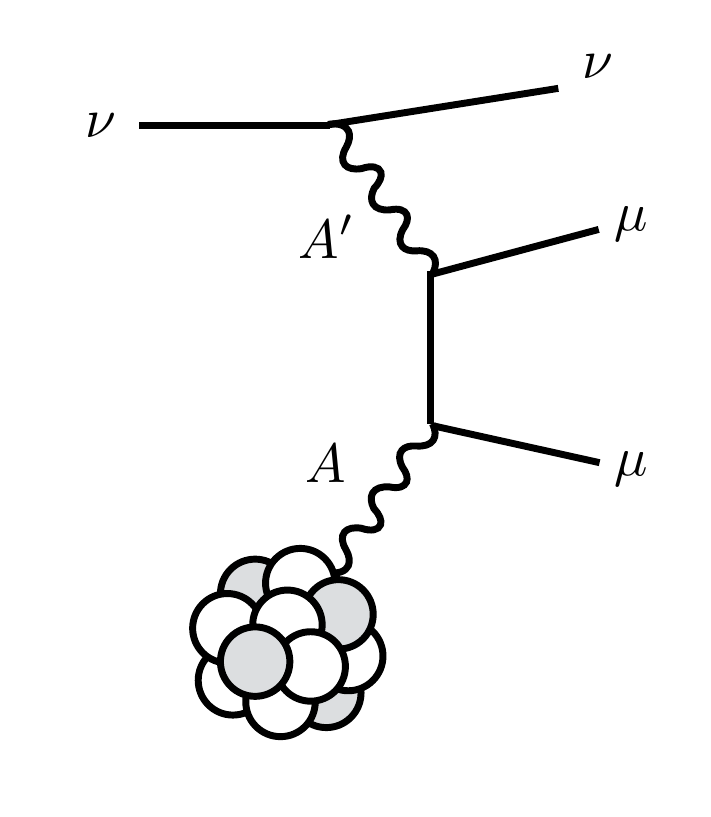}
\end{center}
\caption{\label{fig:trident} Diagram for neutrino trident production of muons.  }
\end{figure}
%

\subsubsection{Borexino}
Borexino is a liquid scintillator experiment measuring solar neutrinos scattering off electrons \cite{Bellini:2011rx}. 
This can be used to probe non-standard interactions between the neutrinos and the target.

The resulting constraints are irrelevant for hidden photons with suppressed couplings to neutrinos, but relevant for any other gauge group we consider. Limits from Borexino for the $U(1)_{B-L}$ gauge boson have been derived in \cite{Harnik:2012ni} and generalized to the case of $U(1)_{L_\mu-L_\tau}$ gauge bosons in \cite{Kaneta:2016uyt}. We adopt the method of \cite{Kaneta:2016uyt} and rescale the constraints on $U(1)_{B-L}$ bosons as
\begin{align}
\alpha_{B-L}^2\rightarrow \begin{cases}
 \Big[\sum_{i,j=1}^3 f_i \,| (U^\dagger Q_{\mu e} U)_{ij}|^2    \Big]^{1/2}\, \alpha_{\mu e}^2\,,&\quad \text{for} \quad U(1)_{L_\mu-L_e}\,,\\[.3cm]
 \Big[\sum_{i,j=1}^3 f_i \,| (U^\dagger Q_{e\tau} U)_{ij}|^2    \Big]^{1/2}\,  \alpha_{e \tau}^2\,, &\quad \text{for} \quad U(1)_{L_e-L_\tau}\,,\\[.3cm]
\Big[\sum_{i,j=1}^3 f_i \,| (U^\dagger Q_{\mu \tau} U)_{ij}|^2    \Big]^{1/2}\, \alpha\, \alpha_{\mu \tau} \, \epsilon_{\mu\tau} (q^2)\,,&\quad \text{for} \quad U(1)_{L_\mu-L_\tau}\,,
\end{cases}
\end{align}
in which $f_1, f_2$ and $ f_3$ denote the fraction of the corresponding mass eigenstates of $^7$Be neutrinos at the earth \cite{Nunokawa:2006ms}, $U$ is the lepton mixing matrix and $Q_{\mu\tau}=\text{diag}(0, 1, -1)$, $Q_{\mu e}=\text{diag}(1, 0, -1)$ and $Q_{e \tau}=\text{diag}(1, 0, -1)$. Mixing suppressed contributions have been omitted.

\subsubsection{Texono}
\begin{figure}[h!]
\includegraphics[width=0.3\textwidth]{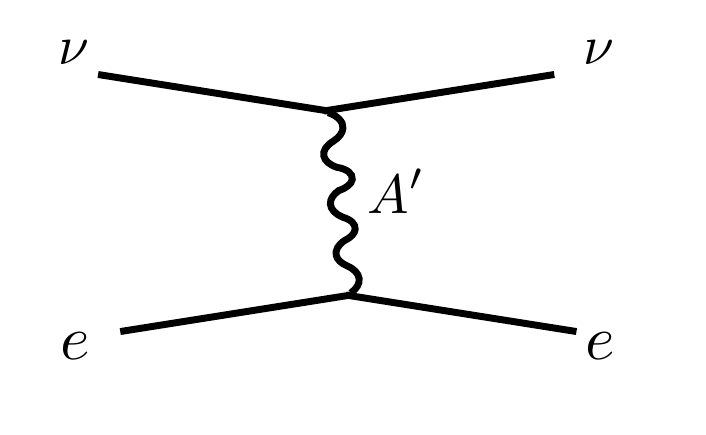}
\centering
\caption{Neutrino-electron scattering via $A'$ exchange.}
\label{fig:enu_scatt}
\end{figure}
The TEXONO collaboration has measured the elastic $\bar\nu_e - e^-$ scattering cross section at the Kuo-Sheng Nuclear Power Reactor with a CsI(TI) scintillating crystal array  \cite{Deniz:2009mu}. The detector is located at a distance of 28 m  from the reactor core such that the flux of incoming neutrinos can be assumed to be pure $\bar\nu_e$.
 
 The experimentally determined $\bar\nu_e - e^-$ scattering spectrum can be used to constrain extra scattering due to the exchange of a new light $A'$ boson  as depicted in Fig.~\ref{fig:enu_scatt}. This has been done in  \cite{Bilmis:2015lja, Lindner:2018kjo} for the case of a gauged $U(1)_{B-L}$ with a particular emphasis on interference effects of the $A'$ with the SM. The determined limit on the gauge coupling $g_{B-L}$ directly applies to the case of $U(1)_{L_\mu-L_e}$ and $U(1)_{L_e-L_\tau}$, where the first generation of leptons also carries a charge of $|Q_{L_e}|=1$.

 \subsubsection{COHERENT}
 
 Another limit for $A'$ couplings to neutrinos can be derived from coherent elastic neutrino-nucleus scattering (CE$\nu$NS).  The high sensitivity of CE$\nu$NS to deviations of the weak mixing angle from the predicted SM value \cite{Akimov:2015nza} can be translated into a bound on the (induced) mixing parameter $\epsilon$, which generates such deviations. \\
 The COHERENT experiment has only recently measured this process for the first time \cite{Akimov:2017ade}. The detector consisted of a CsI target  that was exposed to neutrinos from decays of secondary pions, which were produced from a proton beam dumped into a mercury fixed target.  The observed signal has been used in \cite{Abdullah:2018ykz} to set limits on a secluded hidden photon as well as on a $U(1)_{L_\mu-L_\tau}$ gauge boson. Furthermore, a future accelerator setup with a COHERENT detector consisting of a NaI/Ar target and a total exposure of 10 ton$\cdot$year has been used to derive projected sensitivities.

 \subsubsection{Charm-II}

In the years 1987 -- 1991 the CHARM-II detector has been exposed to the horn focused wide band neutrino beam at CERN in order to study $\nu_\mu(\bar\nu_\mu)e^-$ scattering. The CHARM-II collaboration has published both the measured total number of scattering events \cite{Vilain:1994qy} as well as the differential cross section \cite{Vilain:1993kd}. 

In \cite{Bilmis:2015lja, Lindner:2018kjo}  these results have been used to set a limit on the coupling constant $g_{B-L}$ of a gauged $U(1)_{B-L}$, again considering interference effects. However, from the experimental publications, the exact neutrino fluxes seem to be unknown and the SM prediction of the differential cross section given in \cite{Vilain:1993kd} seems to be determined by a shape fit. Therefore, we are in doubt whether a rigorous calculation of the neutrino rate $R$ at CHARM-II is possible and whether the $\chi^2$-fit used for limit determination in \cite{Bilmis:2015lja}  is applicable. Noting this, we will show the corresponding limits by a dashed line assuming the correct neutrino flux was used for limit calculation.

In the cases  of $U(1)_{L_\mu-L_e}$ and $U(1)_{L_\mu-L_\tau}$ (for $U(1)_{L_e-L_\tau}$ no coupling to $\nu_\mu$ exists) first and second generation leptons carry opposite charges under the new group. Thus, the  interference term changes sign relative to the  $U(1)_{B-L}$ case and the full interference-sensitive limit cannot be obtained by rescaling. Therefore, we extract an upper bound on the change in cross section $\Delta \sigma_\mathrm{lim}$ from the limits on $g_{B-L}$ in \cite{Bilmis:2015lja}, which have been provided for the case of only taking  into account pure $A'$ contributions (no interference). We use this bound  $\Delta \sigma_\mathrm{lim}$ to set limits on $g_{ij}$ where constructive interference is expected ($\nu_\mu e^-$ scattering in both cases). Here a full analysis accounting for interference should yield stronger bounds. Therefore, this approach is  conservative.

\subsubsection{Neutrino matter effects and Super-K}
A very recent paper~\cite{Wise:2018rnb} considered the fact that new leptonic forces modify the matter potentials relevant for neutrino oscillations.
If the matter effects are changed this should be visible in neutrino oscillations and
Super-K provides an interesting limit on the difference between the matter potential for $\nu_{\mu}$ and $\nu_{\tau}$, $|\epsilon_{\mu\mu}-\epsilon_{\tau\tau}|<0.147$~\cite{Mitsuka:2011ty,Ohlsson:2012kf,Gonzalez-Garcia:2013usa}\footnote{Note, that here $\epsilon_{\mu\mu}$ and $\epsilon_{\tau\tau}$ quantify the interaction strength between muon and tau neutrinos ~\cite{Mitsuka:2011ty,Ohlsson:2012kf,Gonzalez-Garcia:2013usa}, not to be confused with $\epsilon_{\mu\tau}, \epsilon_{\mu e}$ and $\epsilon_{e \tau}$ defined in \eqref{eq:lmutaumix}. }. 

In the region of interest to us the matter effects are given by,
\begin{align}
|\epsilon_{\mu\mu}-\epsilon_{\tau\tau}|=\begin{cases}\frac{4\pi\alpha_{\mu e}}{\sqrt{2}G_{F}M^{2}_{A'}}
&\quad{\textrm{for}}\quad U(1)_{L_\mu-L_e}\\
\frac{4\pi\alpha_{e \tau}}{\sqrt{2}G_{F}M^{2}_{A'}}&\quad{\textrm{for}}\quad U(1)_{L_e-L_\tau},\\
\quad\quad0&\quad{\textrm{for}}\quad U(1)_{L_\mu-L_\tau},\\
\quad\quad 0&\quad{\textrm{for}}\quad U(1)_{B-L}.
\end{cases}
\end{align}
For the latter two groups the measurement is insensitive because there is no difference in the matter effects for the two considered neutrino species.

The authors of~\cite{Wise:2018rnb} also consider a potential future measurement in the DUNE-like setup that could improve the limits into the $|\epsilon_{\mu\mu}-\epsilon_{\tau\tau}|\sim 0.01$ range.

\subsection{White dwarf cooling}\label{sec:WD}
 %
\begin{figure}
\begin{center}
\includegraphics[width=.5\textwidth]{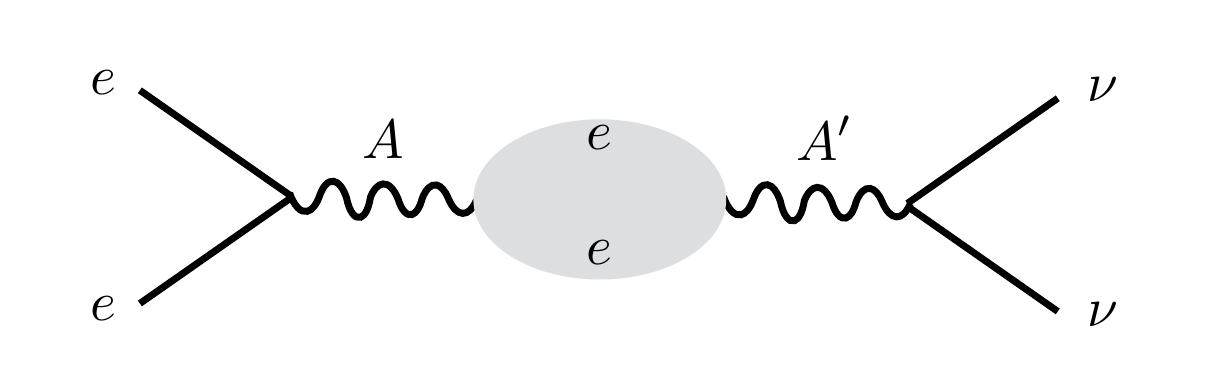}
\end{center}
\caption{\label{fig:WDs} Neutrino production contributing to white dwarf cooling from hidden photon decays.  }
\end{figure}
%

Several constraints on hidden photons arise from astrophysical observations. For example, limits on hidden photons from supernovae constraining very low couplings have been discussed in a number of papers \cite{Bjorken:2009mm,Roy:2013, Rrapaj:2015wgs,Hardy:2016kme,Chang:2016ntp,Mahoney:2017jqk,Knapen:2017xzo}\footnote{For strong constraints at much lower masses see~\cite{Popov:1999,Redondo:2008aa,An:2013yfc,Redondo:2013lna}.} (mostly for the secluded case\footnote{A limit for B-L has been given in \cite{Knapen:2017xzo} but in the region of interest to us it is based on the hidden photon limit.}). However, there seem to be significant differences in the results and corresponding uncertainties in the limits. While there is clear need and motivation for further investigations, this is beyond the scope of this work. Furthermore, the very low coupling regime constrained by supernovae bounds is not the focus of this paper. We therefore prefer not show any limit and instead refer the reader to the corresponding literature. Further constraints arise from the potential impact of hidden photons on cosmic microwave background and big bang nucleosynthesis (BBN) anisotropies \cite{Fradette:2014sza} as well as from a potential hidden photon contribution to the cooling of white dwarfs~\cite{Dreiner:2013tja}.

The most relevant constraints for the parameter space we consider arise from white dwarf cooling, which is measured by observing variations of the white dwarf luminosity function. Following the strategy of~\cite{Dreiner:2013tja}, we consider plasmon decay into neutrinos mediated by hidden photons, which contributes to the cooling. The corresponding process is illustrated in Fig.~\ref{fig:WDs}. The limit on possible additional contributions from hidden photons is reported~\cite{Dreiner:2013tja} as a limit on the Wilson coefficient in
\begin{align}
\mathcal{L}=C_\text{WD}\, (\bar \nu \gamma_\mu P_L \nu)(\bar e \gamma_\mu e)\,,
\end{align}
with
\begin{align}
\frac{1.12\times 10^{-5}}{\text{GeV}^2} <C_\text{WD}< \frac{4.50\times 10^{-3}}{\text{GeV}^2} \,,
\end{align}
in which the upper limit corresponds to an interaction strength that leads to a trapping of the neutrinos, which therefore effectively do not contribute to the cooling of the white dwarf. Note that the trapping requires a sizable interaction with electrons and the above upper limit is probably quite conservative. 

For a secluded hidden photon the contribution to the Wilson coefficient $C_\text{WD}$ is strongly suppressed, because a coupling to neutrinos only arises through mixing with the $Z$ boson,
\begin{align}
C_\text{WD}&= \frac{4\pi}{M_{A'}^2}\,\alpha \epsilon\,\delta = \frac{4\pi}{M_{Z}^2}\,\alpha \epsilon\,,\qquad \text{for}\quad U(1)_X\,.
\end{align}
For the $U(1)_{B-L}$ and $U(1)_{L_\mu-L_e}$, $U(1)_{L_e-L_\tau}$, $U(1)_{L_\mu-L_\tau}$ gauge groups however, the contributions to the Wilson coefficient can become important for small masses and sizable couplings,
\begin{align}
C_\text{WD}&=\begin{cases}\displaystyle\frac{8\pi}{3M_{A'}^2}\,\alpha_{\mu e} \,,\qquad &\text{for}\quad U(1)_{L_\mu-L_e}\,,\\[.3cm]
\displaystyle \frac{8\pi}{3M_{A'}^2}\,\alpha_{e \tau} \,,\qquad &\text{for}\quad U(1)_{L_e-L_\tau}\,,\\[.3cm]
\displaystyle\frac{8\pi}{3 M_{A'}^2}\,\alpha_{\mu \tau} \,\epsilon_{\mu\tau} (M_{A'}^2)\,,\qquad &\text{for}\quad U(1)_{L_\mu-L_\tau}\,,\\[.3cm]
 \displaystyle\frac{4\pi}{M_{A'}^2}\,\alpha_{B-L}\,,\qquad &\text{for}\quad U(1)_{B-L}\,.
\end{cases}
\end{align}

\subsection{Big Bang Nucleosynthesis}\label{sec:bbn}
For the case of $U(1)_{L_\mu-L_\tau}$ an additional constraint arises from the coupling to neutrinos in the early universe. 
The $A'$ gauge boson will stay in equilibrium with $\nu_{\tau}$ and $\nu_{\mu}$. This provides additional energy to these neutrinos and leads to an increase in the effective number of neutrino degrees of freedom at BBN~\cite{Kamada:2015era}. 

In principle, a similar effect is present also for the other gauge groups with neutrino couplings. However, in this case also couplings to the electron and electron neutrino exist. A robust limit would require a more detailed analysis which we leave for future work.
\newpage

\begin{figure}[ht!]
\begin{center}
\vspace{-2.1cm}
\includegraphics[width=0.8\textwidth]{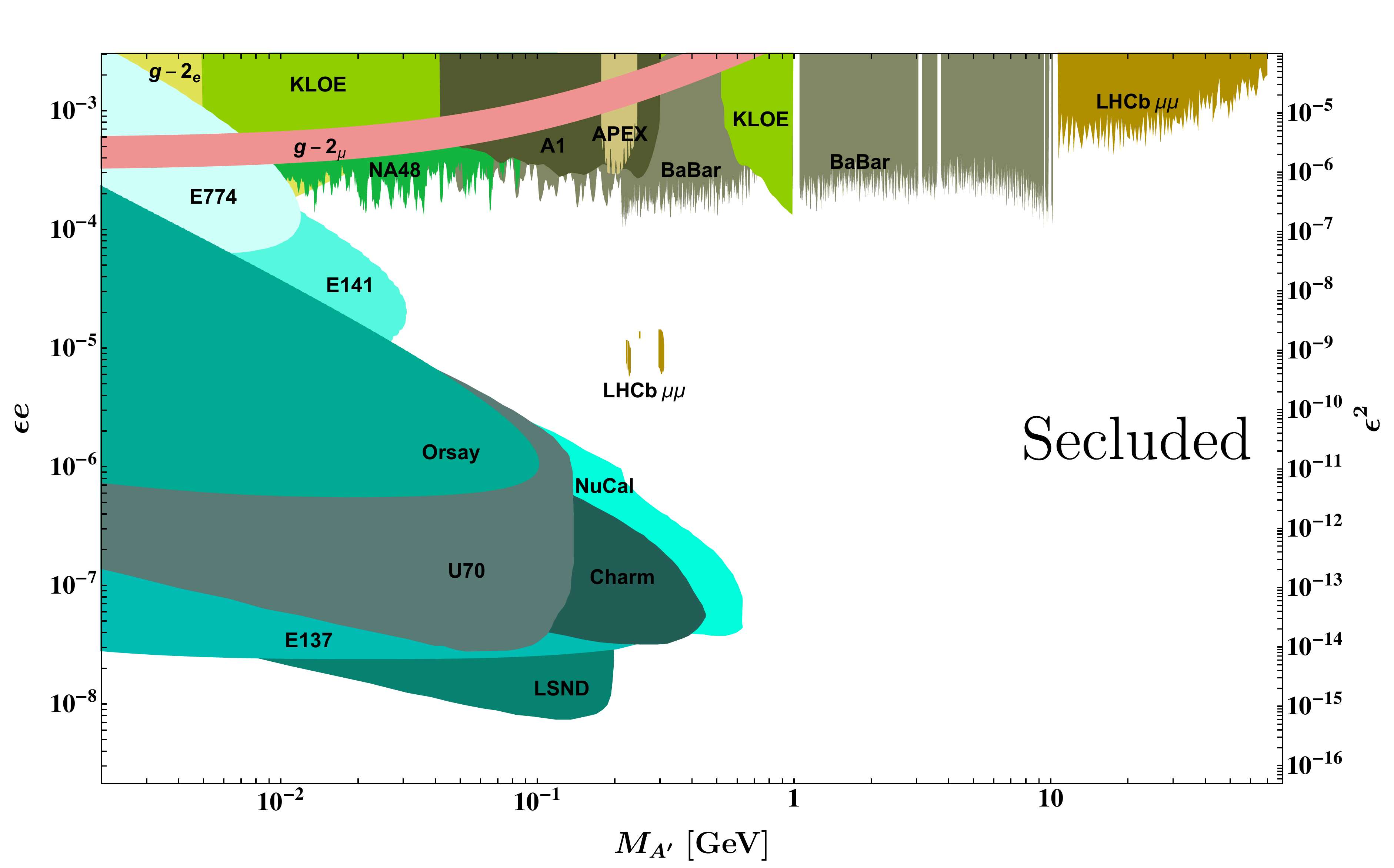}
\includegraphics[width=0.8\textwidth]{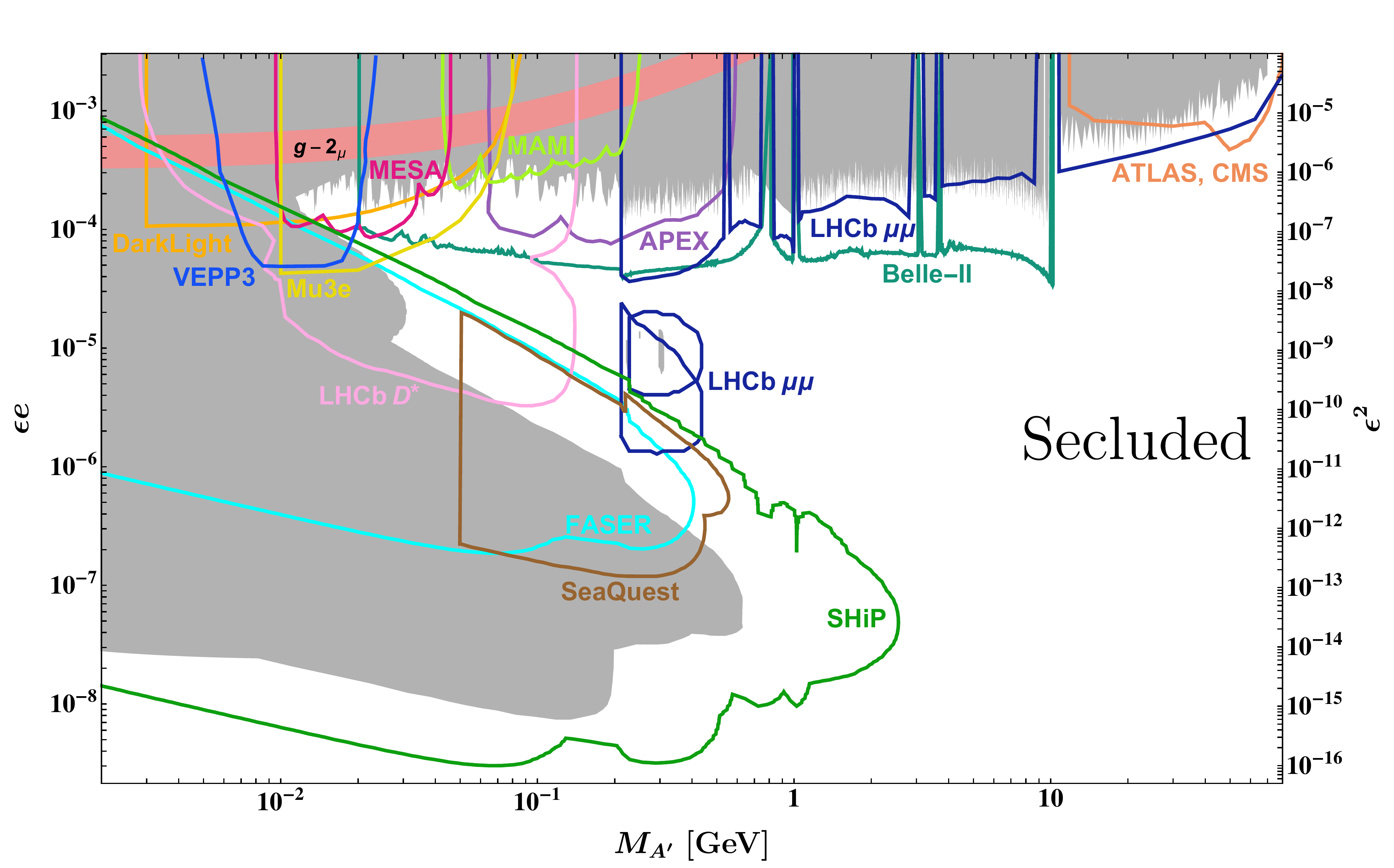}
\vspace{-.7cm}
\end{center}
\caption{\label{fig:universal}  Constraints from current (upper panel) and future (lower panel) experiments on a secluded $U(1)_X$ gauge boson with kinetic mixing parameter $\epsilon$. Additional constraints from supernova cooling are not shown (see Section~\ref{sec:WD}). }
\end{figure}
%
\vspace{-.5cm}
\section{Results}
\label{sec:results}
Our main results are summarized in Figs.~\ref{fig:BL}-\ref{fig:LmLt}, showing exclusion contours for a  $U(1)_{B-L}$, $U(1)_{L_\mu-L_e}$, $U(1)_{L_e-L_\tau}$ and $U(1)_{L_\mu-L_\tau}$,  respectively.
For each of the considered gauge groups we show two plots. One with the existing limits and another one with the planned and future experiments. 

For comparison we show the usual secluded hidden photon case $U(1)_{X}$ in Fig.~\ref{fig:universal}.
Note the features in the projected SHiP reach in Fig.~\ref{fig:universal}; for hidden photons with masses above the pion threshold,  the production through pion decays shuts off and the sensitivity for small gauge couplings is decreased. The dips for sizable masses correspond to hadronic resonances, which increase sensitivity for small gauge couplings and decrease it for sizable gauge couplings, as the hidden photon becomes short-lived. 

Let us now consider each of the different gauge groups and discuss the similarities and changes with respect to the case of a secluded hidden photon. For a detailed discussion of the calculation of beam dump limits and how they are related to the recasted limits we refer to Appendix~\ref{app:comparison}.

\begin{figure}[ht!]
\begin{center}
\vspace{-2.1cm}
\includegraphics[width=0.8\textwidth]{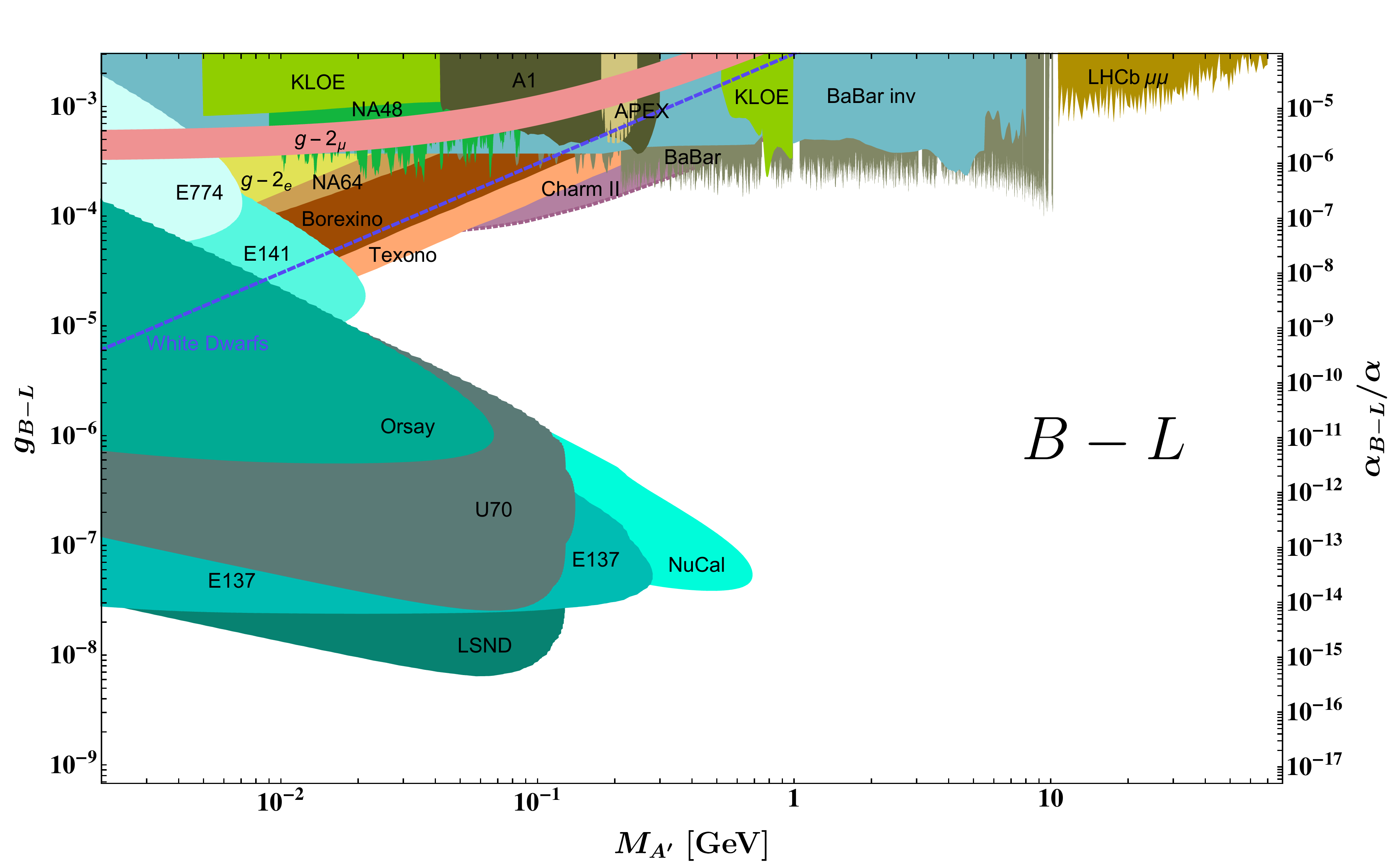}
\includegraphics[width=0.8\textwidth]{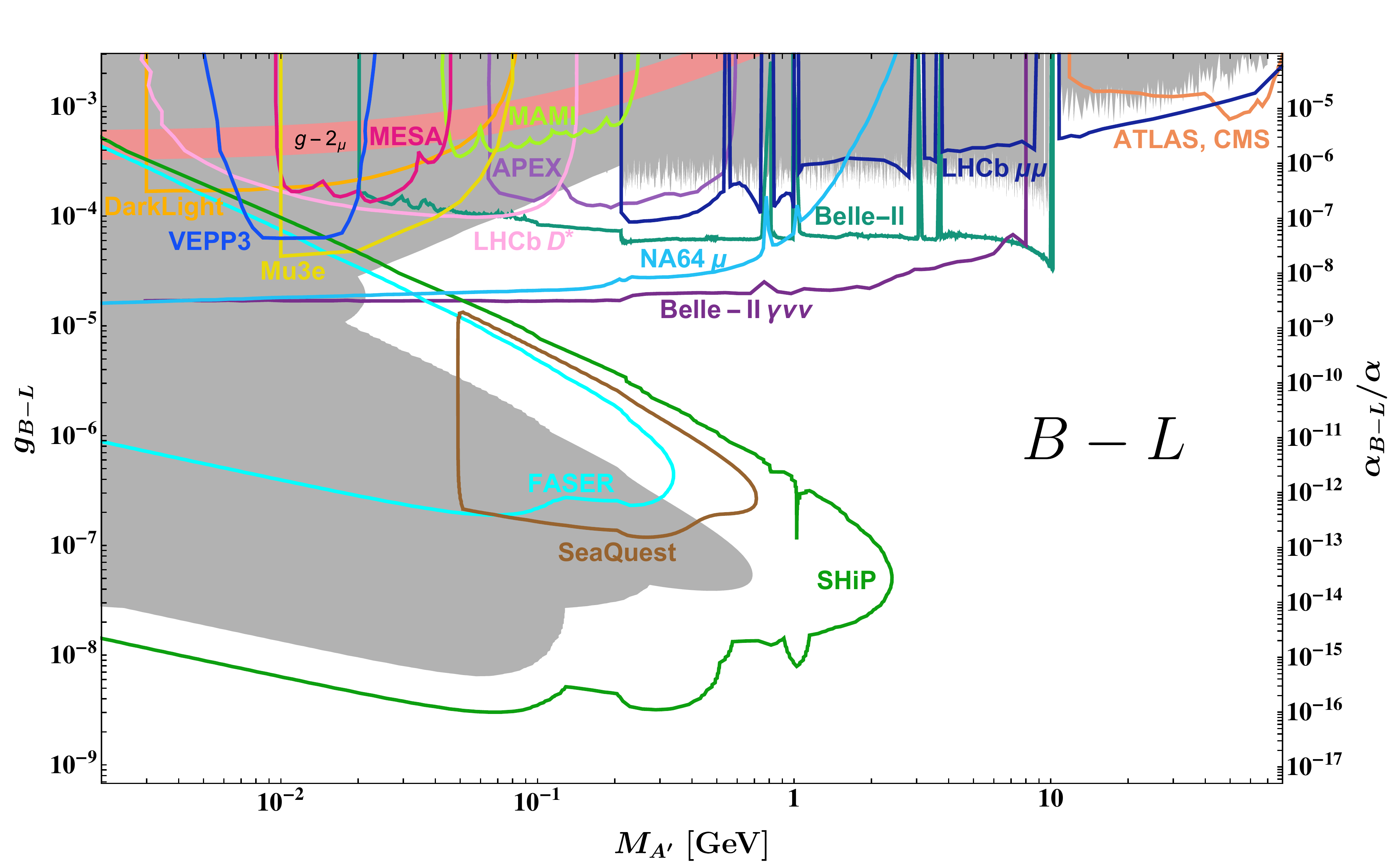}
\vspace{-.8cm}
\end{center}
\caption{\label{fig:BL} Constraints from current (upper panel) and future (lower panel) experiments on a $U(1)_{B-L}$ gauge boson with gauge coupling $g_{B-L}\equiv\epsilon\,e$. Additional constraints from supernova cooling and BBN are not shown (see Sections~\ref{sec:WD} and~\ref{sec:bbn}).}\vspace{-.3cm}
\end{figure}

\subsection{$U(1)_{B-L}$} 
The beam dump, fixed target and collider limits are very similar to the case of a secluded hidden photon. We note that the limit from CHARM and the LHCb displaced searches are absent because we lacked sufficient information to adequately reproduce these limits, not because there is a physics reason that makes these searches insensitive. However, the CHARM region is mostly covered by other experiments as one can also see from the rescaling done in \cite{Ilten:2018crw}.\\
The most notable difference arises from the coupling to neutrinos. This makes the B-L gauge group testable in a variety of neutrino experiments strongly constraining the (10-200)~MeV region. It also leads to constraints from the cooling of white dwarfs.
The most promising future probes are the beam dumps SHiP and SeaQuest, Belle-II, and at LHC, LHCb and FASER (similarly CodexB and MATHUSLA). The projected SHiP reach shows similar features as in the case of a secluded $U(1)_X$ couplings due to the tree-level coupling to hadrons.

\begin{figure}[th!]
\begin{center}
\vspace{-2.1cm}
\includegraphics[width=0.8\textwidth]{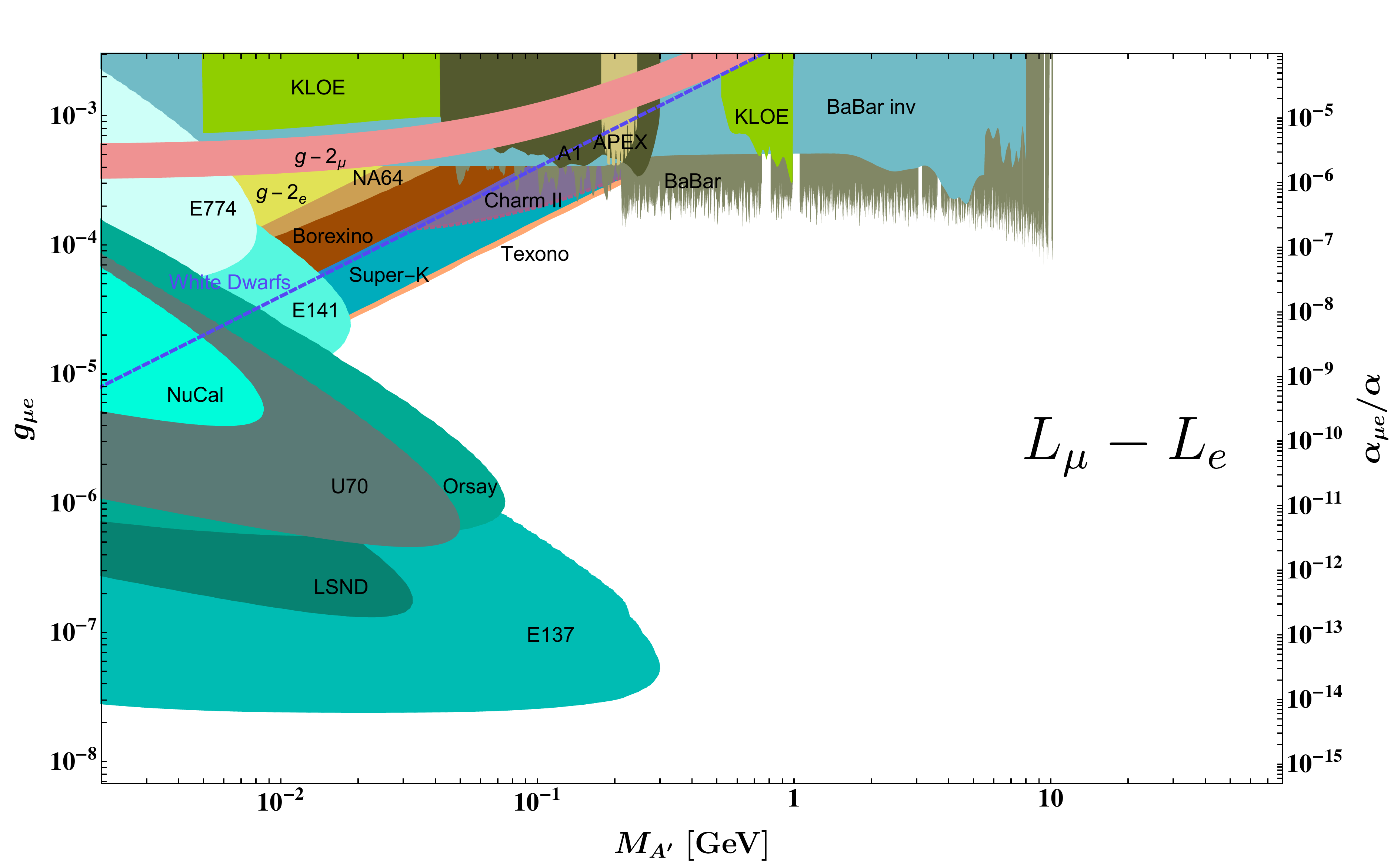}
\includegraphics[width=0.8\textwidth]{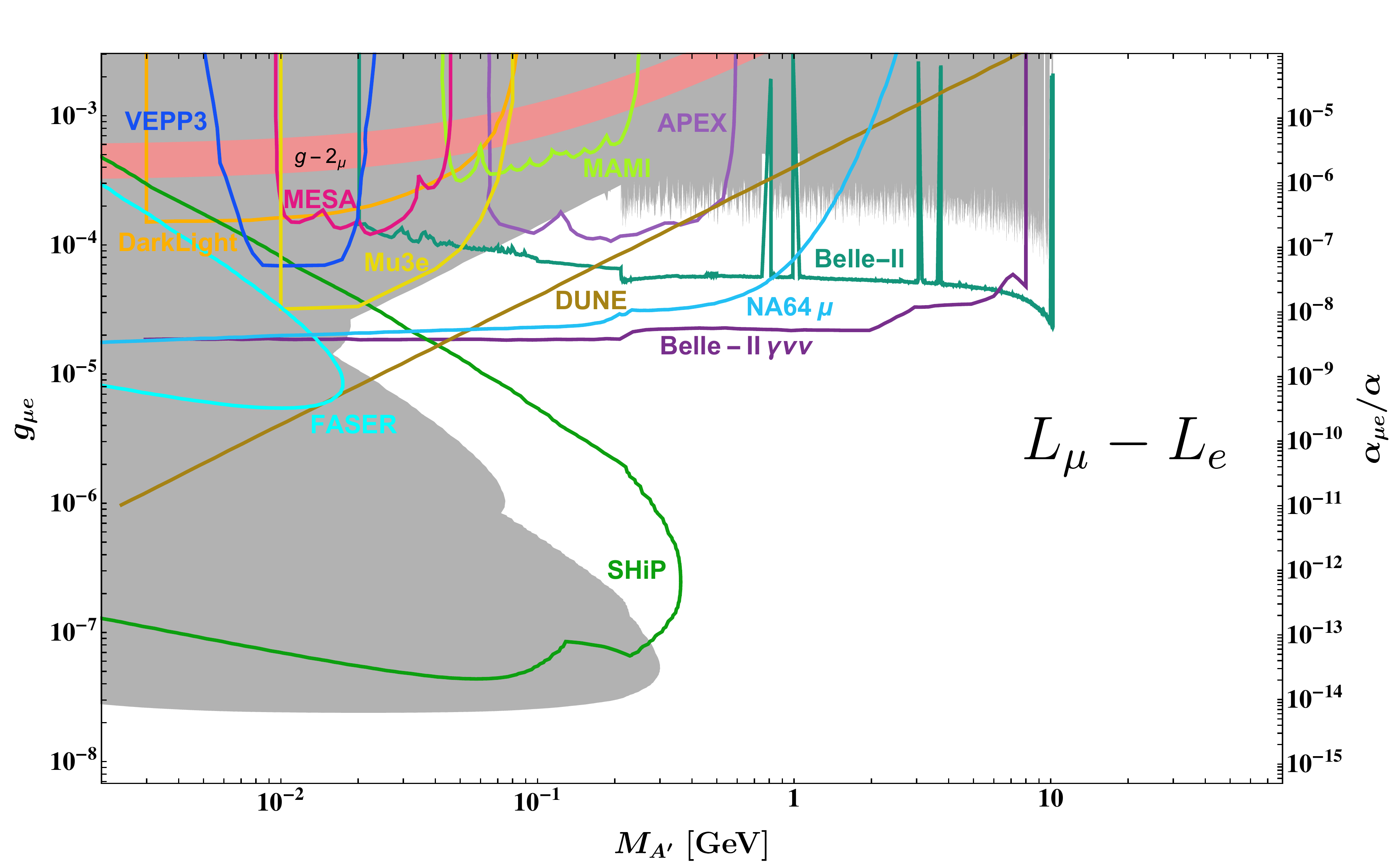}
\vspace{-.8cm}
\end{center}
\caption{\label{fig:LmLe} Constraints from current (upper panel) and future (lower panel) experiments on a $U(1)_{L_{\mu}-L_e}$ gauge boson with gauge coupling $g_{\mu-e}=\epsilon\,e$. Additional constraints from supernova cooling and BBN are not shown (see Sections~\ref{sec:WD} and~\ref{sec:bbn}). }\vspace{-.4cm}
\end{figure}

\subsection{$U(1)_{L_{\mu}-L_e}$}
For this and all the following gauged lepton family number groups one main difference is the weakening of all hadronic collider, beam dumps and fixed target experiments, since the only interaction with hadrons is via a loop-suppressed kinetic mixing. Electron beam dumps are favorable to explore very small couplings. The upper boundaries of the beam dump limits are significantly less affected, because this boundary arises from the premature decay of the produced particles in the shielding. It therefore mostly depends on the total decay width and is less sensitive to the production. Here, a favorable geometry is more important.
Strong limits from neutrino experiments lead to additional constraints.
Especially strong constraints arise from Super-K~\cite{Wise:2018rnb} due to the non-universal coupling of neutrinos to matter that modify the neutrino oscillations and the scattering of electron neutrinos in TEXONO~\cite{Bilmis:2015lja}.

Future interesting probes may be provided by SHiP (in the region where it benefits from a suitable geometry and a high boost factor), Belle-II, DUNE and NA64$\mu$. The reach for small couplings in SHiP and NA64$\mu$ is slightly diminished above the pion and the muon threshold, respectively.  

\begin{figure}[th!]
\begin{center}
\vspace{-2.1cm}
\includegraphics[width=0.8\textwidth]{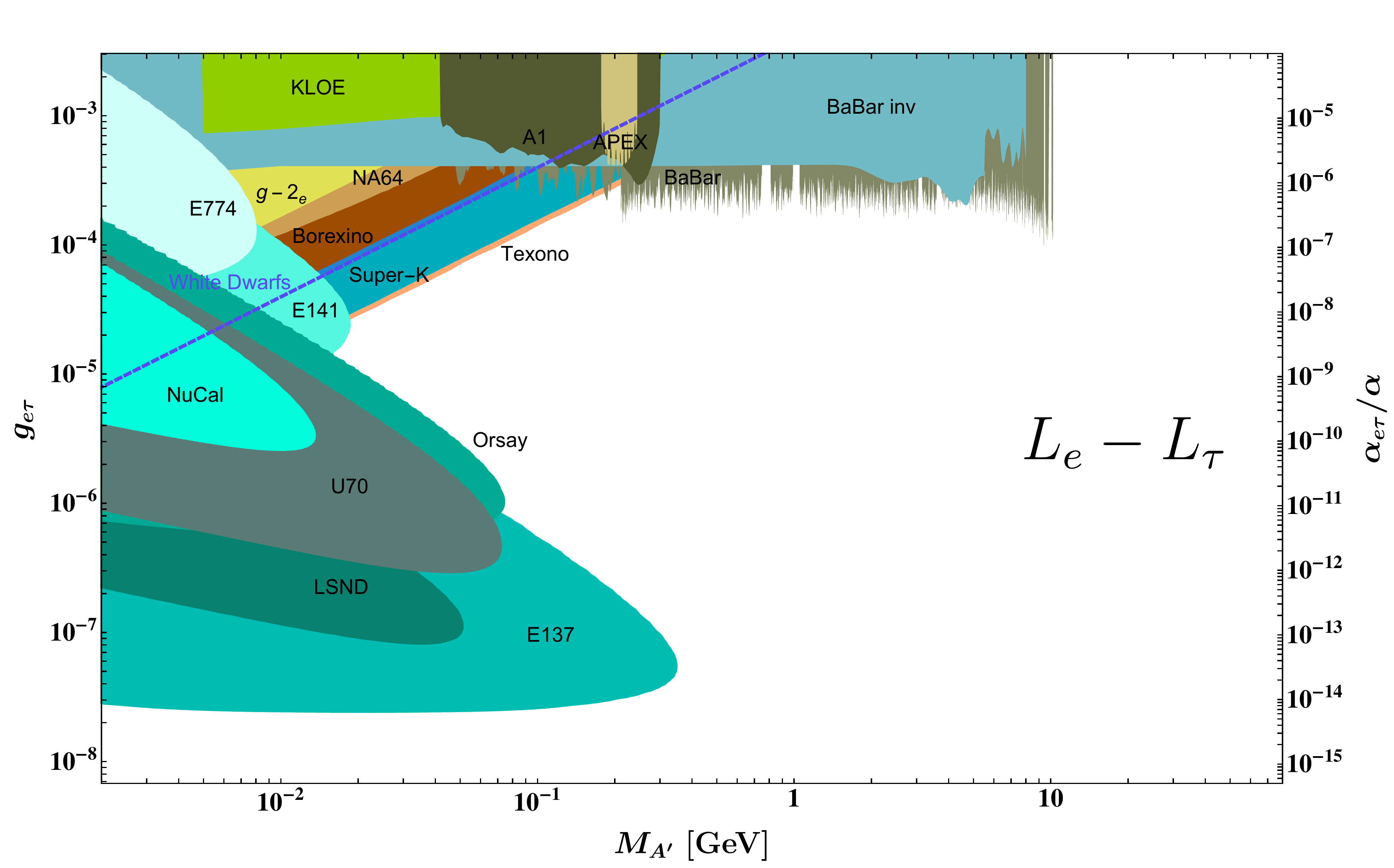}
\includegraphics[width=0.8\textwidth]{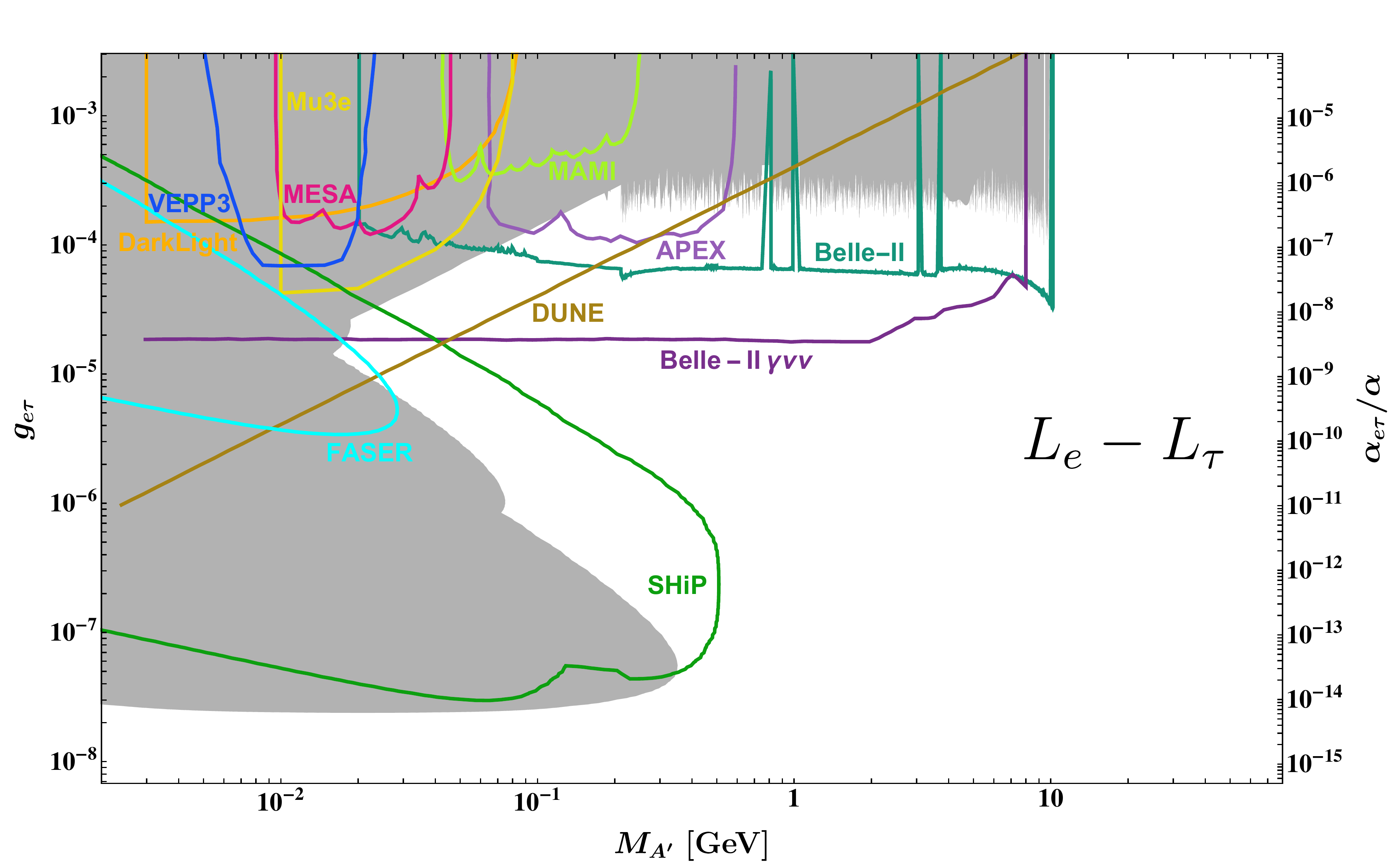}
\vspace{-.8cm}
\end{center}
\caption{\label{fig:LeLt} Constraints from current (upper panel) and future (lower panel) experiments on a $U(1)_{L_{e}-L_\tau}$ gauge boson with gauge coupling $g_{e-\tau}=\epsilon\,e$. Additional constraints from supernova cooling and BBN are not shown (see Sections~\ref{sec:WD} and~\ref{sec:bbn}). }\vspace{-.4cm}
\end{figure}

\subsection{$U(1)_{L_{e}-L_\tau}$}
For $U(1)_{L_{e}-L_\tau}$ the situation is very similar to that of $U(1)_{L_{\mu}-L_e}$.
The most notable difference is the absence of the high-mass KLOE limit based on a muon channel.

In addition to  SHiP, APEX, Belle-II and DUNE, also FASER gains some sensitivity compared to the $U(1)_{L_{\mu}-L_e}$ case. This is because above the threshold for the heavier of the two leptons, i.e. the muon in case of $U(1)_{L_{\mu}-L_e}$ the kinetic mixing is suppressed as it evolves towards zero at large momenta. For $U(1)_{L_{e}-L_\tau}$ this happens only above the tau mass. Therefore the mixing is larger in the relevant region.

\newpage

\begin{figure}[th!]
\begin{center}
\vspace{-2.1cm}
\includegraphics[width=0.8\textwidth]{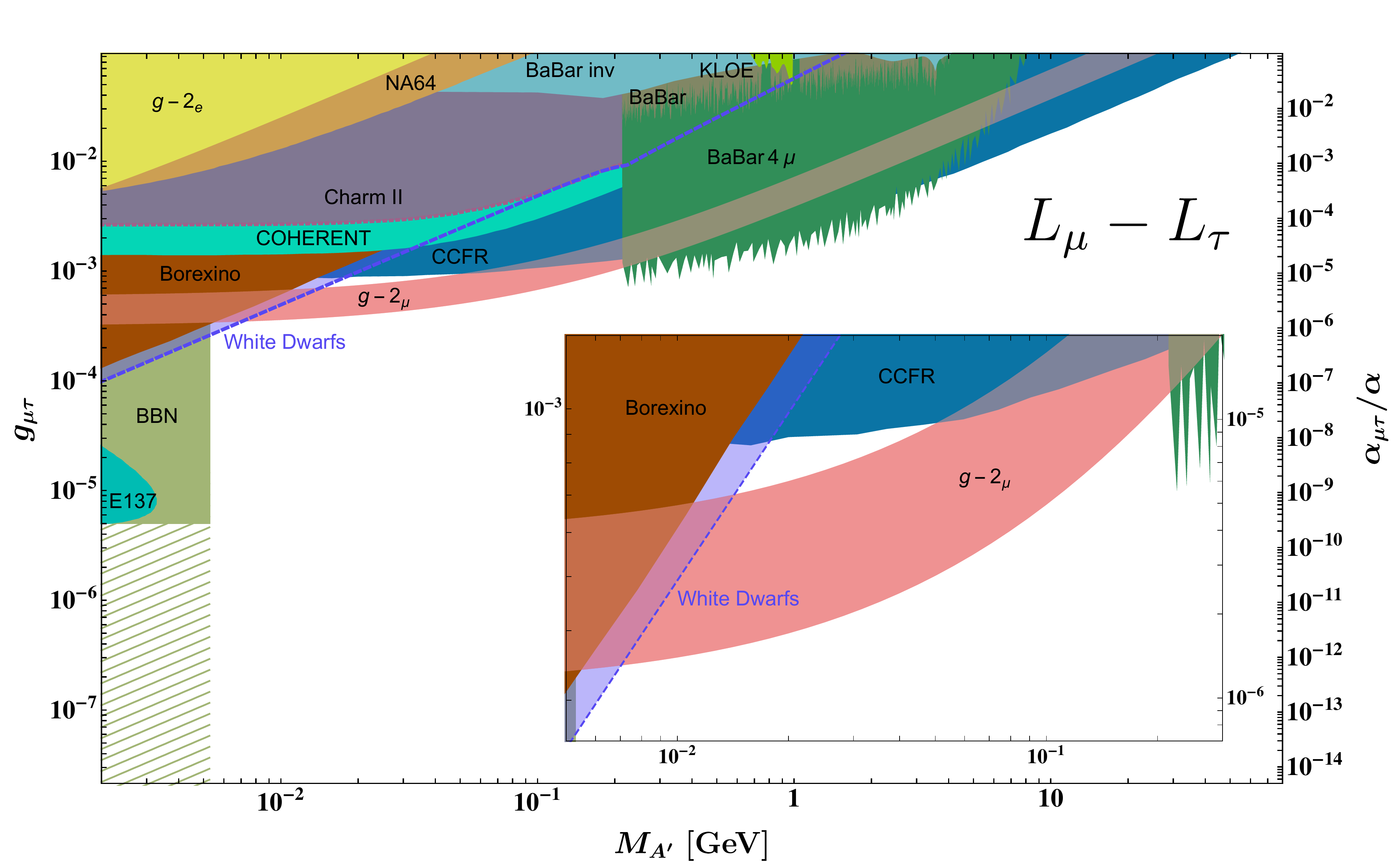}
\includegraphics[width=0.8\textwidth]{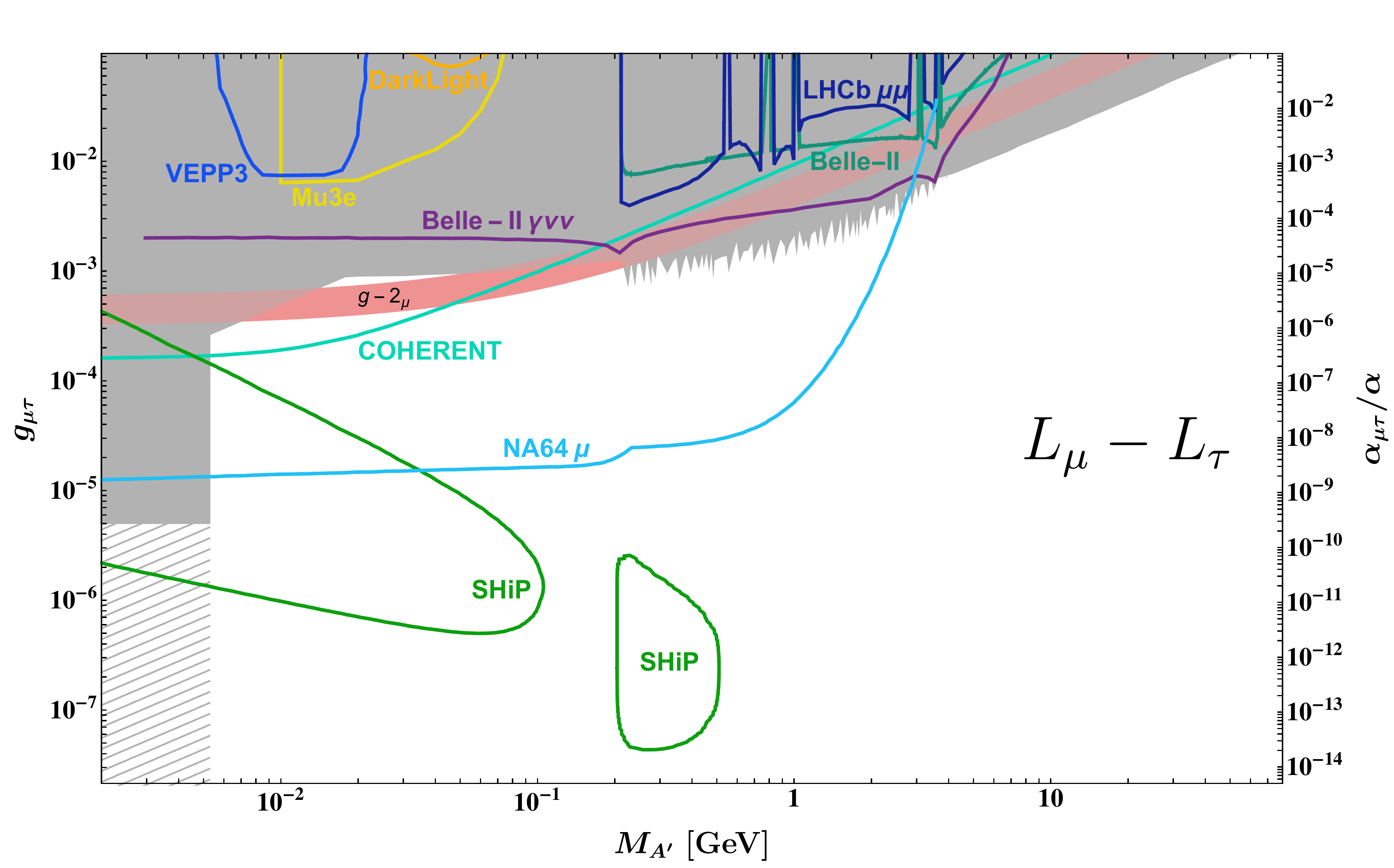}
\vspace{-.8cm}
\end{center}
\caption{\label{fig:LmLt} Constraints from current (upper panel) and future (lower panel) experiments on a $U(1)_{L_{\mu}-L_\tau}$ gauge boson with gauge coupling $g_{\mu-\tau}=\epsilon\,e$. Additional constraints from supernova cooling are not shown (see Section~\ref{sec:WD}). }\vspace{-.4cm}
\end{figure}

\subsection{$U(1)_{L_{\mu}-L_\tau}$}
This group exhibits the biggest changes compared to the case of pure kinetic mixing, due to suppressed couplings to hadrons and electrons. 
The best current limits arise from experiments and observations that only require one kinetic mixing factor. In addition, there is the BBN limit from~\cite{Kamada:2015era}.\footnote{For this limit we show the coupling range displayed in~\cite{Kamada:2015era} as solid. For weaker couplings the region is hatched. A determination of the decoupling of the gauge boson in the early universe would require a more sophisticated analysis.}
Importantly, we note that there is still room for an explanation of the $(g-2)_{\mu}$ anomaly~\cite{Altmannshofer:2014pba}\footnote{For similar discussions around flavor-changing couplings we refer to \cite{Altmannshofer:2016brv, Foldenauer:2016rpi}.}.
This makes it particularly attractive for future experimental probes. While SHiP will cover a large region of parameter space it will not reach the area suggested by $(g-2)_{\mu}$. This area will be probed by COHERENT \cite{Abdullah:2018ykz} but most decisively by the proposed muon run of NA64$\mu$~\cite{Gninenko:2014pea,Gninenko:2018tlp}.
The additional region of projected SHiP sensitivity for $M_{A'} > 2 m_\mu$ is a consequence 
 of high statistics and  the unsuppressed $\text{Br}(A'\to \mu^+ \mu^-)$.

\section{Conclusions}
\label{sec:conclusions}
In this paper we have investigated and collected phenomenological constraints on weakly coupled gauge bosons of the anomaly-free gauge groups  $U(1)_{B-L}$, $U(1)_{L_\mu-L_\tau}$, $U(1)_{L_\mu-L_e}$ and $U(1)_{L_e-L_\tau}$.
For this we have considered a wide variety of constraints from laboratory experiments as well as astrophysical and cosmological observations. We also provide a survey of future possibilities. Our main results are summarised in Figs.~\ref{fig:BL}-\ref{fig:LmLt}.

Important constraints can be translated from experiments and observations limiting hidden photons interacting only via kinetic mixing (see Fig.~\ref{fig:universal} and cf. also~\cite{Ilten:2018crw}). However, there are also a number of significant differences as well as special features that need to be taken into account.
\begin{itemize}
\item{} All the gauge bosons considered in this analysis interact with neutrinos. This makes them amenable to experiments and observations from neutrino physics, which results in important additional constraints. The reactor experiments TEXONO and Super-K provide the leading constraints for a sizable part of the parameter space in the case of $U(1)_{B-L}$,  $U(1)_{L_\mu-L_e}$ and $U(1)_{L_e-L_\tau}$.  DUNE has the potential to significantly increase these limits for $U(1)_{L_\mu-L_e}$ and $U(1)_{L_e-L_\tau}$. Limits from white dwarf cooling, which have not been considered before, provide the leading constraint for a $U(1)_{L_\mu-L_\tau}$ in the low mass region, which is slightly better than the Borexino limit. Morover, a future high-exposure run of  COHERENT will probe substantial parts of the   $(g-2)_{\mu}$ explanation. 

\item{} The gauge bosons of purely leptonic gauge groups interact with hadrons only via kinetic mixing. This kinetic mixing is  automatically generated by the Standard Model particles, and is finite. Taking 
this mixing into account those gauge bosons can also be tested in experiments with protons and other hadrons, providing limits previously not considered.
\item{} A gauge boson of $U(1)_{L_\mu-L_\tau}$ has direct interactions only with the second and third generation leptons. Again the loop-generated kinetic mixing becomes important. But the limits are generally weaker. This makes experiments that directly use muons or taus especially attractive. In particular since this gauge group still allows for a viable explanation of the $(g-2)_{\mu}$ anomaly.
\end{itemize}

Going beyond the existing experiments we can look towards a bright future. Experiments like SHiP, SeaQuest, LHCb, CodexB, FASER, MATHUSLA, Belle-II, a muon run of NA64, Mu3e\footnote{Currently, improvements of the Mu3e reach are under investigation including the sensitivity to displaced decays~\cite{ASprivate}.} as well as neutrino experiments such as DUNE and COHERENT, will explore large and interesting areas of parameter space and thereby provide many opportunities for a discovery.

\section*{Acknowledgements}
We would like to thank A.-K.~Perrevoort, N.~Berger and A.~Sch\"oning for interesting discussions. We are also indebted to B.~Echenard, T.~Ferber, F.~Kling and Y.~Soreq for providing valuable data necessary for the completion of this work.
P.F. thanks the DFG for support via the GRK 1940 ``Particle Physics beyond the Standard Model''. P.F. and J.J. are supported by the European Union Horizon 2020 research and innovation under the Marie Sk\l odowska-Curie grant agreement Number 690575. J.J. is grateful to the DFG for support within the DFG TR33 ``The Dark Universe''.

\appendix
\section{Rotation to Mass Eigenstates}\label{rotation}
Starting from the Lagrangian,
\begin{align}\label{eq:lagapp}
\mathcal{L}=-\frac{1}{4}\hat F_{\mu\nu} \hat F^{\mu\nu}-\frac{\epsilon'}{2}\hat F_{\mu\nu}\hat X^{\mu\nu}-\frac{1}{4}\hat X_{\mu\nu} \hat X^{\mu\nu}-g'\,j_\mu^Y \hat B^\mu -g_x\,j_\mu^x \hat X^\mu+\frac{1}{2}\hat M_X^2 \hat X_\mu \hat X^\mu\,, 
\end{align}
we use this Appendix to provide a step by step rotation to the relevant mass eigenstates.
As a first step, we
introduce the non-orthogonal rotation $G(\epsilon')$ 
\begin{align}
\begin{pmatrix}
                    \hat{B}_\mu\\  
                    \hat W^3_\mu\\
		    \hat{X}_\mu\end{pmatrix} = G(\epsilon')\, \begin{pmatrix}
                    B_\mu  \\[2pt]
                    W^3_\mu\\[2pt]
		    X_\mu   
    \end{pmatrix}  \,,
                   \end{align}
                in which $W^3_\mu$ denotes the third $SU(2)_L$ gauge boson, and
                   \begin{align}
   G(\epsilon')                = 
\begin{pmatrix}
                    1  &   0&       - \dfrac{\epsilon'}{\sqrt{1-\epsilon^{\prime 2}}}    \\
                    0&1&0\\
		   0   & 0&    \dfrac{1}{\sqrt{1-\epsilon^{\prime 2}}}
                   \end{pmatrix}              \,.
\end{align}
The combined mass matrix for the three neutral electroweak gauge bosons
$B_\mu$, $W^3_\mu$, and $X_\mu$ reads in the limit of small $\epsilon'$
\begin{align}
\mathcal{M}^2
=\frac{v^2}{4}
\begin{pmatrix} g'^2  & -g\,g' &-{g'}^2 \epsilon' \\[2mm]
               -g\,g' & g^2    & g\,g' \,\epsilon'\\
               -{g'}^2 \epsilon' \quad  & \quad g\,g'\epsilon' \quad & \quad \dfrac{4M_X^2}{v^2}(1+\epsilon^{\prime 2})+g^{\prime 2} \epsilon^{\prime 2}
\end{pmatrix} +\mathcal{O}(\epsilon^{\prime 3}) ,
\label{eq:matrix_vectors}
\end{align}
where $g$ and $g'$ denote the $SU(2)_L$ and $U(1)_Y$ gauge couplings,
respectively. This mass matrix  can be diagonalized through a
combination of two block-diagonal rotations with the weak mixing angle
$\theta_w$ and an additional angle $\xi$,
$R_1(\xi)R_2(\theta_w) \mathcal{M}^2\,R_2(\theta_w)^TR_1(\xi)^T=
\text{diag}\,(M_\gamma^2,M_Z^2,M_{A'}^2)$, with the rotation matrix

\begin{align}
R_1(\xi)R_2(\theta_w)=\begin{pmatrix} 1&0 & 0\\
0&\cos\xi & \sin\xi \\
0&-\sin\xi & \cos \xi \end{pmatrix}\,
\begin{pmatrix} \cos \theta_w& \sin \theta_w & 0\\
-\sin \theta_w& \cos \theta_w&0\\
0&0&1\end{pmatrix}\,,
\end{align}
and 
\begin{align}\label{eq:tan2xi}
\tan 2 \xi = \frac{2\epsilon' \sin\theta_w}{1-\delta}+\mathcal{O}(\epsilon^{\prime 2})\,.
\end{align}
Here we have defined $\delta=\hat M_X^2/\hat M_Z^2$ and  $\hat M_Z=\sqrt{g^2+g^{\prime 2}} v/2$ is the mass of the $Z-$boson in the SM.
The mass eigenvalues are then given by $M_\gamma^2=0$ and 
\begin{align}
M_Z^2=\hat M_Z^2\big(1+\epsilon^{\prime 2}\sin^2\theta_w(1+2\delta)\big)+\mathcal{O}(\delta^2\epsilon^{\prime 2})\,,\quad  M_{A'}^2=\hat M_X^2 \big(1+\epsilon^{\prime 2}(1-\sin^2\theta_w(1+\delta))\big)+\mathcal{O}(\delta^2\epsilon^{\prime 2})\,.
\end{align}
Couplings between the gauge boson mass eigenstates $A_\mu, Z_\mu$ and $A'_\mu$ and the fermion currents are then given by\footnote{We denote gauge bosons in the non-orthogonal basis by hatted fields and define the neutral gauge bosons in the electroweak symmetric phase by $B_\mu, W^3_\mu, X_\mu$ and in the electroweak broken phase by $A_\mu, Z_\mu, A'_\mu$. }
\begin{align}\label{eq:currentcouplings}
\left(ej_\text{EM} , \frac{e}{\sin \theta_w \cos \theta_w} j_Z, g_{x}j_{x}\right) 
\begin{pmatrix}\hat A\\ \hat Z\\\hat A'\end{pmatrix}
=&\left(ej_\text{EM} , \frac{e}{\sin \theta_w \cos \theta_w} j_Z, g_{x}j_{x}\right) \,K\,\begin{pmatrix}A\\ Z\\  A'\end{pmatrix} \,, 
\end{align}
with
\begin{align}
K=\left[ R_1(\xi)R_2(\theta_w)G^{-1}(\epsilon')R_2(\theta_w)^{-1}\right]^{-1}
 &= \begin{pmatrix}
1 & 0 & -\epsilon'\cos \theta_w \\
0 & 1& 0 \\
0 & \epsilon' \sin\theta_w&  1
\end{pmatrix} +\mathcal{O}(\epsilon' \delta, \epsilon^{\prime 2})\,\notag \\[.2cm]
 &= \begin{pmatrix}
1 & 0 & -\epsilon \phantom{e}\\
0 & 1& 0 \\
0 & \epsilon \tan\theta_w&  1
\end{pmatrix} +\mathcal{O}(\epsilon \delta, \epsilon^{ 2})\,,
\label{eq:KK}
\end{align}
where in the second line we have introduced $\epsilon \equiv \epsilon' \cos \theta_w$.
Couplings of the massless photon are protected by the unbroken electromagnetic gauge symmetry, and the new gauge boson $X$ couples to leading order in $\epsilon$ to the electromagnetic current \cite{Gopalakrishna:2008dv, Davoudiasl:2012ag}. This 
motivates the name \emph{hidden photon} for a secluded $U(1)_X$ gauge boson with couplings to the SM through kinetic mixing.

 The leading terms in $\epsilon$ of the $A'W^+W^-$ coupling follow from replacing the photon by $A\to A-\epsilon A'$ in the $AW^+W^-$ vertex. Couplings of the new gauge boson $A'$ to the $Z$-current only appear at $\mathcal{O}(\delta \epsilon)$ and can be obtained by replacing $Z\to Z-\epsilon\delta\tan\theta_w  A'$. As a consequence, couplings of the Higgs boson $H$ to the new gauge boson are further suppressed,
 \begin{align}
\frac{\hat M_Z^2}{2v}\begin{pmatrix} A& Z & A'\end{pmatrix} \begin{pmatrix}0&0&0\\[5pt]
 0&1& -\epsilon  \tan\theta_w\,\delta\\[8pt]
 0&-\epsilon  \tan\theta_w\,\delta & \epsilon^2  \tan^2\theta_w\,\delta^2 \end{pmatrix} \,H\,\begin{pmatrix} A\\ Z\\ A'\end{pmatrix} ,
\end{align}
where we have only kept the leading terms in the $\epsilon \delta$-expansion for each element. 

These couplings determine the decay modes of the $A'$ boson for a given mass $M_{A'}$. We assume $M_{A'}\ll M_Z$ such that only fermionic decay modes are relevant with a natural hierarchy between couplings to fermions from $j_\mu^x$ and mixing-induced couplings,
\begin{align}\label{eq:Zpwidth}
\Gamma(A'\to f \bar f) =\frac{M_{A'}}{24\pi}C_f\sqrt{1-\frac{4m_f^2}{M_{A'}^2}}\left[(g_L^2+g_R^2)\Big(1-\frac{m_f^2}{M_{A'}^2}\Big)+6\frac{m_f^2}{M_{A'}^2}g_Lg_R\right]\,,
\end{align}
where $C_f = 3(1)$ for quarks (leptons) is a color factor and the couplings can be determined by matching the currents in \eqref{eq:currentcouplings} to $j_\mu = \bar f \gamma_\mu (g_L P_L+g_RP_R)f$ with projectors $P_{R/L}=\frac{1}{2}(1\pm \gamma_5)$.
In addition, there are exotic Higgs decays $H \to Z A'$ and $H \to A'A'$, 
\begin{align}\label{eq:Hwidths}
\Gamma(H \to A'Z)&=\frac{1}{16\pi}\epsilon^2\tan^2\theta_w\frac{M_H^3}{v^2}\frac{M_{A'}^2}{M_Z^2}\left(1-\frac{M_Z^2}{M_H^2}\right)\,,\\[2pt]
\Gamma(H \to A'A')&=\frac{1}{32\pi}\epsilon^4\tan^4\theta_w\frac{M_H^3}{v^2}\frac{M_{A'}^4}{M_Z^4}\,,
\end{align}
where we have neglected higher order corrections in $\delta \epsilon $ and $M_{A'}^2/M_H^2$. Therefore, Higgs decays do not provide relevant constraints, if the only coupling between the Higgs and the hidden gauge boson is mediated by the kinetic mixing term.\\

\section{Beam dump limit calculation}\label{app:comparison}

\begin{figure}[t]
\begin{center}
\includegraphics[width=0.48\textwidth]{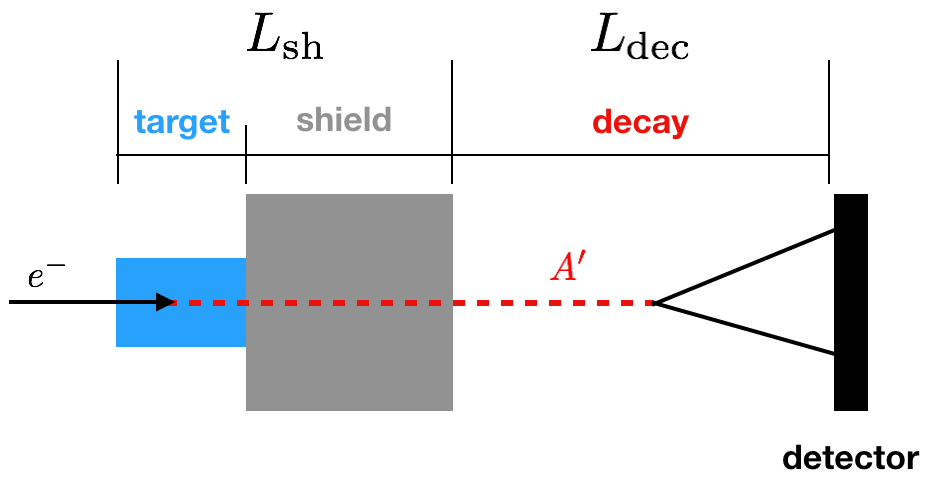}
\end{center}
\caption{\label{fig:BDsketch} Schematic of an electron beam dump experiment like SLAC E137.}
\end{figure}

In this Appendix we want to give an example of a prototypic limit calculation for beam dump experiments. For concreteness, we will consider the electron beam dump experiment E137 operated at SLAC in the 1980s.

The E137 setup is schematically shown in Fig.~\ref{fig:BDsketch}. A total of  $N_e\sim 1.87\times10^{20}$  electrons with momentum $p=20$ GeV have been dumped in an aluminum target followed by 200 m of rock, which served as shielding. In the target material the incoming electrons interact with the nuclei and lose energy via Bremsstrahlung.  The hidden photon of a secluded $U(1)_X$ has the same coupling to the electron as the photon, only suppressed by the mixing parameter $\epsilon$. Hence, it can also be produced in a Bremsstrahlung process. The total number of produced $A'$ decaying visibly within the detector volume is described by \eqref{eq:eevents}.  During the full data taking period no events have been observed in the E137 detector \cite{Bjorken:1988as}. According to Poisson statistics we can therefore exclude any point in model parameter space predicting more than $N_{95}=3$ observed events.

\begin{figure}[t]
\begin{center}
\includegraphics[width=0.48\textwidth]{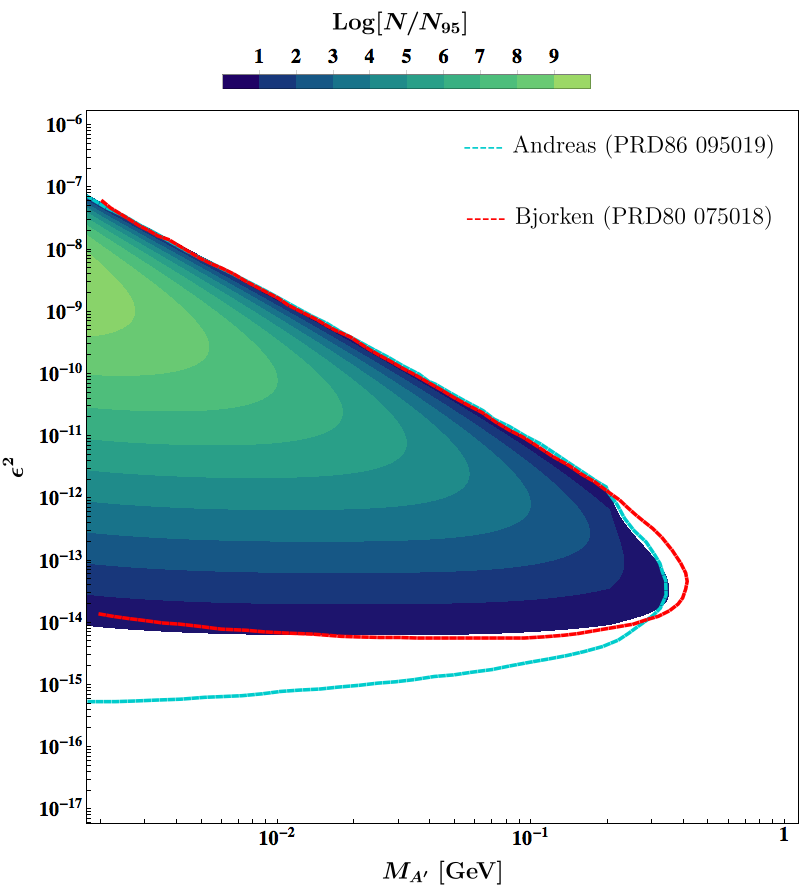}\hspace*{0.5cm}
\includegraphics[width=0.48\textwidth]{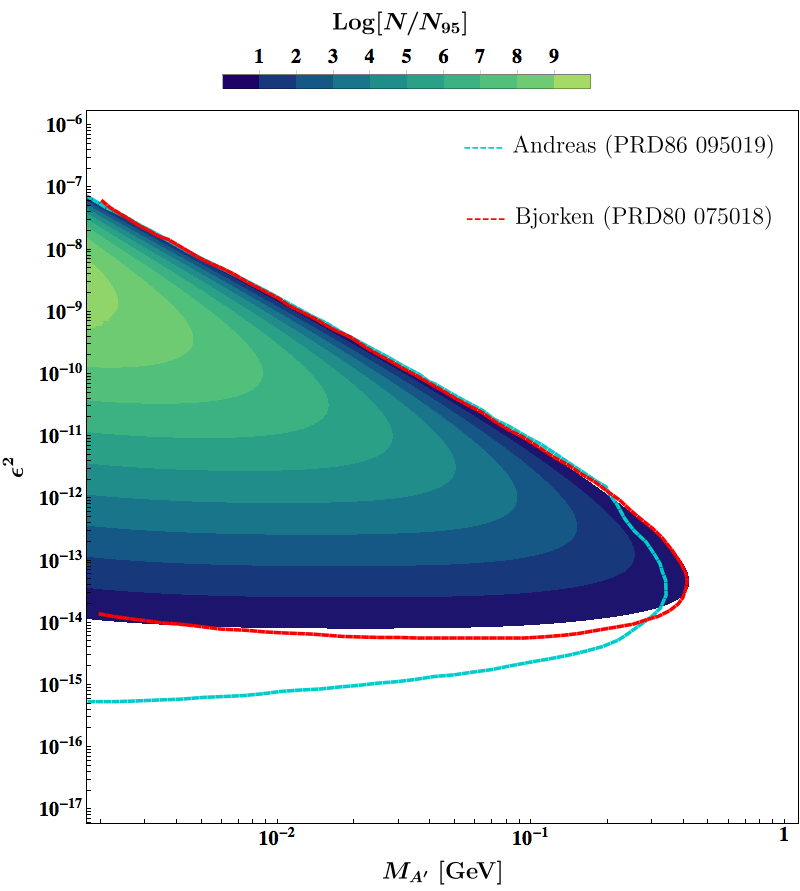}
\end{center}
\caption{\label{fig:E137BJ} Expected number of events $N$ in this work for a secluded hidden photon $A'$ at the SLAC E137 experiment obtained from the \textit{\textbf{Bjorken implementation}} as a function of the mixing parameter $\epsilon^2$ and the mass $M_{A'}$ normalized to the 95\% C.L. limit $N_{95}$.  The red (Bjorken) and cyan (Andreas) lines show the exclusion contours calculated in \cite{Bjorken:2009mm} and in \cite{Andreas:2012mt}, respectively. Left: Including the full $A'$ width, in particular also the partial width into muons. Right: Only including the partial width of the $A'$ into electrons, assuming $N_{95}=10$ and including the extra factor of 2.}
\end{figure}

\subsection{Bjorken implementation}

For E137, limits on hidden photons have been calculated first by {Bjorken \textit{et al.}} \cite{Bjorken:2009mm} using \eqref{eq:eevents} with the approximate differential cross section given in \eqref{eq:diffcross} (which includes an erroneous factor of 2 in \cite{Bjorken:2009mm} that has been corrected in  \cite{Andreas:2012mt}). For the full details of this calculation we refer the reader to Appendices A - C of \cite{Bjorken:2009mm}. 

We have implemented the full calculation of {Bjorken \textit{et al.}} in \texttt{MATHEMATICA}~\cite{mathematica}, which we will refer to as \textit{\textbf{Bjorken implementation}}. In order to derive limits we have discretized the 2D parameter space of the $M_{A'} - \epsilon^2$ plane into a finely-grained grid and calculated the expected number of hidden photon induced events $N$ at each point. The expected number of events normalized to the 95\% C.L. limit $N/N_{95}$ are depicted in Fig.~\ref{fig:E137BJ}. The edge of the outermost blue contour gives our limit. 

The left panel shows the results of the \textit{\textbf{Bjorken implementation}} including the full $A'$ width, which is described in detail in Section~\ref{sec:BRs}. This includes in particular the partial width into muons that becomes equal in size to the one into electrons for $M_{A'}\gtrsim 2  m_\mu$. If we compare our limit to the those of Bjorken and Andreas~\cite{Andreas:2012mt} (for details see Appendix~\ref{sect:andreas}), we see that our limit aligns perfectly with the Andreas limit in the high-$\epsilon^2$ domain, where it also shows the exact same threshold behavior at $M_{A'}\sim 2  m_\mu$. However, we can reconstruct the Bjorken limit nearly exactly (within the limits of numerical integration and discretization) if we only include the partial width of the $A'$ into electrons, assume 10 events as exclusion bound and include the erroneous extra factor of 2. This is shown in the right panel of Fig.~\ref{fig:E137BJ}. 

However, for a secluded hidden photon a muon coupling with the same strength as of the electron coupling is unavoidable. Hence, the limits for E137 derived by {Bjorken \textit{et al.}}, seemingly neglecting the partial width into muons, rather tend to overestimate the hidden photon mass reach. This is not an issue for the E141 and E774 limits as the mass reach is well below the dimuon threshold.

\subsection{Andreas implementation} \label{sect:andreas}

\begin{figure}[t]
\begin{center}
\includegraphics[width=0.48\textwidth]{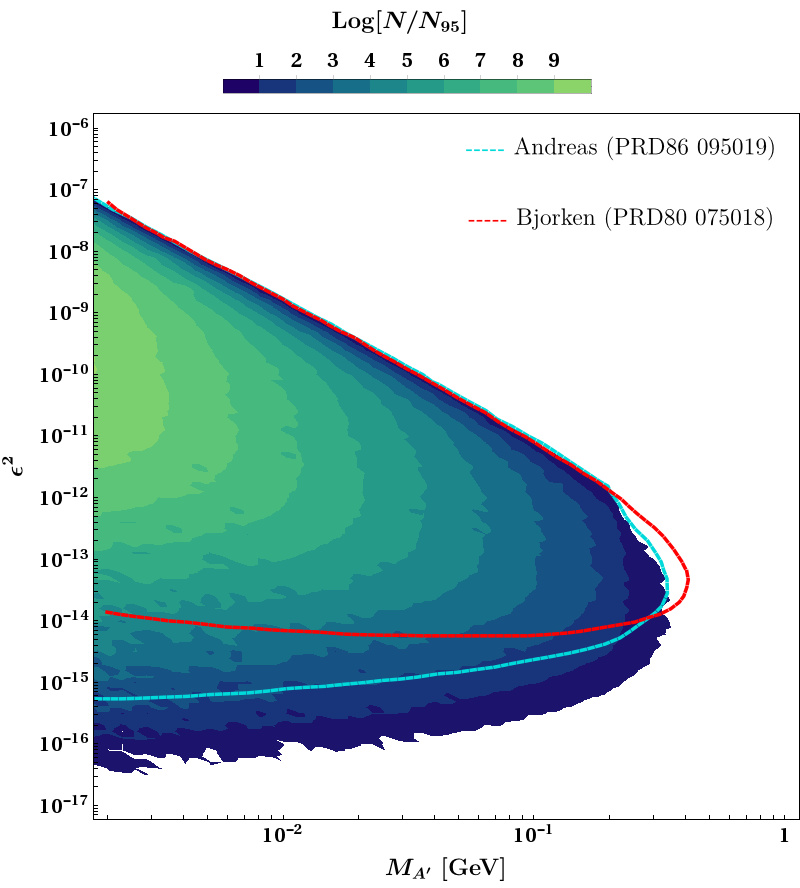}\hspace*{0.5cm}
\includegraphics[width=0.48\textwidth]{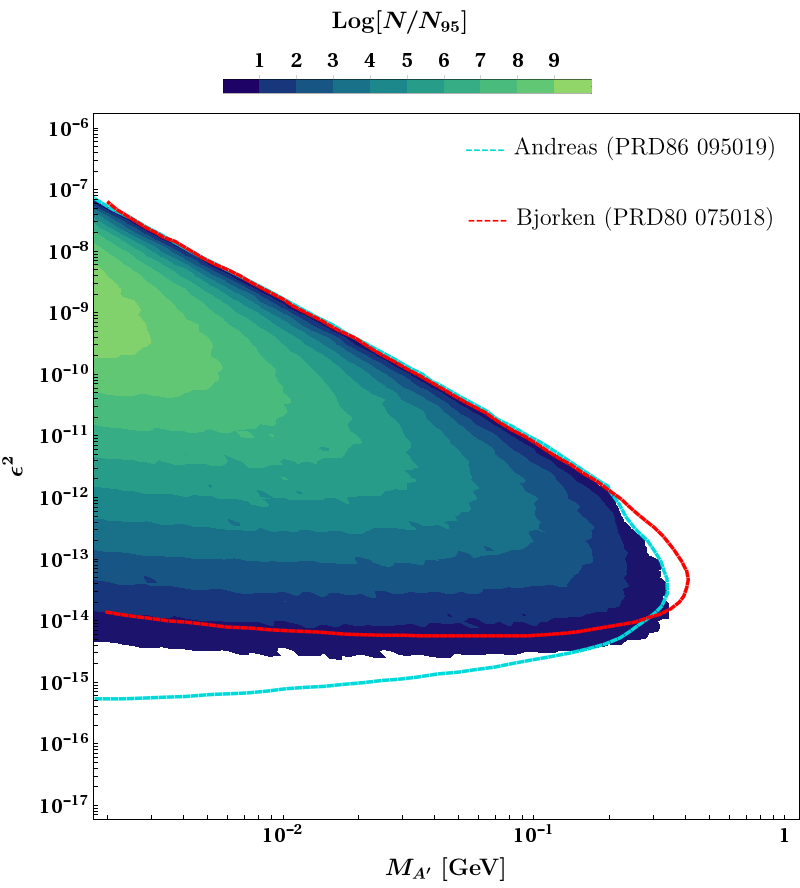}
\end{center}
\caption{\label{fig:E137SA} Expected number of events $N$ for a secluded hidden photon $A'$ at the SLAC E137 experiment obtained from the \textit{\textbf{Andreas implementation}} as a function of the mixing parameter $\epsilon^2$ and the mass $M_{A'}$ normalized to the 95\% C.L. limit $N_{95}$.  The red (Bjorken) and cyan (Andreas) lines show the exclusion contours calculated in \cite{Bjorken:2009mm} and in \cite{Andreas:2012mt}, respectively. 
Left: Using the hidden photon mass as minimum $A'$ energy $E_{A'}^\mathrm{min}=M_{A'}$. Right: Including the experimental minimum cutoff energy $E_{A'}^\mathrm{min}= E_\mathrm{cut}$.}
\end{figure}

As already mentioned, the hidden photon limits of E137 have been rederived in a more rigorous treatment by {Andreas \textit{et al.}} \cite{Andreas:2012mt}. The approximate differential cross section \eqref{eq:diffcross} has been corrected. But in particular a full-fletched Monte Carlo simulation of the $A'$ decays including detector geometry has been done and for the limit determination the full energy dependence of the differential cross section has been taken into account. A very thorough account of the many important details of this calculation can be found in Chapter 3 and Appendix B of  \cite{Andreas:2013xxa}. 

Again we have implemented the full calculation in \texttt{MATHEMATICA}, which we will refer to as  \textit{\textbf{Andreas implementation}}. As before we have discretized the parameter space and calculated the expected number of hidden photon induced events $N$ at each point. The expected number of events from the \textit{\textbf{Andreas implementation}} normalized to the 95\% C.L. limit $N/N_{95}$ are depicted in Fig.~\ref{fig:E137SA}.

The left panel shows the results of the \textit{{Andreas implementation}} using $E_{A'}^\mathrm{min}=M_{A'}$ in \eqref{eq:eevents} for the minimum  energy of the produced $A'$, as suggested in \cite{Andreas:2012mt}. This is a sensible choice as it corresponds to an $A'$ produced on shell in the lab frame, which can then resonantly decay into electrons. Such an $A'$ can be looked for in a dielectron resonance search in the experiment. This choice explains the mass-dependent behavior of the low-$\epsilon^2$ domain both of the Andreas and our derived limit (The fact that our limit is excluding even smaller $\epsilon^2$ is mainly due to our lack of a full Monte Carlo simulation of the $A'$ decay geometry). 
\par

To see this explicitly, let us note that the lower boundary of the beam dump limit is reached when the $A'$ has a typical decay length that is much larger than the experimental setup $\ell_{A'}\gg L_\text{sh},L_\text{sh}$. In this case, we can expand the hidden photon decay probability as
\begin{equation}
P_\text{dec}\ =  \ e^{-\frac{L_\text{sh}}{\ell_{A'}}}\left(1-e^{-\frac{L_\text{dec}}{\ell_{A'}}}\right) \ \approx\ \frac{L_\text{dec}}{\ell_{A'}}\ \propto \ L_\text{dec}\, \frac{M_{A'}^2\,\alpha\, \epsilon^2}{E_{A'}} \,.
\end{equation}
Combined with the estimate for the beam dump cross section in equations (A15) and (A16) of Ref.~\cite{Bjorken:2009mm}, 
 \begin{equation}
\sigma \propto \frac{\alpha^3\,\epsilon^2\,Z^2}{M_{A'}^2}
\,,
\end{equation}
we see that the leading powers of $M_{A'}^2$ cancel in the calculation of the expected number of hidden photon events
\begin{align}\label{eq:ExpScale}
N \propto \int_{E_{A'}^\text{min}}^{E_0} dE_{A'}\, \sigma_{A'}\, P_\text{dec}(E_{A'}) \,.
\end{align}

If we use the hidden photon mass as the lower integration limit $E_{A'}^\mathrm{min}=M_{A'}$ in  \eqref{eq:ExpScale},  as done  in \cite{Andreas:2012mt}, the lower boundary of the beam dump limit has an explicit  dependence on  $M_{A'}$. This is shown in the left panel of Fig.~\ref{fig:E137SA}. In general, we can see that in this regime the total number of events scales with $\epsilon^4$ and therefore has a quite steep fall-off, such that the lower boundary is mainly statistics limited. 
There still might be a residual  logarithmic dependence of $N$ on the mass $M_{A'}$ coming from the integration of $d E_{A'}/E_{A'}$.

\begin{figure}[ht!]
\begin{center}
\includegraphics[width=0.48\textwidth]{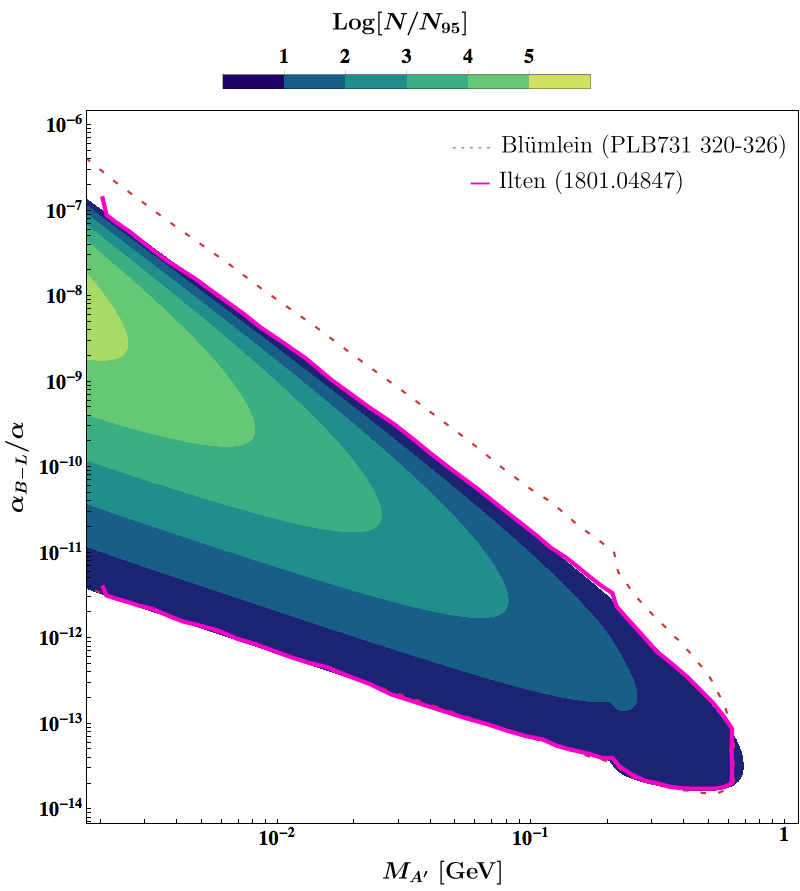}\hspace*{0.5cm}
\includegraphics[width=0.48\textwidth]{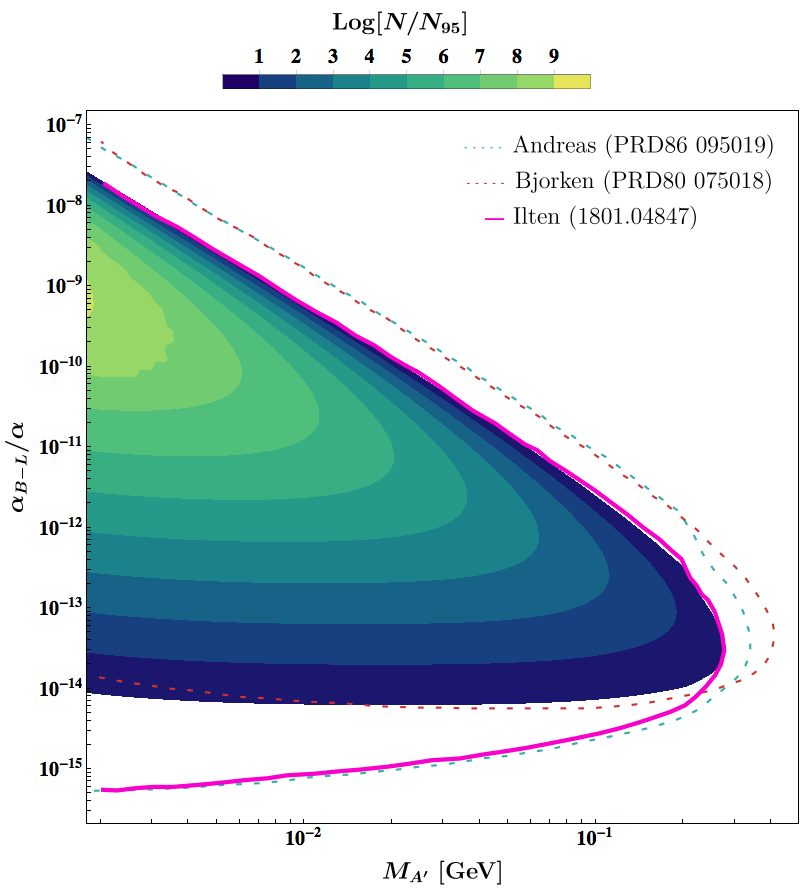}
\end{center}
\caption{\label{fig:BDcomp} Comparison of beam bump limits for the $U(1)_{B-L}$ gauge boson calculated  in this work to the recasted limits of  \cite{Ilten:2018crw}. The calculated limits are given by the border of the outermost blue contour. The recasted limits are depicted by the pink solid line. {\bf NuCal} (left): The red dashed line shows the respective limit for the secluded hidden photon. {\bf E137} (right): The red and cyan dashed lines show the respective limits for the secluded hidden photons discussed previously.}
\end{figure}

However, the experimental analysis searching for resonant dielectron events applied a cut of $E_\mathrm{cut}=3\ \text{GeV}$ to their data~\cite{Bjorken:1988as}. Implementing this experimental cut as  $E_{A'}^\mathrm{min}= E_\mathrm{cut}$ the  \textit{\textbf{Andreas implementation}} yields the results in the right panel of Fig.~\ref{fig:E137SA}. We see that we recover the horizontal scaling (i.e. no mass dependence) of the low-$\epsilon^2$ domain, which is present in the Bjorken limit (again we expect the overshooting of our exclusion contour in this domain to be fixed by including Monte Carlo simulated geometric acceptances).  This is what we expect from \eqref{eq:ExpScale}. With a constant minimum energy $E_\mathrm{cut} \gg M_{A'}$ the exponent only scales with $\epsilon^2$ and shows no mass dependence anymore.

In summary, it seems likely  that  {Andreas \textit{et al.}} have not included experimental cuts on the minimum energy of the observed events.  This would mean that the limit deduced for E137 by {Andreas \textit{et al.}} is too optimistic in the low-$\epsilon^2$ domain. This issue also persist for the derived limits for E141, E774 and Orsay as all the relevant analyses include energy cuts (cf. Table~\ref{tab:Ecut}). 

\begin{table}[h!]
\begin{center}
 \begin{tabular}{c | c c c c} 
& E137 & E141& E774 &Orsay\\[.3cm]
\hline
&&&&\\[-.3cm]
$E_\mathrm{cut}$ [GeV]& 3  & 4.5 & 27.5 & 0.75
\end{tabular}
\end{center}
\caption{\label{tab:Ecut} Cuts on the minimum event energy used in the analyses of E137~\cite{Bjorken:1988as}, E141~\cite{Riordan:1987aw}, E774~\cite{Bross:1989mp} and Orsay~\cite{Davier:1989wz}.}
\end{table}

\subsection{Towards more accurate limits}

We have seen that both the limits derived by {Andreas \textit{et al.}} and {Bjorken \textit{et al.}} can possibly be refined in certain aspects. We therefore adopted the improved approximate equations derived in \cite{Andreas:2012mt} with the inclusion of the experimental energy cut for the calculation of electron beam dump limits in this work. Further improvements could be obtained from a full calculation of these limits as outlined in \cite{Andreas:2013xxa} with the implementation of the energy cuts and a full Monte Carlo simulated detector acceptance.

\subsection{A note on recasting beam dump limits}

In a very recent article, {Ilten \textit{et al.}} have presented a framework for recasting limits on hidden photons~\cite{Ilten:2018crw}. In particular, they have recasted existing  limits on a secluded hidden photon to the $U(1)_{B-L}$ gauge boson by use of the presented framework. In Fig.~\ref{fig:BDcomp}  we compare these recasted limits to those derived  in this work from the full calculation for the case of $U(1)_{B-L}$. 

In the left panel we show the limits obtained from the NuCal proton beam bump experiment. The recasted limits of \cite{Ilten:2018crw} match those obtained from the full calculation over a large range of masses. However, the full calculation improves the mass reach of the $U(1)_{B-L}$ limit, which is due to the higher relative branching fraction into leptons for $A'$ masses of the order of the $\omega$ mass. The full calculation excludes $A'$ masses  of up to 688 MeV compared to 624 MeV for the recasted limits. 

In the right panel, the limits obtained from E137 are shown. In the high-$\alpha_{B-L}$ domain the recasted limits match those derived in this work quite well. However, it seems that the limits derived in~\cite{Ilten:2018crw} are based on the analysis in~\cite{Andreas:2012mt} and therefore exhibit the same mass scaling in the low-$\alpha_{B-L}$ domain as the Andreas limits. This is the main reason for the different behavior at small couplings. The same holds for the recasted electron beam dump limits of E141, E774 and Orsay.

Overall, our comparison shows that the recasted limits~\cite{Ilten:2018crw} match a full implementation to a good level (if the same implementation is used). Yet a full implementation provides quantitative improvements and increased confidence in the results.

\newpage

\section{Relevant processes and couplings for the different experiments}\label{app:longtable}
\begin{table}[h!]
\begin{tabular}{|c c|c||r|r|r|r|}\hline
&&&&&&\\[-6pt]
  \quad\quad\quad\textbf{Experiment}&  &  \textbf{Process} &  $\boldsymbol{B-L}$ & $\boldsymbol{L_\mu-L_e}$& $\boldsymbol{L_e-L_\tau}$&$\boldsymbol{L_\mu-L_\tau}$\\[.3cm]
\hline
&&&&&&\\[-6pt]
 \multirow{3}{*}{$\begin{matrix}\text{E137, E141,  E774,} \\ \text{Orsay, APEX, A1/MAMI}\end{matrix}$ } &prod &$e$-Bremsstrahlung& $-g_{B-L}$& $-g_{\mu e}$&$ g_{e\tau}$ &$ e \, \epsilon_{\mu\tau}(q^2) $\\[.3cm]
&det & $A'\to ee$ & $-g_{B-L}$& $-g_{\mu e}$&$ g_{e\tau}$ &$ e \, \epsilon_{\mu\tau}(q^2) $\\[.3cm]
\hline
&&&&&&\\[-6pt]
\multirow{3}{*}{CHARM} &prod &$\eta/\eta'$-decay& $g_{B-L}$& $ e \, \epsilon_{\mu e}(q^2)$&$ e \, \epsilon_{e\tau}(q^2)$ &$ e \, \epsilon_{\mu\tau}(q^2)$ \\[.3cm]
&det& $A'\to ee $ &$-g_{B-L}$& $-g_{\mu e}$&$ g_{e\tau}$ &$ e \, \epsilon_{\mu\tau}(q^2) $\\[.3cm]
\hline
&&&&&&\\[-6pt]
\multirow{3}{*}{LSND, NA48/2} &prod &$\pi^0$-decay&  $g_{B-L}$& $ e \, \epsilon_{\mu e}(q^2)$&$ e \, \epsilon_{e\tau}(q^2)$ &$ e \, \epsilon_{\mu\tau}(q^2)$  \\[.3cm]
&det& $A'\to ee $ & $-g_{B-L}$& $-g_{\mu e}$&$ g_{e\tau}$ &$ e \, \epsilon_{\mu\tau}(q^2) $\\[.3cm]
\hline
&&&&&&\\[-6pt]
\multirow{6}{*}{U70/NuCal} &prod &$\pi^0$-decay&  $g_{B-L}$& $ e \, \epsilon_{\mu e}(q^2)$&$ e \, \epsilon_{e\tau}(q^2)$ &$ e \, \epsilon_{\mu\tau}(q^2)$  \\[.3cm]
&prod &$p$-Bremsstrahlung & $g_{B-L}$& $ e \, \epsilon_{\mu e}(q^2)$&$ e \, \epsilon_{e\tau}(q^2)$ &$ e \, \epsilon_{\mu\tau}(q^2)$ \\[.3cm]
&det& $A'\to ee $ &  $-g_{B-L}$& $-g_{\mu e}$&$ g_{e\tau}$ &$ e \, \epsilon_{\mu\tau}(q^2) $\\[.3cm]
&det& $A'\to \mu\mu $ &$-g_{B-L}$& $g_{\mu e}$&$ e \, \epsilon_{e\tau}(q^2) $ &$ g_{\mu\tau}$ \\[.3cm]
      \hline
&&&&&&\\[-6pt]
\multirow{4}{*}{DarkLight}&prod &$e$-Bremsstrahlung&  $-g_{B-L}$&$-g_{\mu e}$&$ g_{e\tau}$ &$ e \, \epsilon_{\mu\tau}(q^2) $\\[.3cm]
 & det& $A'\to ee$&  $-g_{B-L}$& $-g_{\mu e}$&$ g_{e\tau}$ &$ e \, \epsilon_{\mu\tau}(q^2) $\\[.3cm]
  & det& $A'\to \mathrm{inv}$  &  $-g_{B-L}$& $g_{\mu e}$&$ g_{e\tau}$ &$ g_{\mu\tau} $\\[.3cm]
  \hline
&&&&&&\\[-6pt]
\multirow{4}{*}{NA64}&prod &$e$-Bremsstrahlung& $-g_{B-L}$& $- g_{\mu e}$&$ g_{e\tau}$ &$ e \, \epsilon_{\mu\tau}(q^2)$ \\[.3cm]
&prod &$\mu$-Bremsstrahlung& $-g_{B-L}$& $g_{\mu e}$&$ e \, \epsilon_{e\tau}(q^2) $ &$ g_{\mu\tau}$ \\[.3cm]
  & det& $A'\to \mathrm{inv}$ &  $-g_{B-L}$& $g_{\mu e}$&$ g_{e\tau}$ &$ g_{\mu\tau} $\\[.3cm]
    \hline
&&&&&&\\[-6pt]
  \multirow{3}{*}{Mu3e}&prod &$\mu$-Bremsstrahlung&  $-g_{B-L}$& $g_{\mu e}$&$ e \, \epsilon_{e\tau}(q^2) $ &$ g_{\mu\tau}$ \\[.3cm]
 & det& $A'\to ee$ &$-g_{B-L}$& $-g_{\mu e}$&$ g_{e\tau}$ &$ e \, \epsilon_{\mu\tau}(q^2) $\\[.3cm]
 \hline
 \end{tabular}
\end{table}

\begin{table}

\begin{tabular}{|cc|c||r|r|r|r|}\hline
    &&&&&&\\[-6pt]
\quad\quad\quad \textbf{Experiment} & &   \textbf{Process} &  $\boldsymbol{B-L}$ & $\boldsymbol{L_\mu-L_e}$& $\boldsymbol{L_e-L_\tau}$&$\boldsymbol{L_\mu-L_\tau}$\\[.3cm]
\hline
 &&&&&&\\[-6pt]
 \multirow{6}{*}{$\begin{matrix}\text{FASER, SeaQuest,} \\ \text{SHiP}\end{matrix}$ } &prod &$\pi^0/\eta/\eta'$-decay&  $g_{B-L}$& $ e \, \epsilon_{\mu e}(q^2)$&$ e \, \epsilon_{e\tau}(q^2)$ &$ e \, \epsilon_{\mu\tau}(q^2)$  \\[.3cm]
&prod &$p$-Bremsstrahlung & $g_{B-L}$& $ e \, \epsilon_{\mu e}(q^2)$&$ e \, \epsilon_{e\tau}(q^2)$ &$ e \, \epsilon_{\mu\tau}(q^2)$ \\[.3cm]
&det& $A'\to ee $ & $-g_{B-L}$& $-g_{\mu e}$&$ g_{e\tau}$ &$ e \, \epsilon_{\mu\tau}(q^2) $\\[.3cm]
&det& $A'\to \mu\mu $ & $-g_{B-L}$& $g_{\mu e}$&$ e \, \epsilon_{e\tau}(q^2) $ &$ g_{\mu\tau}$ \\[.3cm]
   \hline
    &&&&&&\\[-6pt]
\multirow{3}{*}{VEPP-3}&prod &$e^+e^-\to\gamma A'$& $-g_{B-L}$& $-g_{\mu e}$&$ g_{e\tau}$ &$ e \, \epsilon_{\mu\tau}(q^2) $\\[.3cm]
  & det& $A'\to \mathrm{inv}$ &  $-g_{B-L}$& $g_{\mu e}$&$ g_{e\tau}$ &$ g_{\mu\tau} $\\[.3cm]
      \hline
    &&&&&&\\[-6pt]
\multirow{5}{*}{KLOE}&prod &$e^+e^-\to\gamma A'$& $-g_{B-L}$& $-g_{\mu e}$&$ g_{e\tau}$ &$ e \, \epsilon_{\mu\tau}(q^2) $\\[.3cm]
  & det& $A'\to \mu\mu$ & $-g_{B-L}$& $g_{\mu e}$&$ e \, \epsilon_{e\tau}(q^2) $ &$ g_{\mu\tau}$ \\[.3cm]
    & det& $A'\to \pi^+\pi^-$ &  $\frac{1}{3}g_{B-L}$ & $ e \, \epsilon_{\mu e}(q^2)$&$  e \, \epsilon_{e\tau}(q^2)$ &$  e \, \epsilon_{\mu\tau}(q^2) $\\[.3cm]
      \hline
    &&&&&&\\[-6pt]
\multirow{5}{*}{BaBar, Belle-II}&prod &$e^+e^-\to\gamma A'$&$-g_{B-L}$& $-g_{\mu e}$&$ g_{e\tau}$ &$ e \, \epsilon_{\mu\tau}(q^2) $\\[.3cm]
  & det& $A'\to \mu\mu$ &  $-g_{B-L}$& $g_{\mu e}$&$ e \, \epsilon_{e\tau}(q^2) $ &$ g_{\mu\tau}$ \\[.3cm]
&det& $A'\to ee $ &$-g_{B-L}$& $-g_{\mu e}$&$ g_{e\tau}$ &$ e \, \epsilon_{\mu\tau}(q^2) $\\[.3cm]
      \hline
    &&&&&&\\[-6pt]
\multirow{7}{*}{LHCb}&prod &$\pi^0/\eta/\eta'$-decay&  $g_{B-L}$& $ e \, \epsilon_{\mu e}(q^2)$&$ e \, \epsilon_{e\tau}(q^2)$ &$ e \, \epsilon_{\mu\tau}(q^2)$  \\[.3cm]
&prod &$D^*$-decay&  0& $ e \, \epsilon_{\mu e}(q^2)$&$ e \, \epsilon_{e\tau}(q^2)$ &$ e \, \epsilon_{\mu\tau}(q^2)$  \\[.3cm]
&prod &$p$-Bremsstrahlung & $g_{B-L}$& $ e \, \epsilon_{\mu e}(q^2)$&$ e \, \epsilon_{e\tau}(q^2)$ &$ e \, \epsilon_{\mu\tau}(q^2)$ \\[.3cm]
  & det& $A'\to \mu\mu$ & $-g_{B-L}$& $g_{\mu e}$&$ e \, \epsilon_{e\tau}(q^2) $ &$ g_{\mu\tau}$ \\[.3cm]
&det& $A'\to ee $ &$-g_{B-L}$& $-g_{\mu e}$&$ g_{e\tau}$ &$ e \, \epsilon_{\mu\tau}(q^2) $\\[.3cm]
      \hline
    &&&&&&\\[-6pt]
\multirow{5}{*}{ATLAS/CMS}&prod &Drell-Yan&  $\frac{1}{3}\,g_{B-L}$& $ e \, \epsilon_{\mu e}(q^2)$&$ e \, \epsilon_{e\tau}(q^2)$ &$ e \, \epsilon_{\mu\tau}(q^2)$  \\[.3cm]
  & det& $A'\to \mu\mu$ &  $-g_{B-L}$& $g_{\mu e}$&$ e \, \epsilon_{e\tau}(q^2) $ &$ g_{\mu\tau}$ \\[.3cm]
&det& $A'\to ee $ &$-g_{B-L}$& $-g_{\mu e}$&$ g_{e\tau}$ &$ e \, \epsilon_{\mu\tau}(q^2) $\\[.3cm]
\hline
\end{tabular}
\caption{\label{tab:allLimits} Coupling strengths for the different gauge groups relevant for the production and decay of hidden photons in experiments discussed in this paper  compared to the universal $e \, \epsilon \, Q_\mathrm{EM}$ coupling of the secluded hidden photon.}
\end{table}

\begin{table}
\begin{tabular}{|c |c||r|r|r|r|}\hline
    &&&&&\\[-6pt]
  \textbf{Experiment}&    \textbf{Process} &  $\boldsymbol{B-L}$ & $\boldsymbol{L_\mu-L_e}$& $\boldsymbol{L_e-L_\tau}\quad$&$\boldsymbol{L_\mu-L_\tau}\quad$\\[.3cm]
\hline
         &&&&&\\[-6pt]
Borexino &$\nu \,e^-\to\nu\,e^-$&$g_{B-L}^2$& $g_{\mu e}^2$&$ g_{e\tau}^2$ &$ e \, g_{\mu\tau}\,\epsilon_{\mu\tau}(q^2)$\\[.3cm]
      \hline
   &&&&&\\[-6pt]
\multirow{3}{*}{Charm-II} &$\nu_\mu \,e^-\to\nu_\mu \,e^-$& $g_{B-L}^2$& $-g_{\mu e}^2$&$ 0$ &$ e \, g_{\mu\tau}\,\epsilon_{\mu\tau}(q^2) $\\[.3cm]
  &  $\bar\nu_\mu \,e^-\to\bar\nu_\mu \,e^-$ &$-g_{B-L}^2$& $g_{\mu e}^2$&$ 0$ &$ -e \, g_{\mu\tau}\,\epsilon_{\mu\tau}(q^2) $\\[.3cm]
      \hline
         &&&&&\\[-6pt]
Texono &$\bar\nu_e \,e^-\to\bar\nu_e \,e^-$&$-g_{B-L}^2$& $-g_{\mu e}^2$&$ -e \, g_{e\tau}\,\epsilon_{e\tau}(q^2)$ &$ 0 $\\[.3cm]
      \hline
               &&&&&\\[-6pt]
DUNE, Super-K &$\nu_{\mu/\tau} e^-$&$0$& $g_{\mu e}^2$&$  g_{e\tau}^2$ &$ 0 $\\[.3cm]
      \hline
                     &&&&&\\[-6pt]
$\begin{matrix}\text{Charm-II, CCFR,} \\ \text{NuTeV}\end{matrix}$  & $\nu Z \to \nu \mu\mu Z$&$g_{B-L}^2$& $g_{\mu e}^2$&$ 0$ &$ g_{\mu \tau}^2 $\\[.3cm]
      \hline
\end{tabular}
\caption{\label{tab:allLimits} Coupling strengths for the different gauge groups relevant at neutrino experiments. Note that the hidden photon of a secluded $U(1)_X$ does not have any neutrino couplings and therefore is not constrained by these experiments.}
\end{table}

 \end{document}